\documentclass[10pt]{article}
\usepackage{amsmath,amsthm,amssymb}
\usepackage[cp1251]{inputenc}
\usepackage[russian]{babel}
\usepackage{graphicx}
\usepackage[large]{caption2}
\usepackage{indentfirst}
\usepackage{amsmath}
\usepackage{afterpage}

\usepackage{epsfig}
\usepackage{epstopdf}

\usepackage[usenames]{color}

\usepackage{amssymb,amsmath}
\usepackage{amsfonts}
\usepackage{epsfig}
\usepackage{cite}
\let\svdot.
\catcode`.=\active
\def\NoBibDots#1{\let.\relax#1\catcode`.=12}
\catcode`.=12

\def\gee{ \, \lower 1mm\hbox{$\,{\buildrel > \over{\scriptstyle\scriptstyle\sim} }\displaystyle \,$}}
\def\lee{ \, \lower 1mm\hbox{$\,{\buildrel < \over{\scriptstyle\scriptstyle\sim} }\displaystyle \,$}}
\def\|{\partial}
\def\o {\over}
\def\Oo {\displaystyle}
\def\varkappa {{\scriptstyle\partial}\! e}

\def\namepoint{$\!\!\!\!\!\!\!{\textbf{.}}\,$}

\let\b=\baselineskip
 \renewcommand{\baselinestretch}{1.25}
\begin{document}\large
\renewcommand{\captionlabeldelim}{.}
\headheight 1.50true cm \headsep  0.7true cm \righthyphenmin=2

\parindent=10.5mm

\marginparwidth=20true mm

\begin{titlepage}

\noindent
 Zasov A.V., Saburova A.S., Khoperskov A.V., Khoperskov S.A. Dark matter in galaxies. Physics-Uspekhi (Advances in Physical Sciences), 2017, v.60, no.1,  pp. 3-39

 \noindent
 DOI: 10.3367/UFNe.2016.03.037751

\vskip 0.4\hsize

\centerline{{\Large A.V. Zasov, A.S. Saburova, A.V. Khoperskov, S.A. Khoperskov}}


\begin{center}{\bf
    Dark matter in galaxies
}\end{center}

{\large

УДК 524.6

}

 {\large Dark matter in galaxies, its abundance, and its
distribution remain a subject of long-standing discussion,
especially in view of the fact that neither dark matter particles nor dark matter bodies have yet been found. Experts'
opinions range from `a very large number of completely dark galaxies exist' to `nonbaryonic dark matter does not exist at
all in any significant amounts'. We discuss astronomical
evidence for the existence of dark matter and its connection
with visible matter and examine attempts to estimate its mass
and distribution in galaxies from photometry, dynamics, gravitational lensing, and other observations (the cosmological
aspects of the existence of dark matter are not considered in
this review). In our view, the presence of dark matter in and
around galaxies is a well-established fact. We conclude with
an overview of mechanisms by which a dark halo can influence
intragalactic processes.}

\end{titlepage}

{\large \b=1.3\b

\renewcommand{\baselinestretch}{0.90}
\thispagestyle{myheadings}
\renewcommand{\contentsname}{\Large Contents}
\thispagestyle{myheadings} \tableofcontents
\thispagestyle{myheadings}

}

\section{{Introduction}}
\label{sec-UFN_DarkMatter-Introduct}

\begin{figure}[!h]
\centering\includegraphics[width=0.5\hsize]{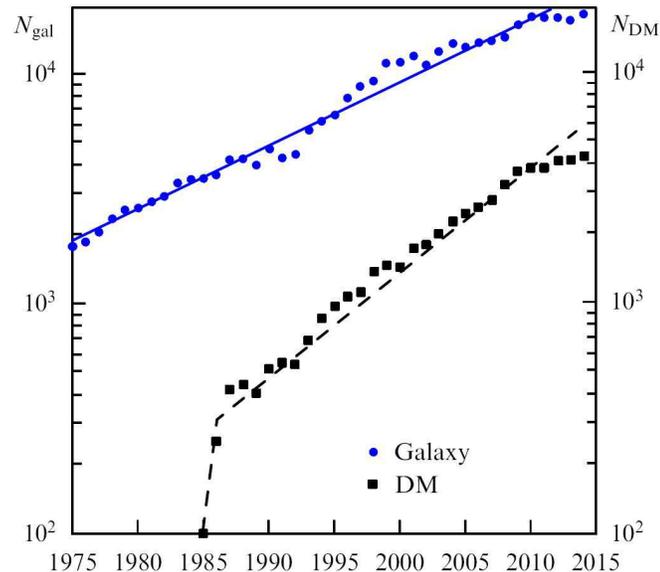}\vskip -0.\hsize
\caption{Dynamics of the number of papers per year devoted to galactic studies, $N_{gal}$ (key word ‘galaxies') and to dark matter (key words ‘dark matter' or ‘dark halo') according to NASA's Astrophysical Data System (ADS). A nearly exponential growth is seen: $N_{gal}\propto \exp(t$[year]\!$/15.8)$, $N_{DM}\propto \exp(t$[year]\!$/9.4)$.
}\label{Fig-Number-articles}
\end{figure}

Dark matter (DM) is one of the fundamental problems of modern astrophysics. It remains unsolved because the nature of DM is unknown and there is no clear understanding of DM relation with observed astronomical objects. Nevertheless, by the present time a lot of information has been obtained enabling the estimation of DM amount and distribution and the role it plays in the evolution of cosmic objects. The number of scientific papers directly or indirectly discussing DM ever increases (Fig.\ref{Fig-Number-articles}), which reflects a big interest to this field and steadily growing amount of observational data. Clearly, no review is able to pretend to grasp the whole problem.

In the first place, we should precise the term ‘dark mass’ or ‘dark matter’.

All cosmic bodies and media are sources of radiation in some spectral range, although they can appear ‘dark’ in the optical (for example, dark nebulae identified with molecular clouds). Dark matter or dark mass is traditionally referred to a medium that manifests itself only through gravitational interaction with ‘visible’ object, although it is not possible to exclude its dim glowing, for example, in gamma-rays due to DM particle annihilation, which is being searched for in observations (see, for example, \cite{2013AcPol..53..545M}). Here the notion of dark matter does not reduce to some exotic or low abundant in nature forms of matter, since its total mass in galaxies and their systems  must significantly exceed that of all visible, ‘light’ matter.

Here and below the term ‘light’ will be referred to the observable matter consisting of atoms (baryonic matter), which includes all types of stars, rarefied  interstellar and intergalactic gas, as well as small bodies, including small dust particles, i.e. all matter consisting of baryons (protons and neutrons). There are various estimates of both DM density and different opinions on its role in the evolution of observed galaxies and their systems. There are all grounds to believe that in many galaxies the DM density much exceeds that of baryonic matter even within the optical boundary limits, although in the literature opinions can be met that DM is unnecessary or at least its role is strongly exaggerated. In any case, the existence of large DM masses can be considered to be a well-motivated hypothesis, which will persist until DM particles (or bodies) will be reliably detected.

The notion of dark (hidden) mass and its historical development have been discussed in the last years by many authors.  Different aspects of this problem can be found in Refs ~\cite{1970Sci...170.1363O...Oort-1970!Galaxies-Universe, 2013BrJPh..43..369E...Einasto-2013!History-DM, 2013pss5.book.1091T...Trimble-2013!History-DM, 2012JMPh....3.1152R-Roos-Cosmological-Probes-DM}.

F. Zwicky was apparently the first to justify the existence of hidden mass in galaxy clusters ~\cite{1937ApJ....86..217Z...Zwicky-1937!Masses-Nebulae}, and more than 60 years ago in paper ~\cite{1953PhT.....6....7Z...Zwicky-1953!dark-matter}, the term ‘dark matter’ in the modern sense was first utilized in the title of the paper.  However, Zwicky considered velocity of galaxies in clusters and not internal motions in individual galaxies. Nearly at the same time, J. Oort found the mass deficit to explain the attraction force to the Galaxy disk, although his estimates were not very reliable from the modern point of view. In addition, Oort could not take into account the mass of gas in the Galactic disk. Later it became clear that the observed mass deficit is the general property of all galactic systems, and the larger the scale of the system the larger should be the mass of dark matter required for its virialization.

 The modern stage of the dark matter concept started in the 1970s caused by the necessity to explain gas kinematics in massive galaxies – the extended rotation curves of galactic gas components that does not decrease even beyond the stellar disk limits~\cite{1974Natur.250..309E...Einasto-etal-1974!massive-coronas,1976ApJ...208..662P...Peterson-etal-1976!Rubin-Vrot,1978ApJ...225L.107R...Rubin-etal-1978!Rot,1985ApJ...289...81R...Rubin-etal-1985!Rot}. The need for introduction of an additional mass arose also from modeling mass distribution in elliptical and dwarf spheroidal galaxies that have no disks (see Sections 2.3, 3.3).

 The discovery of hierarchical structure of the Universe in the 1980s  \cite{1982Natur.300..407Z...Zeldovich-etal-1982!Giant-voids}, including superclusters and giant voids, and numerical modelling of the structure formation in the expanding Universe gave fresh impetus to the development of DM ideas. The bulk of cosmological data in the commonly adopted model of the expanding Universe suggest the dominance of dark mass over the baryonic matter: the DM mass fraction is about a quarter of the total mass, including dark energy, and the baryonic fraction is merely around 4--5\%, which in turn is by an order of magnitude higher than the total mass of luminous matter concentrated in galaxies. Thus, both DM and most of the baryonic matter remains undiscovered by direct observations; apparently, the latter is comprised in hot gas in dark halos and in the intergalactic gas -- there is simply no place to hide.

 Numerical models of galaxy and their system formation assume that galaxies arose from DM dominated medium. A galaxy itself, more precisely, its baryonic part, forms at the ‘bottom’ of the potential well  created by collisionless DM inside a three—dimensional dark halo. Therefore, what we are used to call galaxy is located deep inside an extended DM halo. The theoretical size of this halo (the so-called virial radius $R_{vir}$) is about one order of magnitude as large as the observed optical diameter.  Therefore, the modern galaxy models also include an extended massive halo in addition to baryonic components, although part of DM must be present within  the optical limits of galaxies, including inside the star-gas disks.

 In the present review we will mainly focus on astronomical evidences of DM related to gravitational action of DM on baryonic matter, and on the DM mass estimates, leaving apart the issue of the nature of non-baryonic DM and experimental searches for DM particles and their possible annihilation radiation; the interested reader is referred to, for example, reviews \cite{2015EPJWC..9601027S...Schumann-2015!Dark-Matter-Hadron-Fusion-Physics, 2013IJMPA..2830022B-Dark-Matter-Investigation-Dama, Blinnikov-2014Mirror-matter-DM, 2008PhyU...51.1091R-dark-matter-particles, 2014PhyU...57....1B-clumps-DM}.

We note the so-called alternative hypotheses aimed at explaining the observations without invoking DM hypothesis. First of all, these are theories based on a modified Newtonian gravity, in which the gravitational acceleration $a_g$ deviates from the classical Newtonian expression $a_N = GM/r^2$, and at large $r$ significantly exceeds $a_N$, asymptotically approaching the value $(a_N\cdot a_0)^{1/2}$, where the constant $a_0\approx 10^{-8}cm/s^2$. This is the so-called Modified Newtonian Dynamics (MoND) (see \cite{2011IJMPD..20.2749D...Dodelson-2011!Problem-MOND, 2013AJ....146...48C...Carignan-etal-2013!NGC3109-Kinematics-Mass, 2016MNRAS.455..476S...Santos-Santos-etal-2016!distribution-mass-components, 2007PhyU...50..380O...Okun-06}  and references therein). An attractiveness of this approach is based on the natural explanation of the dependence of the luminous (i.e. baryonic) mass of galaxies on the radial velocity corresponding to the rotation curve plateau (the so-called Tully-Fisher baryonic relation), which indeed is close to $M_{bar} \propto V^4$. Other forms of the gravitational potential different from the classical expression have also been proposed \cite{2012MNRAS.423..141C...Cardone-etal-2012!MOND}: $\Phi(r)=-Gm/r + \Phi_{NN}(r; m)$ or $\Phi(r)=-Gm/(r(1+\delta(r)))$.

From the theoretical point of view, the main problem of these models is the use of additional hidden parameters which are absent in the classical theory. The MoND uses the parameter $a_0$, Horava-Lifshitz theory ~\cite{2012MNRAS.423..141C...Cardone-etal-2012!MOND} uses four parameters for function $\Phi_{NN}(r; m)$; the similar situation occurs in the conformal gravity theory \cite{2012PhRvD..85l4020M...Mannheim-OBrien-2012!conformal-gravity} and in other theories exploiting diverse gravitational potentials. However, only observations of galaxies and their systems can play a decisive role in testing different approaches, which in some cases revealed serious discrepancies of data with MoND predictions.

The discussion and critical analysis of the modified models from both theoretical and observational point of view can be found in papers \cite{2013PhRvD..88j3501C...Chan-2013!MOND, 2011IJMPD..20.2749D...Dodelson-2011!Problem-MOND, 2013AJ....146...48C...Carignan-etal-2013!NGC3109-Kinematics-Mass}. In particular, the MoND approach has met serious problems when interpreting gravitational fields derived from weak lensing observations of galaxy clusters filled with X-ray emitting gas, in which accelerations are very small. The triaxiality or, generally, lack of spherical symmetry in the gravitational potential distribution at large distances from galaxies are also difficult to bring in agreement with the simple modification of the gravity law. Note that the apparently increasing inconsistency between the Newtonian and real motion of bodies with decreasing velocities can be reproduced in numerical simulations of galaxy formation in the standard $\Lambda$CDM (Lambda-Cold Dark Matter) model without using additional universal constants ~\cite{2016MNRAS.455..476S...Santos-Santos-etal-2016!distribution-mass-components}.

Among non-standard DM theories, the mirror DM theory should be mentioned. In this theory, DM particles are identical to those of the ordinary matter but have ‘mirror quantum properties’ without being anti-particles. The ‘mirror partners’ cannot interact directly to the ordinary matter but via gravitational attraction; however, with themselves the mirror particles should interact exactly as ordinary particles (see the discussion in  \cite{2007PhyU...50..380O...Okun-06, Blinnikov-2014Mirror-matter-DM}).Note also a theoretical possibility considered by S.I. Blinnikov and co-authors that DM exists in the form of large amount of collapsed bodies from anti-matter (collapsed low-mass stars from anti-matter), which could be produced in the very early Universe under certain assumptions   \cite{2015PhRvD..92b3516B...Blinnikov-2015}. Some papers consider DM models consisting of primordial black holes (see the discussion of this model and its applications in \cite{2013PhRvD..87l3524Capela-BH}). Other exotic DM models are also considered in the literature, however their observational tests are difficult.

Different papers examine the possibility that a significant fraction (if not all) of DM inside galaxies could comprise unobservable and thus unaccounted for baryonic mass related to very cold molecular hydrogen \cite{1994A&A...285...79P...Pfenniger-Combes-1994!dark-matter-cold-gas}, or another medium distributed like the interstellar gas (see, for example,  \cite{2001MNRAS.323..453H...Hoekstra-etal-2001!neutral-hydrogen-DM}). This conclusion about the relation of DM to observable gas in galaxies was, however, criticized in paper  \cite{2013AstL...39..291Z...Zasov-Terekhova-2013!neutral-hydrogen-dark-mass}. The unaccounted for and hence not directly registered interstellar gas indeed exists in our and other galaxies (the so-called dark molecular and atomic gas), as suggested by indirect observations, but the amount of this gas is far insufficient to explain the required mass of DM.

We stress, at first, that the analysis of observations in the present review is carried out exclusively using the classical Newtonian potential. At second, the nature of the unseen component in our discussion of observational manifestations of DM is not very important. The additional gravity from DM can be partially due to unseen baryonic components, but cosmological data suggest that the bulk of DM should have non-baryonic origin.

Vast observational data obtained by ground-based space telescopes in the last decade suggest that in the frame of fundamental physical models the conclusion of the existence of DM in galaxies and beyond seems to be inevitable. However, the existing estimates of DM properties and  its mass and density distribution in galaxies have been quite contradictive so far, and the DM effects on internal processes in galaxies, although being extensively discussed, remain poorly understood. Observations of galaxies showing DM are complemented by another approach based on the construction of numerical dynamical models of stellar, gas and star-gas gravitating systems with ever increasing reliability and credibility. In the so-called  $N$-body models, the number of particles already approaches the number of stars in galaxies and is far above the number of stars in large globular clusters and dwarf galaxies ($N\sim 10^7 -10^9$ and more). We show in the present review that the use of numerical simulations enables us to better understand the mass distribution in galaxies and to follow their dynamical evolution, where DM should play a large, if not decisive, role.

In the present review, we will consider the main arguments in favor of the existence of non-baryonic, non-relativistic DM and different methods of its finding in galaxies, as well as its relation to characteristics of galaxies, without delve into cosmological aspects of the problem, which are widely discussed in the literature.

\section{Mismatch between kinematic and photometric mass estimates}
\label{sec-UFN_DarkMatter-1Curves}

The problem of existence of dark or hidden mass in nature became actual when massive measurements of galactic velocities made it possible to apply the virial theorem (the modulus of the gravitational energy of a system is equal to the total kinetic energy of its components) for pairs, groups and clusters of galaxies. However the fact that we measure only one velocity component of each system’s member (excluding several nearest galaxies in which proper motions of stars are measured) makes the mass estimate model-dependent. Nevertheless, the difference between the dynamical mass of the entire system and photometrically determined mass of stars is too large to be ascribed to the estimate errors.

Measurements of the velocity of motion of galactic satellites in pairs confirmed that the dark halo can extend far beyond the optical boundary and fully dominate in the regions almost free of stars and gas. The integral mass of galaxies with account for DM in these remote regions can be by an order of magnitude larger than total visible mass. For example, the mean ratio of  mass to the total visible luminosity in the K infrared band (2.2~$\mu cm$) in galactic pairs,  determined for more than 500 pairs, turns out to be  very high for a stellar population –- only 11 solar units \cite{2008AstBu..63..299K-Karachentsev-Makarov}, and for galaxy groups --– more than two times as high: $M/L_K$=26 solar units \cite{2008IAUS..244..370M-Makarov-Karachentsev}. For comparison, models of a purely stellar  old population  give $M/L_K \lesssim 1$.

The mass of the Local group of galaxies, apparently, should be determined more precisely. This mass is mainly comprised in two galaxies: the Andromeda nebula and the Milky Way, however the motion of  Sun relative to the Local group barycenter is not known precisely. By assuming that total momentum of galaxies in the Local group is close to zero, Diaz et al  \cite{2014MNRAS.443.1688Diaz-mass-Local-Group} obtained for the total mass of the Milky Way $0.8\pm 0.5\cdot10^{12}M_\odot$ and that of  М\,31 $1.7\pm 0.3\cdot10^{12}M_\odot$, respectively, which much exceed the total mass of stars and gas in these galaxies. This conclusion for nearby galaxies is confirmed by diverse estimates of DM to baryonic mass ratio  \cite{2014AJ....148...50Karachentsev-Masses-Nearby-Galaxies}.
 Various estimates of the mass of the Local group are given in \cite{2009MNRAS.393.1265K-Karachentsev-Hubble-flow}. From the analysis of the velocity field, the authors  $1.9\pm 0.2 \cdot 10^{12} M_\odot$.
This is apparently the most accurate mass estimate at present. It exceeds by many times the total mass of observed stars and gas in the Local group of galaxies.

Estimate of the dark matter to stellar mass ratio within the optical limits of an individual galaxy is less uncertain than the integral mass estimates, however here we a priori do not know neither the dark halo form nor its radial density profile.

In this Section, we will consider first of all  galaxies with comparatively thin rotating star-gas disks. These galaxies include lenticular, spiral and irregular galaxies (of types S0, S, and Irr, respectively), excluding dwarf irregular galaxies (dIrr), in which the disk thickness is comparable to their radii and the stellar and gas velocity dispersion is comparable to the rotational velocity.

The disk density distribution in the considered galaxies is the sum of the surface density of stars, gas and DM inside the disk: $\Sigma(r) = \Sigma_*(r)+\Sigma_g(r)+\Sigma_{dm}(r)$. The last term is taken into account only if the disk comprises a sizable fraction of DM.

The DM mass estimate in a galaxy reduces to separating from the total galaxy mass its ‘luminous’ (i.e. baryonic) components, stars and gas. The density of cold gas  $\Sigma_g(r)$ can be estimated the most reliably, since it is derived directly from the gas radio emission. Here the problem of account for the ‘dark’ gas remains, whose radiation is not detected due to its low temperature or large optical depth of clouds in radio lines, and for molecular gas – also due to photodissociation of molecules which are used to estimate its amount.  However, in most cases stars and not gas mainly contribute to the disk density. Photometry enables the determination of brightness and color distribution in a galaxy: from the center, near which the bulge mostly contributes to the brightness and stellar mass, to a far disk periphery, where the surface brightness is several hundred or thousand times as small as in the center.

The radial brightness distribution in the stellar disk $I(r)$, as a rule, is described by the exponential law,  $I(r)=I(0)\cdot \exp(-r/r_d)$, where $r_d$ is the radial disk scale. The disk size, as well as the size of the entire galaxy, is a conventional quantity, since there are no sharp boundaries. For certainty, the galaxy radius  is usually assumed to be the effective radius $R_e$, comprising half integral luminosity, or the so-called photometric (optical) radius $R_{opt}$ corresponding to a certain isophote. The latter is usually related to the 25th  stellar magnitude per square second in the B-band (this is close to the galaxy size visible on a good image), although stellar disks extend far beyond.

In the normal-brightness galaxies $R_{opt}$ is $(3-4)\cdot  r_d$, however there is a whole class of low-surface brightness disk galaxies (LSB-galaxies) with a much smaller $R_{opt}/r_d$ ratio. Frequently, beyond the optical radius the stellar density decreases with  $r$ much steeper than at smaller $r$, nevertheless there are objects for which the exponential brightness law holds up to 10 radial scales $r_d$ (for example, for NGC 300 \cite{2011IJMPD..20.2749D...Dodelson-2011!Problem-MOND}), or galaxies in which, in contrast, the brightness decrease at large $r$ becomes even more flat  \cite{2012A&AT...27..313Ilyina}. The rotation curves obtained from optical observations rarely reach $R_{opt}$,  but in gas-rich galaxies, it is possible to follow the disk rotation by the neutral hydrogen line (HI) up to several $R_{opt}$.

In the 1980s, when the after not very precise at that time radio measurements the progress in optical observations enabled estimations of rotation velocities of several galaxies at different distances from centers, it became evident that the rotational velocities remain high even at far disk outskirts. This led to the conclusion that there is a missing mass, apparently forming a dark halo, whose gravitational field is responsible for the high circular velocities. Note that observations carried out on the 6-m BTA telescope of the Special Astrophysical Observatory of The Russian Academy of Sciences (SAO RAS) significantly contributed to studies of disk kinematics of S-galaxies  \cite{2011BaltA..20..363A}.

 The wide-spread opinion is that the extended flat parts of the rotation curves $V(r)$, found in many disk galaxies, themselves directly suggest  the presence of dark halos. In fact, it is not exactly the case, and theses flat parts can be explained without invoking the dark matter hypothesis. A smooth rotation curve without sharp local velocity gradients of any form can be always explained either by assuming that almost all the mass is comprised in a spherical halo, or by assuming that the halo is absent, and there is only one axially symmetric disk with certain radial density law (any intermediate variants are also allowed).

 The first case is trivial: for a spherically symmetric mass distribution, the total mass $M(r)$ within radius  $r$ reflects the change of the radial velocity $V_c(r)$ along the radius: $M(r)=rV_c^2/G$, and for any radial velocity distribution, unless it drops faster than $r^{-1/2}$, the corresponding mass distribution can be found.

 The second case is not so obvious: the relation between  $M$ and $r$ for the disk is more complicated, as the gravitational potential estimate at a given radius requires the knowledge of density distribution at all  $r$, and then the disk parameters can be chosen such that the resulting rotation curve be close to the observed one. The classical example is the so-called Mestel disk. This is a thin disk in which the local surface brightness decreases inversely proportional to the distance to the center: $\Sigma (r) = \Sigma (r_d)r_d/r$, where $r_d$ is the radial disk scale. For the Mestel disk, the calculated circular velocity does not change with radius  $r$, i.e. the rotation curve represents a horizontal line.

 Why then a dark halo is required to interpret rotation curves demonstrating flattening at large distances? The answer is simple: the surface density of the luminous baryonic matter (stars + gas) in real disks does not follow the $1/r$, law but decreases much faster, exponentially as a rule. Thus, the presence of dark matter is suggested not by the form of the rotation curve itself but by the inconsistency with the expected form if all gravitating matter in the galaxy were comprised only in stars and gas.  To estimate the DM contribution to the mass and rotation curve it is necessary to find the method of ‘calculation’ of the baryonic mass contribution to the rotation curve. This important problem can be solved in different ways and frequently with not desired accuracy.

\subsection{Rotation curves of gas beyond the optical radius}\label{sub-UFN_DarkMatter-Curves-out}

As most of the disk mass is comprised in stars, in the first approximation we can assume that the surface density follows the surface brightness, i.e. that the stellar mass  ${M}$ to light ${L}$ ratio is approximately constant: ${M}/{L}\approx\textrm{const}$. This relation the most precisely holds for near infrared (IR) luminosity, since most of the stellar population mass (for very rare exceptions) is comprised in old stars that mainly contribute at these wavelengths. Therefore, the surface density of the stellar disk is well approximated by the same law as for the surface brightness: $\Sigma_*(r)=\Sigma_0\exp(-r/r_d)$ (the discussion of the observed deviations from the exponential law can be found in \cite{2006A&A...454..759P...Pohlen-2006!structure-galactic-disks, 2015MNRAS.453.2965B...Barbosa-etal-2015!photometry-nonexp}). With account for the gas component, if its mass is significant, this makes it possible to use photometry to calculate the rotation curve of the baryonic disk component and, by matching it with observations, to estimate the effect of dark matter on the disk dynamics.

\begin{figure}[!h]
\centering{\includegraphics[width=0.5\hsize]{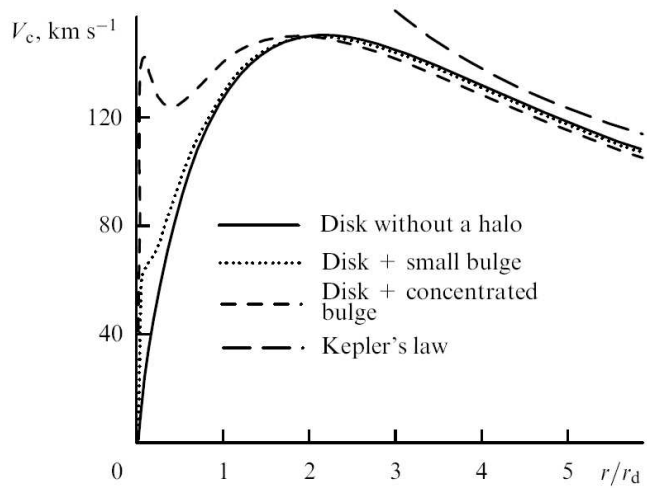}}
\vskip -0.\hsize
\caption{Rotation curves in the absence of a halo }
\label{fig-Vrot(r)-noHalo}
\end{figure}

 The calculation of the rotational velocity $V_c(r)$, by assuming that the gravitational potential $\Psi(r)$ is produced by a disk with exponential density drop, shows that the circular rotation curve demonstrates a specific maximum at the distance $r\simeq 2.2\,r_d$ from the galaxy center and monotonically decreases to periphery (Fig.~\ref{fig-Vrot(r)-noHalo}). In fact, the rotational velocity at large $r$ in real galaxies does not follow this curve: as a rule, it remains at nearly the same level or continues increasing. The contribution from the bulge into the gravitational potential of the galaxy increases the rotational velocity in the inner part of the disk, but to maintain the high velocity of rotation at large $r$ compared to that expected for an exponential disk, additional mass is required  which is not related to stars or observed gas. It is this mass that is ascribed to DM.

In the absence of significant non-circular gas motions, the equilibrium condition can be written in the form
\begin{equation}\label{eq-1-balans-forces}
    \frac{V_c^2}{r} = \frac{d\Psi}{dr}+c_s^2\frac{d\Sigma_g}{\gamma\Sigma_g\,dr}\,,
\end{equation}
where $V_c$ is the circular velocity (which is usually close to the gas rotational velocity $V_g$), $\Sigma_g$ is the gas surface density, $c_s$~is the adiabatic sound velocity, $\gamma$~is the adiabatic exponent, $\Psi(r)$ is the gravitational potential that is determined from the known mass distribution in the galaxy. The last term in the right hand side of equation (\ref{eq-1-balans-forces}) takes into account the deviation of the observed rotational velocity from the circular one, which arises due to thermal velocity of gas atoms or molecules. In fact, the local gas velocity dispersion is mainly caused by not thermal motion of particles but by the turbulent gas motions or, if we consider the rotation of the stellar and not gas disk, by the local stellar velocity dispersion. In these cases, the sound velocity square $c_s^2/\gamma$ is changed by the velocity dispersion square along the radial coordinate $c_r^2$. For the rotating gas the contribution of this term in (\ref{eq-1-balans-forces}), is small as a rule since  $c_r^2\ll V_g^2$ (the characteristic value $c_r/V_g\simeq 0.05-0.1$), and then the gas rotation velocity $V_{g}$ is almost equal to $V_c$. Therefore, one often neglect the difference between these two velocities (however, this is not always true for stars \cite{Fridman-Khoperskov-2011!book}).

\begin{figure}[!h]
\centering{\includegraphics[width=0.7\hsize]{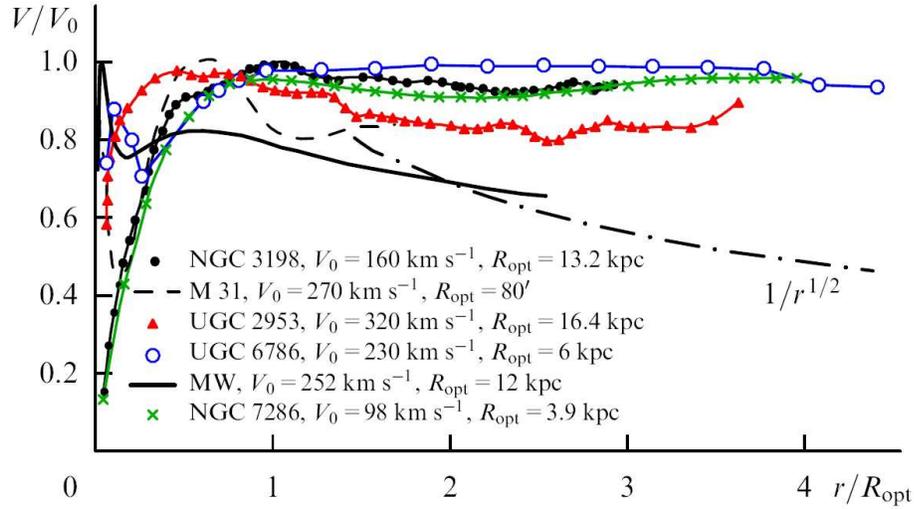}}
 \vskip -0.\hsize
\caption{(Color online.) Rotation curves of spiral galaxies normalized to the rotation velocity maximum. The distance from the center is in units of the optical radius $R_{opt}$. The long-dashed line shows the Keplerian curve corresponding to the case where all the mass is in the central region. Numerous examples of extended rotation curves can be found in \cite{2015PASJ...67...75S...Sofue-2015!DM-M31, 1983IAUS..100....3R...Rubin-Athanassoula-1983!book, 2014IJMPD..2330005D...DelPopolo-2014!Dark-Matter-Cosmology} color on-line}
\label{fig-Vrot(r)-normal_1}
\end{figure}

In Fig.~\ref{fig-Vrot(r)-normal_1} we show examples of rotation curves $V_g(r)$ with plateau for S-galaxies as constructed from gas velocity measurements. Without massive dark halos, to explain the rotation curve of a Milky-Way (MW) like galaxy, for which $r_d\simeq 3$~kpc, $V_{\max}\simeq 220$~km/s we should assume very strong deviations from the exponential density distribution in the outer disk zone at $r>2r_d$, which increase from the center.

\begin{figure}[!h]
\centering\includegraphics[width=0.6\hsize]{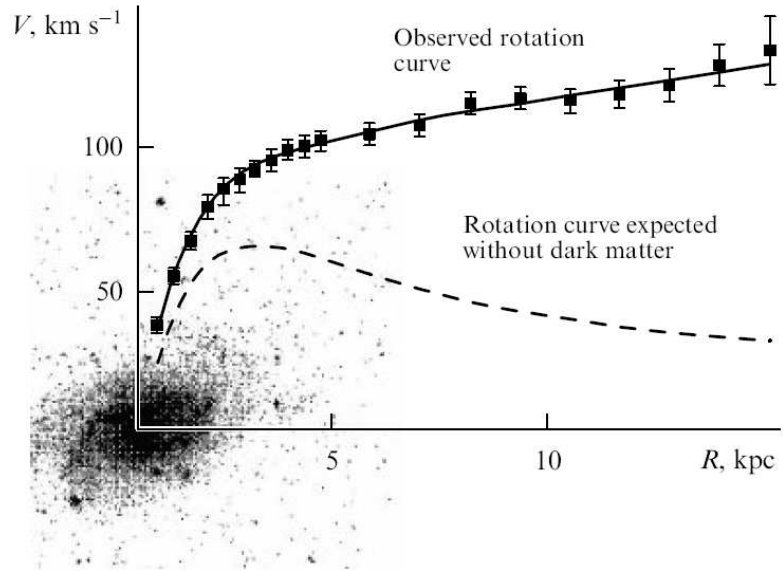}\vskip -0.\hsize
\caption{Decomposition of the rotation curve of the M\,33 galaxy suggesting the DM dominance in the region inside the optical radius.}
\label{fig-m33}
\end{figure}

In particular, rotation curves of two nearby spiral galaxies (M31 and M33) behave differently: the rotation curve of M31 (the Andromeda Nebula) does not decrease up to 150 angular minutes from the center \cite{2006PhT....59l...8R...Rubin-2006!Dark-Matter-Andromeda} with the optical radius of the galaxy being about 80 arc minutes, and remains constant up to 35 kpc  \cite{2009IAUS..254P..73T...Tamm-etal-2009!M31}, whereas the rotation curve of M33 (the Triangle Nebula) keeps slowly increasing even at distances of several $r_d$ (Fig.\,\ref{fig-m33}), so that the contribution of dark halo into the rotation curve of the galaxy becomes dominant already within the observed optical limits \cite{2012AstL...38..139S}.

\begin{figure}[!h]
\centering\includegraphics[width=0.5\hsize]{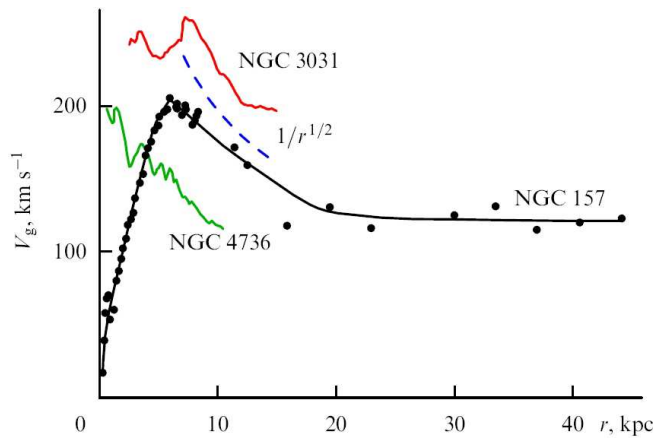}\vskip -0.\hsize
\caption{Examples of rotation curves of spiral galaxies with a decreasing rotational velocity inside the optical boundaries: NGC\,157 \cite{Ryder1998}, NGC\,4736, NGC\,3031  \cite{2008AJ....136.2648D...deBlok-etal-2008!Rotation-Curves-THINGS}. The dashed line is the Keplerian curve $\propto 1/\sqrt{r}$. }
\label{fig-v-ngc157etal}
\end{figure}

\begin{figure}[!h]
\centering\includegraphics[width=0.4\hsize]{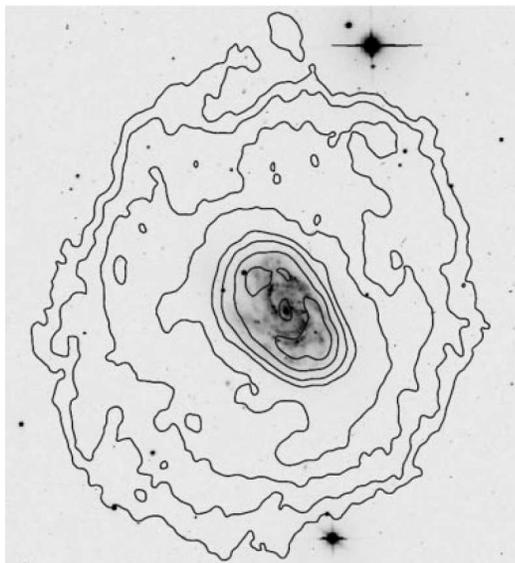}\vskip -0.\hsize
\caption{Gas disk of the galaxy NGC\,157 (shown by the surface isodensities) extends far beyond the stellar disk (the dark central region) and lies in the dark halo gravitational field.}
\label{fig-N157HI}
\end{figure}

There are only a few galaxies with rotation curves monotonically decreasing after the maximum (for example,  NGC\,157, NGC 4736, NGC 3031,  Fig.~\ref{fig-v-ngc157etal}), and this decrease, as a rule, remains more flat than the Keplerian law $V_g(r)\propto r^{-1/2}$, which could be asymptotically expected in the absence of dark matter (Fig.\,\ref{fig-N157HI}).

\subsection{Problem of the rotation curve decomposition: the maximum disk model and lower estimates of the halo mass}\label{sub-maxDisk}

The rotation curve  $V(r)$ can be decomposed into diverse galactic components. The bulge, stellar disk and halo mostly contribute to the gravitational potential of a galaxy; when necessary, the gas disk and central core can be also taken into account. The model rotation curve   $\displaystyle V_c(r)=\sqrt{\sum (V_{c}^{(k)})^2}$, where $V_c^{(k)}$ is the circular velocity of the $k$-th component, is matched to the observed rotation curve. The halo contribution is the most difficult to account for. As a priori halo density is unknown, different halo models are assumed, and the results are compared to the observed rotation curves. The following halo models are the most frequently used.

A quasi-isothermal halo model is characterized by the radial density profile:
\begin{equation}\label{4--Eq-densityhalo}
\varrho_h(\xi) = \frac{\varrho_{h0}}{(1+\xi^2/a^2)} \,,
\end{equation}
Here $\xi=|\vec{r}|=\sqrt{r^2+z^2}$~is the radial spherical coordinate, $\varrho_{h0}$~is a constant. This model corresponds to a nearly constant velocity dispersion of self-gravitating halo particles.

The Navarro-Frenk-White (NFW) ~\cite{1997ApJ...490..493N...Navarro-etal-1997!Model-halo}, model, which is based on the analysis of cosmological halo formation models:
\begin{equation}\label{Eq-3-halo-Navarro}
  \varrho^{(NWF)}_{h}(\xi) = \frac{\varrho_{h0}}{(\xi/r_s)(1+\xi/r_s)^2} \,,
\end{equation}
where $r_s$~is halo scale; the model has a cusp – a singularity at the center,  $\varrho^{(NWF)}_{h} \rightarrow \infty$.

The Burkert  model  \cite{1995ApJ...447L..25B...Burkert-1995!Structure-Dark-Matter-Halos}
 \begin{equation}\label{Eq-3-halo-Barket}
    \varrho_h^{(B)}(\xi) = \frac{\varrho_{h0}}{(1+\xi/b)(1+(\xi/a)^2)}.
\end{equation}

\begin{figure}[!h]
\centering\includegraphics[width=0.55\hsize]{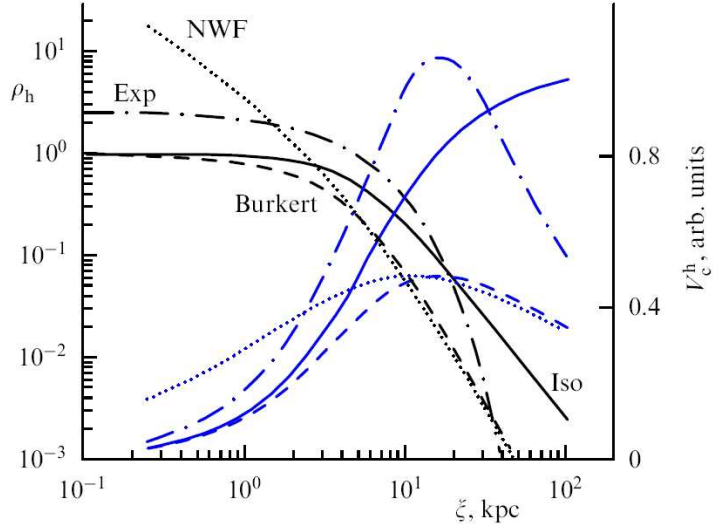}\vskip -0.\hsize
\caption{(Color online.) Comparison of different profiles of the volume density $\varrho_h(r)$ of a spherical halo (the halo scale $a=5$\,kps). The exponential profile (Exp) corresponds to Eqn (\ref{eq-rho-halo-einasto}) with $\alpha=1$. The Iso curve corresponds to an isothermal halo. The blue curves show the corresponding profiles of the circular velocity $V_c^{(h)}$.
}
\label{fig-denshalo(r)}
\end{figure}

  An exponential profile of the halo space density \cite{1997A&A...327..983F...Fux-1997!MW-I}
\begin{equation}\label{eq-rho-halo-exp}
    \varrho_h = \frac{M_{h}}{8\pi e b^3 }\exp\left\{ - \frac{\xi}{b} \right\}
\end{equation}
provides convergence of the integral  $M_h=\int_0^\infty \varrho_h\,4\pi
 \xi^2\,d\xi$ unlike distributions  (\ref{4--Eq-densityhalo}), (\ref{Eq-3-halo-Navarro}),  (\ref{Eq-3-halo-Barket}).
Sometimes the Einasto profile is used:
\begin{equation}\label{eq-rho-halo-einasto}
    \varrho_h = \varrho_0 \exp\left\{ -(\xi/r_s)^\alpha \right\} \,,
\end{equation}
where $\alpha$ is another (third) free model parameter. With decreasing $\alpha$ ($\alpha\ll 1$) the central density increases, which enables description of not only models with smooth central density increase, but also models with cusp. Different halo density profiles are presented in Fig.\, \ref{fig-denshalo(r)}.

Each of the halo models well describes the flat rotation curves with account for the baryonic components contribution. However, models deviate at small $r$, by showing diverse character of DM density increase towards the galaxy center. Models with cusp exhibit rapid central density increase, while the isothermal halo and Burkert halo models demonstrate a flat profile at the center. The quasi-isothermal halo density decreases the most slowly far beyond the optical radius, and therefore the difference of this model from other models is the most prominent at large distances, where there are no (for rare exceptions) observational data.

The usually smooth transition of the rotation curve from the inner region, where rotation is determined by the baryonic components, to the outer zone, where rotation is halo dominated, remains unexplained. Such a disk and halos self-consistency is observed in many cases, and, astonishingly, in galaxies with different disk and halo concentrations. This problem, also known as the ‘disk-halo conspiracy’, suggests a gravitational coupling between the disk and halo leading to their parameters correlation. Paper \cite{2010A&A...519A..47A...Amorisco-Bertin-2010!halos-embedding-thickness-disks}, discusses in detail this feature and proposes a self-consistent dynamical model consisting of a thin exponential disk and a non-spherical halo, which reproduces the flat rotation curve when transition between the components. In this model, in the transitional region between the disk and dark halo, the dark halo should be compressed towards the disk, i.e. should have a significantly non-spherical form.

We note that the numerical models of galaxy formation in the modern cosmological assumptions reproduce extended plateau in the rotation curves of galaxies without assuming dark halo non-sphericity (see, for example,  \cite{2016MNRAS.455..476S...Santos-Santos-etal-2016!distribution-mass-components}), although in paper \cite{2016MNRAS.455..476S...Santos-Santos-etal-2016!distribution-mass-components} the smooth transition of the rotation curve is apparently due to the halo dominance even near the maximum of the disk component.

To interpret rotation curves, one frequently uses the maximum disk solution method (MDS) aimed at determining the upper mass limit of the disk component at which its contribution to the circular rotation still does not contradict to the observed dependence $V(r)$, or the best fit method which minimizes the difference between the model and observed rotation curves. These approaches could provide information on the relative disk and bulge mass, however an unseen component (dark halo) with a priori unknown density distribution significantly complicates the problem.

\begin{figure}[htbp!]
\centering\includegraphics[width=0.7\hsize]{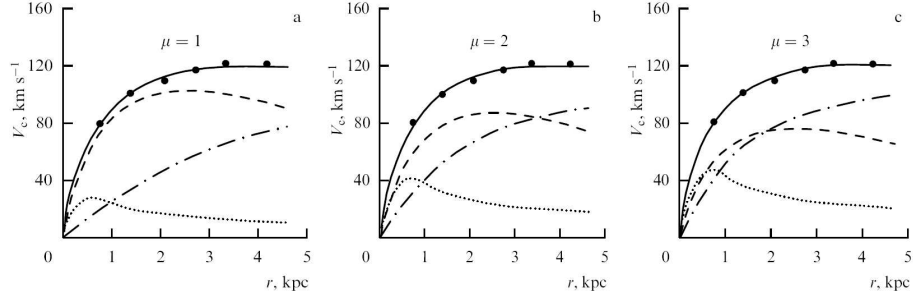}\vskip -0.\hsize
\caption{Radial model (solid curve) and observed (black dots) rotational velosity for NGC~6503 for various relative halo masses $\mu$:
 \textit{a}~---
$\mu = 1$; \textit{b}~--- $\mu = 2$; \textit{c}~--- $\mu = 3$.
Contribution of subsystems to the rotation curve are shown by the dotted line (bulge), dashed line (disk), and dashed-dotted line (halo)~\cite{2001ARep...45..180K}.
}
\label{fig-Vrot(r)-ngc6503}
\end{figure}

When decomposing the galactic rotation curve, the maximum disk model, as a rule, suggests that the disk can well dominate inside the optical radius of the galaxy (Fig. ~\ref{fig-Vrot(r)-ngc6503}а), however it is impossible to ignore dark matter within the optical boundaries of the galaxy, unless a model of halo with empty central part is seriously considered.

 As noted above, the disk component contributes maximally at the radius $r=2.2\,r_d$, where the circular velocity of the exponential disk $\Sigma_d\propto \exp(-r/r_d)$ reaches maximum. For the maximum disk model, this velocity is on average $\approx 85\pm 10$~\% of the circular rotational velocity at this distance: $V_c^{disk}/V_c=0.75\div 0.95$ \cite{1987AJ.....93..816K...Kent-1987!Vrot,1997ApJ...483..103S...Sackett-1997!MW-Vrot-maxdisk}. MDS can slightly overestimate the disk mass, and therefore it gives the lower limit of the halo mass. Inside the optical radius of the galaxy, $r\le R_{opt}$ the relative halo mass $\mu=M_h/M_d\simeq 0.4-0.7$. But already inside the radius $r\approx 2R_{opt}$, which is usually reachable by the measured rotation curve (see Fig.~\ref{fig-Vrot(r)-normal_1}), $\mu$ can be as high as 2-2.5 already in the maximum disk model. As a rule, at least in most cases, we cannot explain the observed rotation curves within the optical radius without invoking dark mass.

Non-uniqueness of the rotation curve decomposition is a serious problem when estimating galactic component masses from the observed rotation curve $V(r)$ even if the functional density radial dependence is known for each component. The point is that generally the observed rotation curve $V(r)$ can be explained using different relations between the disk and spherical component masses (see Fig.~\ref{fig-Vrot(r)-ngc6503}), and therefore their masses cannot be determined uniquely. Each galactic component is described as a minimum by two parameters with the dimension of density and length characterizing the conventional size of the component.  Three main components (bulge, disk and halo) require, as a minimum, six parameters that must correspond to the observed rotation curve. Other parameters can include, for example, the degree of oblateness of the components or parameters that describe a complex density profile of the star-gas disk. The larger the number of parameters, the more precisely the theoretical rotation curve can fit to the observed one, but then the multidimensional parameter space increases.

This problem can be circumvented or at least smoothened by taking into account in the model, in addition to the obvious requirement of increasing the precision of the curve form, the additional a priori information about the galaxy, which is done in practice in almost all cases. First of all, it is the relation of the galaxy brightness distribution with the surface (column) density of the stellar disk population. The proportionality coefficient is the ratio  $M/L$,which is assumed to be known or is determined from the stellar population model from spectral or color measurements of the galaxy (after taking into account the selective light absorption by dust). In the best-fit model, the $M/L$ ratio when interpreting galactic rotation curves is treated as free parameter (see, for example, decomposition of rotation curves from the THINGS (The HI Nearby Galaxy Survey) galaxy survey for  fixed and ‘free’ $M/L$ ratio in \cite{2008AJ....136.2648D...deBlok-etal-2008!Rotation-Curves-THINGS}).

For gas-rich galaxies, in addition to the brightness distribution, the gas density distribution in the disk obtained from radio observations should be given. To diminish uncertainties, estimates of the stellar velocity dispersion, disk thickness, its gravitational stability condition and other information are used, which narrow the range of possible solutions.

With account for the gas density distribution and additional information on stellar velocity dispersion in the disk,  it was confirmed that the disk mass is as a rule smaller than follows from the MMD model. For example, according to paper \cite{2013A&A...557A.131M...Martinsson-2013!dark-matter-spiral-galaxies}, in which rotation curves of 30 galaxies mostly of Sc and later types were analyzed with account for photometry, velocity dispersion and gas layer contribution to the rotation curve, the ratio between the ‘baryonic’ and total rotation curves near the maximum disk contribution to the rotation curve is less than 0.75 (the mean value is 0.6). Assuming that for MMD this ratio is on average 0.85, the disk masses in this model are overestimated  (the halo masses inside galaxies are underestimated, correspondingly) by about factor two.

The rotation curves in most cases are derived from gas velocity estimates. The rotational  velocities of the stellar disk are usually made only in the central part of the galaxy due to complex procedure of the Doppler shift estimate using absorption lines, which requires a high signal-to-noise ratio, since absorption lines are less contrast on the continuum background and are broader than emission lines. Nevertheless, in the inner disk region the accuracy of determination of radial velocity curve and hence rotational velocities from stars is sometimes better than from gas, since gas velocities are more frequently distorted by non-circular motions.

 Recently, using long-slit spectroscopy on the MMT (Multiple Mirror Telescope), ‘stellar’ rotational velocities were measured for more than 100 massive galaxies  \cite{2015MNRAS.451..878K...Kauffmann-2015!inner-rotation-curves}. The obtained statistical material enabled the authors \cite{2015MNRAS.451..878K...Kauffmann-2015!inner-rotation-curves} to conclude that the inner DM density profiles are different, and in DM-dominated galaxies the rotation curves are well explained by baryonic components and the NWF halo profiles (Fig\,\ref{fig-Vrot(r)-GASS}).

\subsection{Dwarf galaxies and low-surface brightness galaxies}

\begin{figure}[!h]
\centering\includegraphics[width=0.8\hsize]{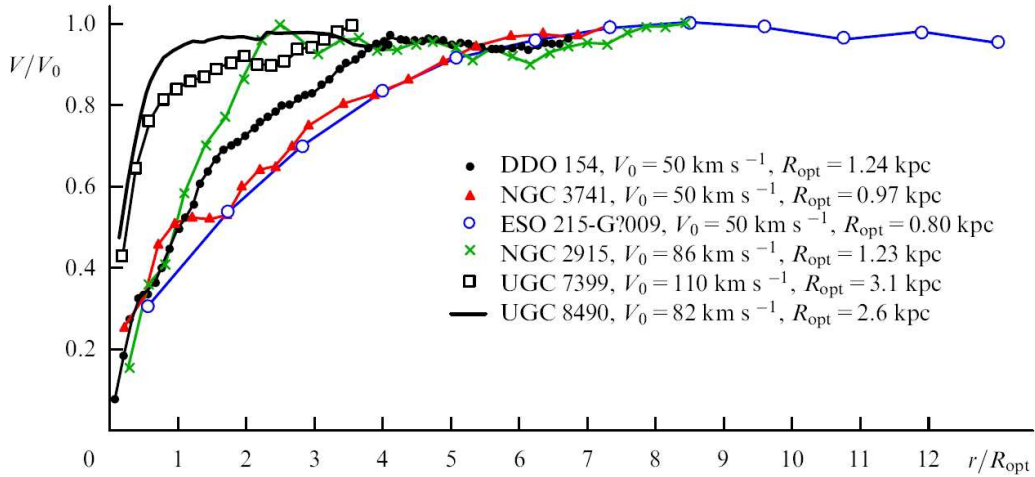}\vskip -0.\hsize
\vskip -0.\hsize
\caption{(Color online.) Rotation curves of dwarf galaxies according to \cite{2013MNRAS.429.2537Meurer-Disc-stability}. Shown are the rotation velocity $V_0$ and the optical radius $R_{opt}$.}
\label{fig-Vrot(r)-Dwarf}
\end{figure}

  The mass and luminosity are usually expressed in solar units. Without dark mass, the $M/L$ ratio for galactic stellar population, which is dominated by old stars, is a few (up to 10) in the B-band and (according to different estimates) and 0.5-1.0 in the IR K-band (2.2 mcm). Intensive star formation can decrease by factor of a few the $M/{L}$ in the B-band but remains almost intact the near IR luminosity, which is preferably used in stellar population mass estimates from photometric data. By comparing the dynamical and photometric integral estimates of the galaxy mass (the latter is obtained by multiplying the integral luminosity by the $M/L$), it is possible, by adding the interstellar mass gas, if present, to the stellar mass, to quantitatively estimate the DM contribution to the galaxy mass without analyzing its rotation curve. If there is gas in a dwarf galaxy, its rotational velocity can be estimated from the integral width of the HI line (with account for the projection effect) and to estimate $M/L$ inside the optical limits of the galaxy. In the absence of information about rotation curves, the error of such estimates is comparable to the estimated value itself, however the accuracy can be increased by considering a larger number of galaxies.

  The results show that the relative amount of DM in diverse objects is strongly different.  Rich statistical data can be found in the Updated Nearby Galaxy Catalog (UNGC) by Karachentsev et al. \cite{2013AJ....145..101K...Karachentsev-2014}, which contain data for several hundred galaxies at distances closer than 11 Mpc. More than in half dwarf galaxies containing gas the $M_{opt}/L_K$ ratio exceeds unity, suggesting large DM contribution, although the number of galaxies in which DM mass is much stronger than the stellar mass is small: just a few objects show $M_{opt}/L_K \ge 15$.

\begin{figure}[htbp!]
\centering\includegraphics[width=0.5\hsize]{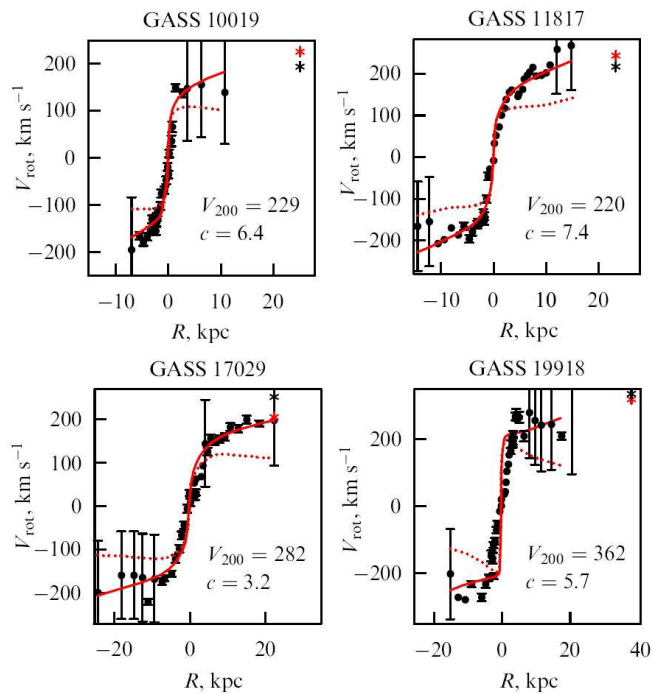}\vskip -0.\hsize
\caption{(Color online.) Decomposition of rotation curves for several DM-dominated galaxies. The dotted red lines show the baryonic matter contribution to the rotation curve. The black dots show the measured values, the dashed red lines mark the baryonic matter contribution, the solid curve is the model rotation curve with an NFW halo, and $c$ is the model parameter \cite{2015MNRAS.451..878K...Kauffmann-2015!inner-rotation-curves}.
}
\label{fig-Vrot(r)-GASS}
\end{figure}

The characteristic feature of many dwarf galaxies is the increase of the rotation curve of gas subsystem even beyond the optical radius (\cite{2013MNRAS.429.2537Meurer-Disc-stability, 2014AstBu..69....1Moiseev, 2012MNRAS.420.2034Salucci-Dwarf-kinematics-spiral}, Fig.\,\ref{fig-Vrot(r)-Dwarf}), which points to DM domination.

 DM mostly contributes into the integral mass in dwarf spheroidal galaxies (dSph) with a size of several hundred parsecs. Dwarf spheroidals belong to the smallest galaxies (except for rare and poorly studied ultracompact dwarfs). Low-luminosity dSph galaxies have the characteristic dynamical mass $M_{opt}\sim 10^7M_\odot$ as derived from the stellar velocity dispersion.  .  Large values of $M/L\sim 1/L$, which amount to several dozens and sometimes hundreds, definitely suggest dark matter domination. These galaxies have no rotating thin disks, and their density distribution is found from dynamically equilibrium models based on the velocity dispersion\,$c_{obs}$ measurements.  The typical values of the central stellar velocity dispersion $c_{obs}$ in such galaxies range from $c_{obs}\simeq 3-10$\,km/s (Segue 1 и 2, Hercules, LGS\,3, Carina, Sextans, Sculptor, Leo\,II, Leo\,IV, Bootes~1, Coma) to 20--30~km/s \cite{1998ARA&A..36..435M...Mateo-1998!Dwarf-Galaxies,2009AJ....137.3100W...Walker-etal-2009!dSph}. A more precise and less model dependent is the mass inside the region within the effective radius $R_e$, that comprises half of the integral luminosity: $M(R_e) \propto \nu~ R_e  c_{obs}^2$ , where the proportionality coefficient is $\nu\approx 580 M_\odot$ [$pc^{-1}km^{-2}s^2$]  \cite{2009ApJ...704.1274Walker-Universal-Mass-Profile-Dwarf}.

   The ${M}/{L}$ ratio in dSph-galaxies is much higher than that expected for galaxies without dark halo, with the empirical relation  $ log M/L_V = 2.5 + 10^7/L_V$ \cite{1998ARA&A..36..435M...Mateo-1998!Dwarf-Galaxies}, where $L_V$~is the luminosity in the $V$ photometrical band, and can lie within the range 10-1000 \cite{2006ApJ...650L..51M...Munoz-etal-2006!Dwarf-Spheroidal, 1998ARA&A..36..435M...Mateo-1998!Dwarf-Galaxies}. That the  $M/L$ ratio is inversely proportional to $L$ means that masses of these dwarfs, mostly due to DM, lie within a much narrow range than their luminosities  \cite{2009ApJ...704.1274Walker-Universal-Mass-Profile-Dwarf}.

Low Surface Brightness (LSB) galaxies also have very low mas disks compared to the total mass, assuming that the initial stellar mass function in these galaxies is about the same as in galaxies with normal surface brightness, which is generally not obvious   \cite{2011ARep...55..409S...Saburova-2011!LSB, 2003Ap&SS.284..719F...Fuchs-2003!Massive-disks-low-surface}. These galaxies are surrounded by very massive extended dark halos, although part of DM can be in the disks as well \cite{2011ARep...55..409S...Saburova-2011!LSB, 2003Ap&SS.284..719F...Fuchs-2003!Massive-disks-low-surface}. Neutral hydrogen in these galaxies, as a rule, dominates over the stellar component.

\begin{figure}[!h]
\centering\includegraphics[width=0.4\hsize]{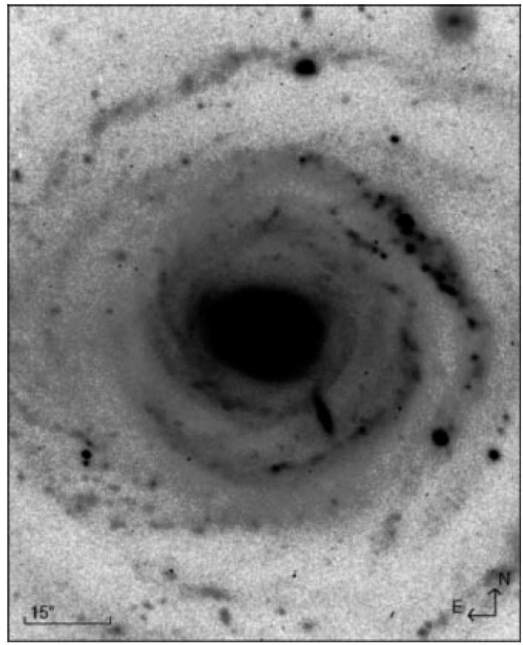}\vskip -0.\hsize
\caption{One of the largest LSB galaxies Malin-2 (Gemini-North telescope)(see \cite{2014MNRAS.437.3072K...Kasparova-2014})}
\label{Fig-malin2}
\end{figure}

LSB-galaxies include both dwarfs and giant systems with radius as large as 100 kpc and larger. Despite large difference in sizes, LSB-galaxies have, as a rule, the rotation curves extending far beyond the optical disk limits \cite{2001AJ....122.2396D...deBlok-etal-2001!Rot-LSB}. Of special interest are LSB-galaxies with giant mass and size, such as Malin-1, Malin-2 (Fig.\,\ref{Fig-malin2}). Their formation and origin are challenging for the modern galaxy formation theories. The analysis of observations of Malin-2 \cite{2014MNRAS.437.3072K...Kasparova-2014}, whose mass exceeds $10^{12}M_\odot$, within the limits of the measured rotation curve, revealed the presence of a very massive halo with unusually low central density. The authors \cite{2014MNRAS.437.3072K...Kasparova-2014} also assumed the presence of large masses of the so-called dark gas in the disk of this galaxy, which is not observed in radio lines.

\section{Velocity dispersion of stars in the disk: constraints on dark matter}\label{sec-UFN_DarkMatter-Dispersion}

Extended rotation curves with $V\simeq\rm{const}$  imply the presence of hidden mass (provided that the baryonic matter density spatial distribution is known), but do not suggest the location and space structure of DM. In principle, DM can be in a spheroidal dark halo but also can be comprised in the disk thus producing its rapid rotational velocity. Studies of dynamical processes in the disk enables setting additional independent constraints on thee gravitating mass distribution.  First of all, this relates to the analysis of dispersion of old disk stars with an age of several billion years, which constitute most of the stellar population.

Until the mid-1990s, the number of galaxies with known radial velocity dispersion along the line of sight beyond the bulge was less than a dozen. Over last 15-20 years, using modern observational methods stellar velocity dispersion in disks was measured in many galaxies. These data relates not only to the central parts of the disk where bulge dominates, but also to very remote disk regions, which plays the key role in constructing realistic galaxy models with account for dark matter and its spatial distribution.

Velocity dispersion of old stars are crucial not only for estimates of disk and halo masses and measurements of the ${M}/{L}$, ratio, but also for the large-scale structures in the disks (spiral pattern, bars, twists, rings, etc.  \cite{Athanassoula1987, 2003ARep...47..357K...Khoperskov-etal-2003!QT, 2012MNRAS.427.1983K, 2011MNRAS.415.1259G}). This is also important to determine the depth of the potential well due to the disk, in which interstellar gas is present concentrating towards the galactic plane. In turn, the potential distribution significantly determines formation conditions of  giant molecular clouds (GMC) in the gas-dust component \cite{2012ApJ...761...37M, 2013MNRAS.428.2311K-GMO, 2015MNRAS.447.3390Dobbs-star-formation, 2016MNRAS.455.1782K-GMO}, regions of star formation and its efficiency \cite{1984ApJ...282...61S...Sellwood-1984!instab, 2008MNRAS.383..809B...Begum-etal-2008, 2011ApJ...729..133M, 2012AstL...38..755A, 2013seg..book..491S, 2014MNRAS.437L..31D, 2014MNRAS.442.1230R, 2014MNRAS.442.3407B}.

\subsection{The disk gravitational stability condition}\label{sub-UFN_DarkMatter-Dispersion-MargInstab}

The star disk dynamics in characterized by both rotation curve and velocity dispersion of stars along three directions: $c_r$, $c_\varphi$, $c_z$. The gas disk, unlike the stellar one, has an almost isotropic local velocity dispersion, because it is a collisional dissipative medium. The gas  velocity dispersion usually relate to the velocity of turbulent motions. Stellar disks, due to their being collisionless, can show various velocity dispersion along different directions. In the epicyclic approximation, in which the velocity dispersion in the disk plane is low compared to the circular velocity, the Lindblad relation  $c_r/c_\varphi=2\Omega/\varkappa$ holds, where $\varkappa$ is the epicyclic frequency (the low-amplitude oscillation frequency relative to the circular orbit) and $\Omega=V/r$ is the disk angular velocity. For disk regions with flat rotation curve $\varkappa = \sqrt{2}\Omega$.

Most spiral galaxies have dynamically cold stellar disks, since the radial velocity dispersion even for old stars constituting most of the stellar disk population is small compared to the rotational velocity: $c_r/V_{с}\simeq 0.3\div 0.1$ (except for near central regions), which justifies the epicyclic approximation. In the absence of non-disk components (dark halo, bulge) the disk could be self-gravitating, and, as numerical simulations show, the velocity dispersion of stars would significantly exceed the observed one because of gravitational instability in the self-gravitating system.

The condition of gravitational stability relative to radial perturbations limits from below the velocity dispersion $c_r$ in a stellar disk  \cite{1964ApJ...139.1217T...Toomre-1964!Criterion-Toomre}:
\begin{equation}\label{Eq-Q_T-def}
c_r \geq c_T = \frac{3.36\,G\,\Sigma_{*} }{ \varkappa},
\end{equation}
 or, using the Toomre parameter $Q_T=c_r/c_T\geq 1$. This relation is valid for thin isothermal disks in the epicyclic approximation. The final disk thickness makes it more stable. However, the presence of colder components (for example, a gas disk), as well as non-radial perturbations and global modes facilitates the disk instability. These facts complicate condition  (\ref{Eq-Q_T-def}), and result in the marginal stable value of $Q_T$ for real disks can be both close to unity and larger than unity, especially far away from the center. The Toomre criterion should be used with caution in central parts of galaxies as well, where the amplitude of epicyclic motions becomes comparable to the mean orbital radius, which violates the epicyclic approximation.

 Different criteria of gravitational stability of galactic disks are discussed, for example, in \cite{1986Afz....24..467M...Morozov-Khoperskov-1986!grav-instab, 2003ARep...47..357K...Khoperskov-etal-2003!QT, Fridman-Khoperskov-2011!book, 2013MNRAS.433.1389R...Romeo-Falstad-2013!Q-stability-parameter}. These criteria give consistent results and show that at the stability limit $Q_T^{(crit)}\sim 1-3$ depending on the local disk parameters \cite{2001MNRAS.323..445R...Rafikov-2001!instability-criterion, 2003ARep...47..357K...Khoperskov-etal-2003!QT, 2014AJ....147..132J...Jog-2014!Q-criterion-disk-stability}. However, there is no simple analytical expression for the critical velocity dispersion that takes into account all the above facts affecting the stability, therefore in the general case numerical modelling is required.

A sufficiently cold disk ($Q_T<1$) is always gravitationally unstable, and perturbations increase in it, which at the non-linear stage leads to the stellar disk heating – its velocity dispersion growth. The efficiency of heating of young stellar disks by short-living (transient) spirals is well studied in numerical experiments of N-body modelling \cite{1984ApJ...282...61S...Sellwood-1984!instab, 2001ARep...45..180K, 2003ARep...47..357K...Khoperskov-etal-2003!QT, 2003ARep...47..443K...Khoperskov-etal-2003!MW}. After the transient spirals have dissipated, the disk becomes stable  ($Q_T>1$). As a  more massive disk should be more ‘hot’ to be stable, the maximal disk model (see Section \ref{sub-maxDisk}) requires higher velocity dispersions: if most of the galaxy mass inside two-three radial disk scales is comprised inside the disk, the relative velocity dispersion at the double disk radial scale, where the disk mostly contributes to the circular velocity, will be   $c_r(2r_d)/V_{\max}\sim 0.5-0.8$, in contradiction to observations, which show $c_r(2r_d)/V_{\max}\le 0.5$ for many disk galaxies.

\begin{figure}[!h]
\centering\includegraphics[width=0.5\hsize]{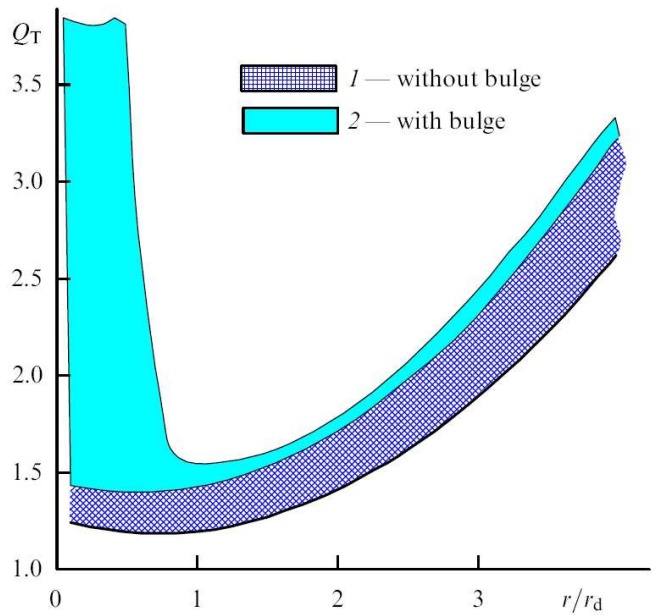}\vskip -0.\hsize
\caption{(Color online.) Gravitational stability boundaries for a variety of N-body models of a marginally stable stellar disk: \textit{1} --- models without a bulge, \textit{2} --- models with a bulge with a size $r_{bulge}< r_d$ }
\label{fig-QTcrit}
\end{figure}

 Under real conditions of inhomogeneous differentially rotating 3D stellar disk $Q_T^{(crit)}$ changes along the radial coordinate (Fig.~\ref{fig-QTcrit}).Numerical models reveal that the development of gravitational instability in the region comprising most of the disk mass stops at $Q_{\rm T} \propto 1 - 1.5$, but if the gravitational field is mainly due to spheroidal components (bulge in the inner galaxy and halo in the outer parts), the minimal velocity dispersion for a purely stellar disk continues increasing up to $Q_{\rm T}\geq 3$.

The analysis of velocity dispersions enables narrowing possible galactic models that explain the observed rotation curve. The main idea is in using in the mass distribution modelling an additional condition that can be formulated as follows: for a stellar disk that reached steady state, the observed velocity dispersion at different distances from the center must be equal to the model values calculated for the threshold (marginal) stability or exceed them, then the disk will be stable.

\begin{figure}[!h]
\centering\includegraphics[width=0.7\hsize]{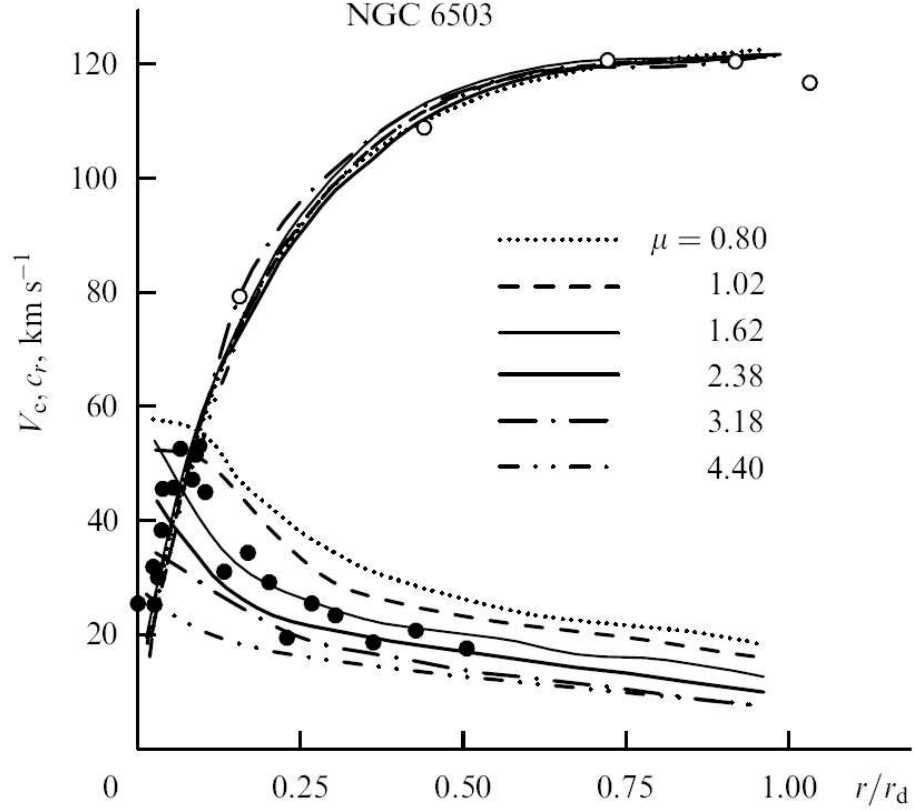}\vskip -0.\hsize
\caption{Radial profiles of the circular velocity and velocity dispersion along the line of sight for different models of NGC\,6503~\cite{2001ARep...45..180K}. The black dots show observations. In models with $\mu<1.6$, the observed velosity dispersions of stars is less than the calculated ones. }\label{Fig-NGC6503-model-cobs-mu}
\end{figure}

In a marginally stable disk, i.e. with parameters  $\Sigma_{*}(r)$, $V(r)$, $c_r(r)$ corresponding to the gravitational stability limit, the estimate $\Sigma_{*}(r)$, found from this condition enables setting upper limits on the real disk mass, irrespective of which matter it consists of, and hence constraining the mass of other components. The above method was applied to the galaxy NGC 6503 Fig.\,\ref{Fig-NGC6503-model-cobs-mu}. The observed rotation curve  $V^{obs}(r)$ can be explained by models with different ratios between the spherical and disk subsystems $\mu = M_s/M_d = 0.8\!-\!5$. The comparison of the observed radial velocity dispersion of stars $c^{obs}(r)$ with marginally stable galactic model for diverse values of $\mu$ enables setting more stringent constraints on the halo mass (see Fig. ~\ref{Fig-NGC6503-model-cobs-mu}).

  If in the region comprising most of the disk mass the critical value $Q_T$ and the rotational velocity $V$  can be assumed approximately constant, the expected ratio of the velocity dispersion to the velocity $V$ turns out to be dependent on the relative mass of the spherical component (i.e. bulge or halo) $\mu=M_s/M_d$. This follows from the simple considerations given below.

\begin{figure}[!h]
\centering\includegraphics[width=0.5\hsize]{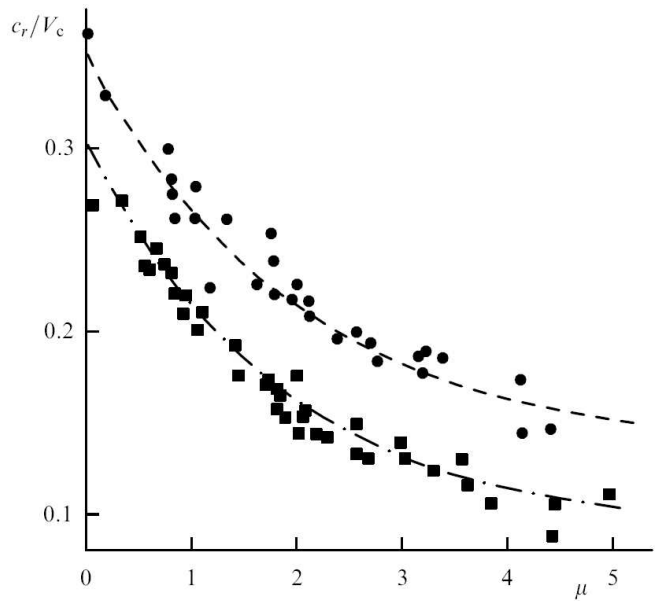}\vskip -0.\hsize
\caption{The ratio $c_r/V_c$ at the radius $r=2r_d$ as a function of the relative mass of the spheroidal subsystem
$\mu$ from dynamical simulations for marginally stable disks with different parameters. Each symbol (black dot or square) corresponds to one numerical model. The dots correspond to stellar disks with a small initial thickness. All models evolve from an initially unstable state.  }
\label{Fig-5-cr/V-from-mu-on-2L}
\end{figure}

  Let the interesting disk region is limited by the radius  $R$, which is $n$times as large as the radial disk scale, $R=n\cdot r_d$. By assuming the balance between the centrifugal and gravitational force, the total mass of the galaxy inside $R$ is $M_{t}= M_s+M_d \approx V^2R/G$. Taking into account that the disk mass is  $M_d\propto \Sigma_*R^2$, for the marginally stable disk: $c_r\propto \Sigma_*/\varkappa\propto \Sigma_*\cdot R/V  \propto {M_d}/{RV}$ (see equation  (\ref{Eq-Q_T-def})), which yields the dependence  $c_r/V\propto M_d/(M_t)\propto (1+\mu)^{-1}$.
This effect is presented in Fig.~\ref{Fig-5-cr/V-from-mu-on-2L}, , which is constructed using results of numerical simulations. The Figure shows the calculated sequences of points (each point for separate model), which corresponds to values  $c_r$/$V$ for galaxies with marginally stable disks of different initial thickness. The Figure suggests that at low mass of spheroidal components the velocity dispersion of stars in the disk must be a substantial fraction of the rotational velocity.

The above relations suggest one interesting constraint on the geometry of marginally stable disks. The vertical size (semi-thickness) of the stellar disk is determined by its surface density and velocity dispersion along the $z$-coordinate: $z_*\sim c_z^2$/$\Sigma_d$.  The velocity dispersion $c_z$, whose minimum value is related to the bending instability, and the radial velocity dispersion $c_r$ are approximately proportional to each other (observations suggest $c_z/c_r =0.5-0.8)$. Therefore, by assuming that $c_z$  is proportional to the critical value of $c_r$, we obtain: $z_*\sim \Sigma_d\cdot R^2$/$V^2$, and, hence $z_*$/$R\sim\Sigma_dR^2$/$(V^2R)\sim M_d$/$M_t$. This implies that there is a limit on the relative disk thickness below which the disk will be unstable, and the larger the relative dark halo mass, the lower this limit. Observations indeed suggest a correlation between the relative disk thickness and its relative mass \cite{2002AstL...28..527Z...Zasov-2002}.

Relative thickness of disks of late-type galaxies (almost without bulges) seen ‘edge-on’ were measured by Bizyaev and Mitronova \cite{2009ApJ...702.1567B...Bizyaev-2009} using IR-photometry  with account for the interstellar light extinction in galaxies. This analysis confirmed the correlation between $z_*/r_d$ and deprojected surface brightness of the disks consistent with the marginal disk stability assumption. The relative disk thickness  $z_*$/$r_d$ was found to little change with radius and is on average 1/6, which corresponds to dark halo mass to disk mass ratio $M_h$/$M_{opt}\approx 1.3$ inside the optical radius (assumed to be 4$r_d$).
 Without the dark halo assumption, it is impossible to explain the small relative thickness $z_*/r_d$ of such stellar disks.

However, the approximate character of these relations should be taken into account. For example, the presence of a massive compact bulge, which is frequently found in S0-Sb galaxies, affects the warp instability of the disk and complicates the relation between the velocity dispersions $c_r$ and $c_z$ (see the discussion in paper \cite{2015MNRAS.451.2376M...Mosenkov2015} and references therein). Note, however, that the accurate two-dimensional photometry of galaxies seen edge-on has not revealed any dependence between the relative disk thickness $z/r_d$ and the bulge contribution to the galactic integral luminosity\cite{2015MNRAS.451.2376M...Mosenkov2015}.

 Thus, the existence of a stellar disk with low velocity dispersion  $c_r\simeq (0.1-0.3)V_c$ requires that a significantly large fraction of the total mass to be outside the disk. A stellar disk without halo ($\mu=0$) is dynamically hot $c_r/V_c\gee 0.3-0.4$ (see Fig.~\ref{Fig-5-cr/V-from-mu-on-2L}). In addition, if  $\mu \ll 1$, the difference between the velocity of rotation of the gas-dust disk $V_c(r)$ and that of the stellar disk $V_{*}(r)$ becomes significant, amounting to $V_*/V_c \simeq 0.5-0.7$, in the outskirts, which occurs very rarely in galaxies.

The approach to the disk and halo mass decomposition with account for the stellar velocity dispersion applied to several galaxy samples~\cite{2001ARep...45..180K, 2004AstL...30..593Z...Zasov-etal-2004!Mass-Estimation, 2011AstL...37..374Z, 2013AN....334..785S...Saburova-Zasov-2013!mass-estimation}, confirmed that the stellar disk gravitational stability condition for Sab-Sd-galaxies assumes the presence of a more massive dark halo than follows from the maximum disk model.

An important conclusion is that to make disk stable with the observed stellar velocity dispersion, most of DM inside the galaxy must form a spheroidal or at least strongly oblate subsystem and cannot be concentrated inside the stellar disk in the form of, for example, dark gas. Otherwise, a lot of DM inside the disk would correspondingly contribute to the disk self-gravity by increasing the critical stellar velocity dispersion required for its stability.

We emphasize that the condition of the threshold (marginal) gravitational stability in the general case sets only an upper limit on the disk density for a given stellar number density (or a lower limit on the velocity dispersion for a given density). In principle, the disk can be dynamically overheated, i.e. can have the marginal stability \cite{2012AstBu..67..362Z}. Apparently, frequently disks are marginally stable in the wide range of galactocentric distances but is ‘overheated’ at the periphery – thanks to low density, it can be easily dynamically heated.

To test to which degree the marginal stability conditions can be applicable to real disk galaxies, in paper  ~\cite{2011AstL...37..374Z} the surface densities of S and S0 galaxies found from the threshold (marginal) stability condition at the fixed distance $r\approx 2r_d$,were compared to the luminosity of the stellar population per unit disk area at the same distance from the center. The resulting $M/L$ ratios were matched to those expected from the model stellar population with known color index. It turned out that for spiral (S) galaxies, the color--$M/L$ ratio dependence derived from the stability condition, is in good agreement with 	the relation suggested by photometrical stellar population models, and consequently the threshold stability condition can indeed be used to estimate the mass of galactic components.  However, for lenticular (S0) galaxies the situation is more diverse: about half of them have ‘overheated’ disks  \cite{2012AstBu..67..362Z, 2011AstL...37..374Z}. This means that their disks, with the same density, have higher velocity dispersions and, apparently, higher thickness than follows from the marginal stability condition.

Clearly, the disk mass and the degree of its dynamical heating are connected to the related dark halo mass in the optically observed region of the galaxy and, therefore, to the formation history end evolution of the galaxy.

 \subsection{Disk density estimates from vertical velocity dispersion measurements}\label{sub2verticaldispersion}

\subsubsection{Local DM density}

 The observed thickness of stellar disks is primarily determined by the vertical velocity dispersion
 $c_z$ and the volume disk density. Therefore, the disk densities can be estimated from measurements of $c_z$ with known vertical star density profile. For our Galaxy near the Sun this approach was applied as early as in the first quarter of the 20th century \cite{Opik-1915!vert-disk, 1922ApJ....55..302K...Kapteyn-1922!MD-vertical-disk}, and has been used up to now. The comparison of the total matter density with independent measurements of the baryonic Galactic components enables the estimate of the local DM density in our vicinity (see below).

 The vertical velocity of stars characterized by the velocity dispersion $c_z(z)$, are related to the matter density on the disk $\varrho(z)$.
 Vertical balance of forces is derived from the $z$-component of the Jeans equation for collisionless stellar disk (the Poisson equation), which together with conditions  $\varrho(z=0) =
\varrho_0$, $d\varrho(0)/dz=0$,
$\int\limits_{-\infty}^{\infty}{\varrho(z;r)\,dz}=\Sigma(r)$
 for a given surface density distribution $\Sigma$ determines the vertical disk structure at the given radius $r$.
For
$c_z=\textrm{const}$ we get the solution
\begin{equation}\label{4--Eq-varrho-ch(z/z0)}
\Oo\varrho(z)= {\Sigma\o 2z_0}
\textrm{ch}^{-2}(z/z_0)
\end{equation}
with the characteristic vertical scale
$z_0=\sqrt{c_z^2/2\pi G\varrho(0)}=
 c_z^2/\pi G\Sigma$.
    Equation (\ref{4--Eq-varrho-ch(z/z0)}) along with the exponential density distribution in the disk plane with the vertical scale
\begin{equation}\label{4--Eq-varrho-exp(-z/hexp)}
 \varrho(z) = \varrho(0)\cdot \exp(-z/h_{exp})
\end{equation}
     is frequently used to fit the observed brightness profiles of galaxies seen edge-on
\footnote{\normalsize For some galaxies the distribution $\varrho\propto \textrm{ch}^{-1}(z/h_{ch})$ (where $h_{ch}$ is the vertical scale) was also used  \cite{1996A&AS..117...19D...Grijs-Kruit-1996!Edge-on-inAASS}.}
    \cite{1981A&A....95..105V...Kruit-van-der-Searle-1981!Edge-on-photometry, 1981A&A....95..116V...Kruit-Searle-1981!Edge-on-II, 1997PhDT........24D...Grijs-disser}.

For our Galaxy, the analysis of the observed stellar density along the vertical coordinate perpendicular to the disk plane together with their velocity distribution enables the local disk density near the Sun to be measured, which in turn enables more precise decomposition of the Galactic rotation curve and dark halo contribution separation. To this goal, data on red dwarf stars are frequently involved. The red dwarfs are objects that mostly contribute to the stellar population mass, although it is also possible to use less numerous red giants or mains sequence stars with better known distances. In any case the resulting density estimates obtained from the stellar kinematics  will relate to the total disk density. By comparing the found local density of stars and gas it is possible to constrain the DM density in the disk.

\begin{figure}[!h]
\centering\includegraphics[width=0.35\hsize]{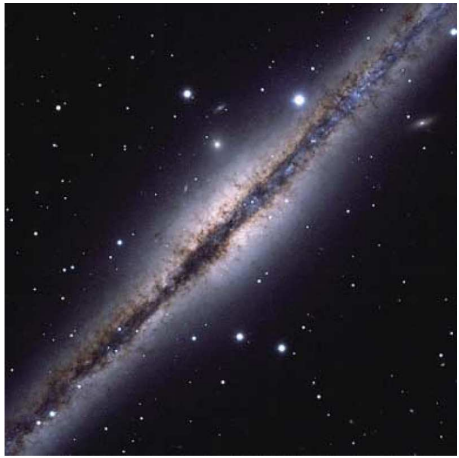}\vskip -0.\hsize
\caption{The galaxy NGC\,891, seen edge-on (Hubble Space Telescope). A thinner gas-dust disk is visible inside the stellar disk.
 }
\label{Fig-N891}
\end{figure}

Taking into account the small disk thickness compared to its radial size, it is convenient to restrict the analysis by the balance
\begin{equation}\label{eq-Jeans-z-balance-disc}
    \frac{d (\nu(z) c_z^2)}{dz} = - \nu\, \frac{d\Phi}{dz}  \,,
\end{equation}
where $\nu$~is the relative density of stars in the sample  (Fig.\,\ref{Fig-N891}).

But even for the Solar vicinity the results of various authors give different local DM density estimates. For example, paper by Korchagin et al. \cite{2003AJ....126.2896K...Korchagin-etal-2003!MW-density} estimates the surface density inside the disk thickness $\pm 350$~pc to be $\Sigma(|z|<350$пк$)=42\pm 6\, M_\odot$pc$^{-2}$. With account for the baryonic density estimates uncertainty, this value is consistent with the absence of noticeable amount of DM in the disk. Bienaime et al. \cite{2014A&A...571A..92B...Bienayme-2014} found a similar value of the surface density $44\pm 4\, M_\odot$pc$^{-2}$, but for a thicker layer of  $\pm 1$\,kpc. The volume DM density in the solar vicinity in is found to be $\rho_{dm}=0.0143\pm 0.0011\, M_\odot$pc$^{-3}$. A close estimate  $\rho_{dm}=0.016\,M_\odot$pc$^{-3}$ is obtained from the spectral survey of G and K main-sequence stars \cite{2015arXiv151006810X...Xia-2015}. McKee et al. \cite{2015ApJ...814...13McKee-DM-Solar-Neighborhood} by comparing the baryonic components density with dynamical density estimates, found the total density of matter in the Solar vicinity $0.097\pm 0.013\,M_\odot$pc$^{-3}$, with the DM density being $0.013\pm 0.003\, M_\odot$pc$^{-3}$. Therefore, the fraction of halo DM in the solar neighborhood is slightly higher than 10$\%$ of the total matter density.

The review of different estimates of the local DM density and the total mass of dark halo in the Galaxy, as well as of all difficulties in their determination, can be found in paper by Famaey \cite{2015arXiv150101788F...famaey-2015}.

\subsubsection{Disk mass estimate from vertical velocity dispersion}

In other galaxies, there is no possibility to estimate the $z$-component of stellar velocities with simultaneous measurement of the space density of stellar population. Therefore, to estimate the local disk density it is needed to measure both the galaxy thickness and velocity dispersion. Here two methods are used.

At first, observations of galaxies seen edge-on enables direct measurements of the stellar disk thickness, and the velocity dispersion can be taken, for example, to be equal to its marginal gravitational stability value or can be estimated from the empirical relation between the velocity dispersion and rotational velocity of the galaxy \cite{1993A&A...275...16B...Bottema-1993!Stellar-kinem-many}. However, the accuracy of this method is low.

At second, the local disk density can be estimated from direct local velocity dispersion measurements, but here the disk thickness should be a priori given (usually it is assumed constant and corresponding to a certain fraction of the radial scale or photometrical disk size, although this is rather crude approximation). This approach is the most convenient for galaxies with inclination angles $i=20-30^\textrm{o}$ \cite{2011ApJ...739L..47B...Bershady-2011}.  For these objects, it is possible to obtain with sufficient accuracy the rotation curve (at $i=0$ the rotational velocity is directed perpendicular to the line of sight and cannot be measured), and the velocity dispersion of stars will be close to that perpendicular to the disk plane.

Both these approaches lead to the conclusion that in most galaxies the disks are ‘submaximal’: the disk contribution to the mass at two radial scales, where it mostly contributes to the rotation curve, is slightly above 50\%  \cite{2005MNRAS.358..503K...Kregel-2005, 2011ApJ...739L..47B...Bershady-2011}. The fraction of baryonic matter inside these limits increases from small galaxies to massive ones and rapidly rotating systems \cite{2013A&A...557A.131M...Martinsson-2013!dark-matter-spiral-galaxies}.

For edge-on galaxies, it is more preferable to consider an equilibrium gas disk instead of stellar one, if the angular resolution of the radio telescope enables estimating the gas layer thickness, since the gas velocity dispersion weakly changes both along the radius and from galaxy to galaxy.  Here the gas velocity dispersion is measured either directly \cite{1996AJ....112..457O...Olling-1996!NGC4244}, or is assumed to be constant (around 10\,km/s). As the gas layer is inside the disk gravitation field, its thickness and velocity dispersion enable the local space density of the disk to be measured.  If the stellar disk thickness is known, it is possible to transit to its surface density at diverse distances from the center and thus to estimate the total disk mass. By separating the disk contribution from circular velocity curve, from the observed rotation curve it is possible both to estimate the halo mass  \cite{2006AstL...32..649Sotnikova, 2010AN....331..731K}, and oblateness \cite{2010A&A...515A..62O...Brien-etal-2010}.

\subsection{Estimates of masses of elliptical galaxies}
Elliptical (E) galaxies almost never reveal extended gas disks, and their rotation velocities are much smaller than stellar velocity dispersion, therefore the mass estimation methods described in Section 3.2 are inapplicable. Stars move in elongated non-closed orbits, therefore the notion of circular velocity $V_c$, which is frequently applied to these galaxies, has a purely formal sense of being the velocity of conventional test particles in circular orbits. Some E-galaxies can significantly deviate from axially symmetric shape, which complicates the dynamical mass estimate.

The mass of E-galaxies can be estimated by the following main methods:
\begin{itemize}
\item Based on measured radial velocity dispersion of stars or globular clusters at various distances from the center, dynamically equilibrium models are constructed. Numerical models depend on a priori unknown character of stellar motions: nearly circular orbits, orbits with isotropic velocity distribution, and nearly radial orbits give significantly different results. However, assuming  $V_c(r)$
 to be constant or slowly changing, it is possible to circumvent these uncertainties by jointly analyzing the brightness and radial velocity dispersion distributions. As shown by Lyskova et al.  \cite{2014MNRAS.441.2013L-Lyskova-Churazov-X-ray} (see also references therein), observations allows to choose the radial distance interval (near the effective radius  $R_e$), in which the estimate $V_c$, and hence masses very weakly depend on the character of motion of stars (see  the description of the method and results of its application for several galaxies in \cite{2014MNRAS.441.2013L-Lyskova-Churazov-X-ray}.

\item When hot gas observed by X-ray emission is present, equilibrium models for gas coronae in gravitational field of galaxies are constructed, which are described by the hydrostatic equilibrium equation:
$$ -\frac{1}{\rho}\frac{dP}{dr} = \frac{V_c^2}{r} =  \frac{GM(r)}{r^2},$$
where $P=\rho k_B T/\mu m_p $ is the thermal gas pressure, $\mu \approx 0.6 $ is the mean atomic weight.

\item
 Methods of weak and strong gravitational lensing (see last Section) are applied, which are very effective for galaxy mass estimates.
\end{itemize}
The above methods of mass estimate suggest that the distribution of gravitational potential in E-galaxies, like in the disk galaxies, at large $r$, is close to the isothermal one, which means the increase in the ratio of DM to baryonic mass with distance from the center. In the central regions of galaxies the DM mass contribution is small and the $M/L$  ratio is close to that expected for purely stellar models, but at large distances this ratio strongly increases implying the presence of extended dark halos with radial density distribution satisfactorily described by the NFW-profile (\ref{Eq-3-halo-Navarro}) at least at large  $r$ (see, for example, \cite{2006ApJ...636..698F-Fukazawa-Elliptical-Galaxies-Chandra, 2013AIPC.1551...84K-Kalinova-CALIFA-DM, 2007ApJ...667..176G-Gavazzi-Sloan-Lens}.  There are grounds to believe that the DM content in E-galaxies within optical limits is higher than in spirals \cite{2013AIPC.1551...84K-Kalinova-CALIFA-DM}.	
However, the situation can be different from galaxy to galaxy. In rare cases (for example, in  NGC\,7507 --- the isolated E-galaxy with two-component stellar halo) the dynamical modelling did not reveal at all signatures of dark halo \cite{2015A&A...574A..93Lane-DM-NGC7507}.

\section{Dark matter from gravitational lensing and X-ray and gamma-ray observations }

Progress in observations led to the appearance of new methods of mass distribution studies, including of invisible component. The most impressive results were obtained from gravitational lensing of background sources by masses located between the sources and the observer. Here we will not delve into methods of restoring gravitational field of astronomical objects from gravitational lensing images. The interested reader is referred to papers  \cite{1996ApJ...466..623B...Brainerd-etal-1996!Weak-Gravitational-Lensing, 2012AmJPh..80..753T...Treu-etal-2012!Gravitational-Lensing}.

\subsection{Gravitational lensing effects}

\begin{figure}[!h]
 \centering\includegraphics[width=0.95\hsize]{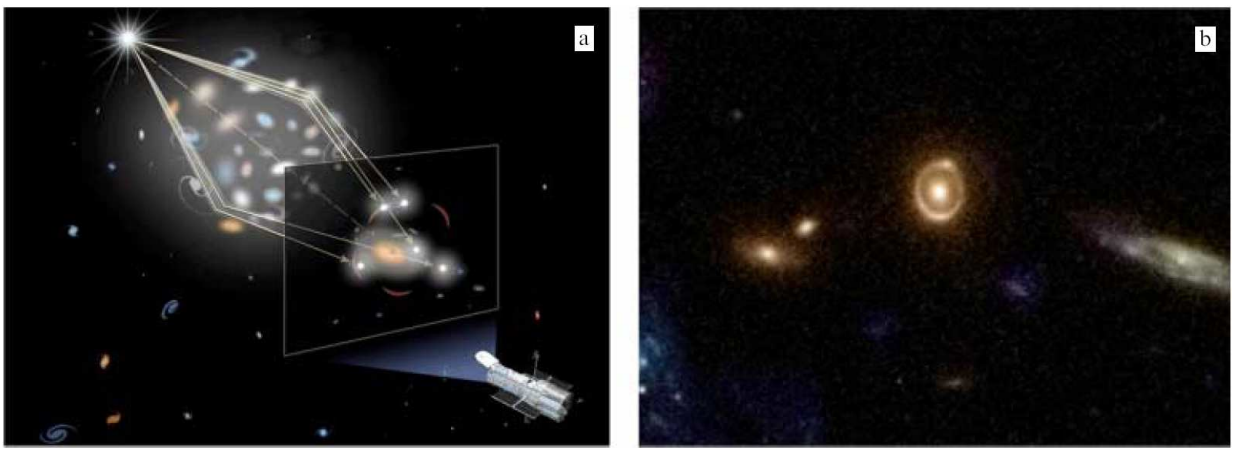}\vskip -0.\hsize
  \caption {\hyphenpenalty=10000
(a) Schematics of rays passing through a gravitational lens. (b) Image of an almost ideal Einstein ring on the gravitational lens 0038+4133 (Hubble Space Telescope).
 }\label{Fig-gravlens}
\end{figure}

Gravitational lensing offers the possibility not only to observe remote objects with multiple brightness amplification, but also to study the lens itself, since its properties are determined by  the spatial distribution of gravitating matter, including DM (Fig. ~\ref{Fig-gravlens}). A gravitational lens has no the focal point at which parallel rays converge; it is more appropriate to consider the focal straight line passing through the  lens and the observer which the rays from the remote object far behind the lens cross at diverse distances depending on how far the rays passed from the lens center. The space bending caused by gravitational field of the lens results in the appearance of several images of the source, which are very different in brightness as a rule. In the ideal case of axially symmetric lens and the object lying strictly behind it, the source image stretches in a ring (the Einstein ring). Such rings were first discovered in both radio and optical observations. The first optical image of the Einstein ring formed by a galaxy was obtained in 1997 by the Hubble Space Telescope \cite{1998MNRAS.295L..41K}.

There are several different gravitational lensing effects: the microlensing effect, when lenses are low-mass bodies (their nature is unimportant), strong and weak lensing. The microlensing offers unique possibility to probe DM halo in our Galaxy. The method includes the search for and statistical analysis of lensing effects from individual stars by objects of arbitrary nature, which in principle is able to reveal the presence of unseen compact bodies with masses of planetary mass and much larger. Search for microlensing effects from single brightness enhancement of a star that spuriously lies in line with a gravitating body is a complicated process, which requires separation of physically variable stars and long-term monitoring of several million stellar images in densely populated patches of the sky. This method and its possibilities are discussed in papers \cite{1997SvPhU..40..869G...Gurevich-etal-1997!UFN-microlensing, 1998SvPhU..41..945Z...Zakharov-Sazhin-1998!Gravitational-microlensing, 2008Ap&SS.317..181B...Bogdanov-Cherepashchuk-2008!microlensing-stars, 2006MNRAS.371.1259T...Tuntsov-Lewis-2005!microlens-gal}. Note only that the results of several independent groups searching for microlenses suggest that their total mass can amount only to a small fraction of the required mass of the Galactic dark halo; apparently, the discovered microlenses are due to ordinary low-mass stars, which confirms non-baryonic nature of DM \cite{2012RAA....12..947M}. The only non-accounted fact could be the presence of unusually heavy bodies, with a mass of more than $100\,M_\odot$ (primordial black holes?), which cannot be found by microlensing due to the long time-scale of  the brightness variations \cite{2015arXiv151108801F}. However, there are no solid observational support of the existence of such exotic components of dark halos.

In the strong lensing, the original image of the lensed source (a galaxy, or, less frequently, a quasar) is split into several images in the form of arcs and sometimes of more complicated figures. The strong lensing image enables the lens mass estimate or its density distribution, if the distance to the lens and to the remote source is known (from redshift measurements). As a rule, a galaxy cluster serves as a strong lens, therefore the lensing allows the central mass of the cluster to be measured.  In some cases, the lens is a galaxy  \cite{2004ApJ...611..739T...Treu-etal-2004!Lens-gal} or its central part \cite{2010ApJ...719.1481V...Ven-etal-2010!Einstein-Cross}.

 Note an interesting possibility of determination of the internal structure of gravitating matter by the strong lensing (quadrupole lenses) of quasars with anomalous flux ratio from the observed multiple images, which are not described by a simple mass distribution model but can be explained by microlensing effect on inhomogeneities (stars?) inside the galaxy-lens. This offers possibility to estimate the mass contribution from stars into the total mass of the galaxy-lens. The analysis of 14 X-ray quadrupole images of quasars by Pooley et al. ~\cite{2012ApJ...744..111P...Pooley-etal-2012!DM-E-gal} suggests a substantial excess of the distributed DM component at the mean projected distance from the center of galaxies of 6.6 kpc.  This, however, is inconsistent with other results suggesting that in the inner regions of galaxies with normal brightness the dark halo contribution to the total mass is significantly smaller than the stellar mass.

 In the weak lensing effect for a large number of background galaxies, statistical methods are used to discover systematic distortions of the shape of the remote background galaxies due to lensing by a closer galaxy (tangential stretching of images which is less than 1\% in each individual case). Thanks to the weak lensing effect, it is possible to reconstruct the mass distribution around individual galaxies as well as in clusters and to estimate the mass of DM, which turns out to be dominant over other forms of matter, as expected \cite{2005astro.ph..9252S...Schneider-2005!Weak-Gravitational-Lensing}.

For weak gravitational lensing important are large sky surveys, such as CFHTLS (Canada-France-Hawaii Telescope Lensing Survey) ~\cite{2012MNRAS.427..146H...Heymans-etal-2012!LenS-Canada-France-Hawaii-Telescope}, Red-Sequence Cluster Survey \cite{2005ApJS..157....1G...Gladders-Yee-2005!Red-Sequence-Cluster-Survey}, SDSS~\cite{2000AJ....120.1579Y...York-etal-2000!Sloan-Digital-Sky-Survey}, COSMOS Survey~\cite{2007ApJS..172..196K...Koekemoer-etal-2007!COSMOS-Survey} etc. Unlike dynamical methods (see Sections 2, 3), weak lensing allow to determine properties of lenses (i.e. spatial distribution of gravitating matter) at larger distances from the center. This relates not only to galaxy clusters, but also to individual massive galaxies.

 To estimate the characteristic parameters of galactic halos, effects of weak lensing by one galaxy (with known redshift) of background dimmer galaxies (the galaxy-galaxy lensing) are statistically analyzed by stacking a large number of images \cite{2009MNRAS.393..377M...Mandelbaum-etal-2009!DM-lensing}. In \cite{2009MNRAS.393..377M...Mandelbaum-etal-2009!DM-lensing} this approach was applied to the SDSS data and revealed, in particular, that galaxies with optically active nuclei do not differ from ordinary galaxies by halo masses, however radio-loud active galactic nuclei were found to have about two times more massive halos for the same stellar galactic masses. This conclusion is far from being undoubted if the radio-loud quasars represent a short active stage of evolution. Another conclusion obtained by the same method relates the difference between the dark halo masses of ‘active’ star-forming galaxies (the blue group) and ‘passive’ (the red group), which predominantly include elliptical galaxies and some disk galaxies: with the same stellar mass, the halo mass in passive galaxies is about two time as high \cite{2015arXiv150906762M...Mandelbaum-2015!DM-lensing}.

Importantly, even if the original density profile in a forming galaxy corresponded to the NFW-profile, during the formation of the ‘baryonic’ galaxy inside the dark halo the  integral density distribution, as well as the dark halo form, can change. The authors  \cite{2015arXiv150906762M...Mandelbaum-2015!DM-lensing} applied the weak lensing method to elliptical galaxies from the SDSS survey to conclude that density profiles in the halo outskirts are consistent with the NFW profile. Here the comparison if the NFW model with kinematic estimates of stellar population in the inner, i.e. ‘baryonic’, part of the galaxies shows disagreement suggesting that galactic dark halos, apparently, experienced adiabatic compression due to gravitation of baryonic matter.

Earlier attempts to match the density in the inner parts of elliptical galaxies (from strong lensing) and in outer parts of the same galaxies (from weak lensing) were done by Gavazzi et al. \cite{2007ApJ...667..176G...Gavazzi-et-al-2007-SLACS}. The authors usually employed for a pseudo-isothermal halo.  concluded that the integral density profile can be described by a two-component model: the de Vaucouleurs profile that well describes the inner galaxy, plus the NFW-profile for outer parts of the galaxy. Here the total profile satisfies the empirical relation $\varrho\propto r^{-2}$, usually employed for a pseudo-isothermal halo.

\begin{figure}[!t]
 \vskip 0.\hsize \hskip 0.30\hsize
            \includegraphics[width=0.4\hsize]{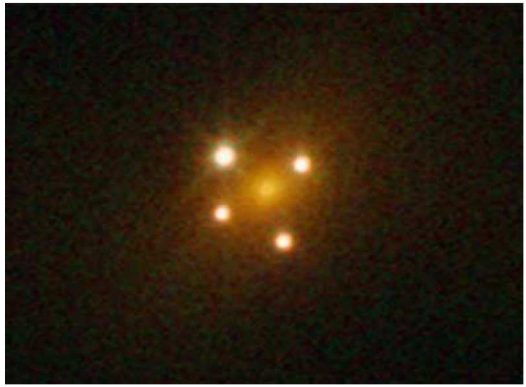}

\vskip  0.0\hsize \hskip 0.0\hsize
  \vbox{\hsize=0.999\hsize
  \caption{\hyphenpenalty=10000
Example of the Einstein cross for the quasar QSO 2237+0305 (Hubble Space Telescope).
 }\label{fig-Einsteins-Cross2} }\vskip 0.0\hsize
\end{figure}

 Application of different methods, including weak lensing, generally consistently suggests a significant dominance (by an order of magnitude and more) of dark halo mass inside the virial radius over the baryonic mass (of stars and gas)
\cite{2010MNRAS.407....2D...Dutton-etal-2010!kinematic-galaxies-dark-matter-haloes, 2007ApJ...667..176G...Gavazzi-et-al-2007-SLACS,2010MNRAS.408.1463S...Schulz-etal-2010!adiabatic-contraction-SDSS}.
 In central parts of galaxies, gravitational lensing  points to insignificant traces of dark mass, in agreement with dynamical modelling of the galaxy rotation curves. For example, parameters of the Einstein cross shown in Fig.\,\ref{fig-Einsteins-Cross2} suggest that inside  $r=2/3R_e$ the DM fraction is less than 20\% of the total mass  \cite{2010ApJ...719.1481V...Ven-etal-2010!Einstein-Cross}. According to  \cite{2007ApJ...667..176G...Gavazzi-et-al-2007-SLACS}, ], inside the effective radius of E-galaxies, dark mass is on average about 1/4 of the total matter mass. Here the virial mass of the halo is several ten times as large as the baryonic mass (although in the standard cosmological model the baryon to dark matter mass ratio should be close to 1/6).

Note that all lensing-related methods are not applicable to low-luminosity galaxies with stellar masses  $\sim 10^{10}M_\odot$ and below.

\begin{figure}[!t]
 \vskip 0.\hsize \hskip 0.0\hsize
            \includegraphics[width=0.49\hsize]{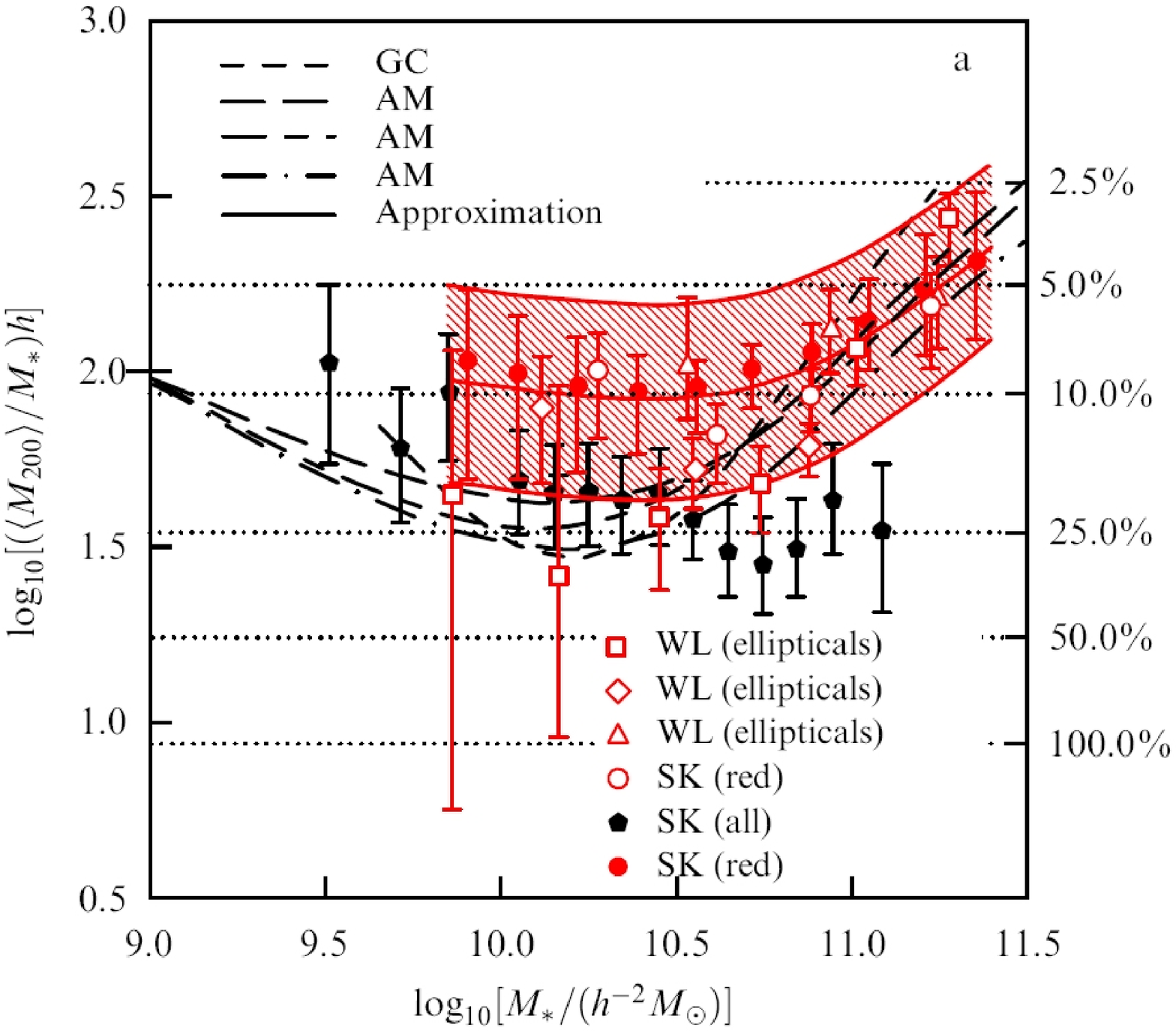} \vskip -0.44\hsize \hskip 0.5\hsize
            \includegraphics[width=0.49\hsize]{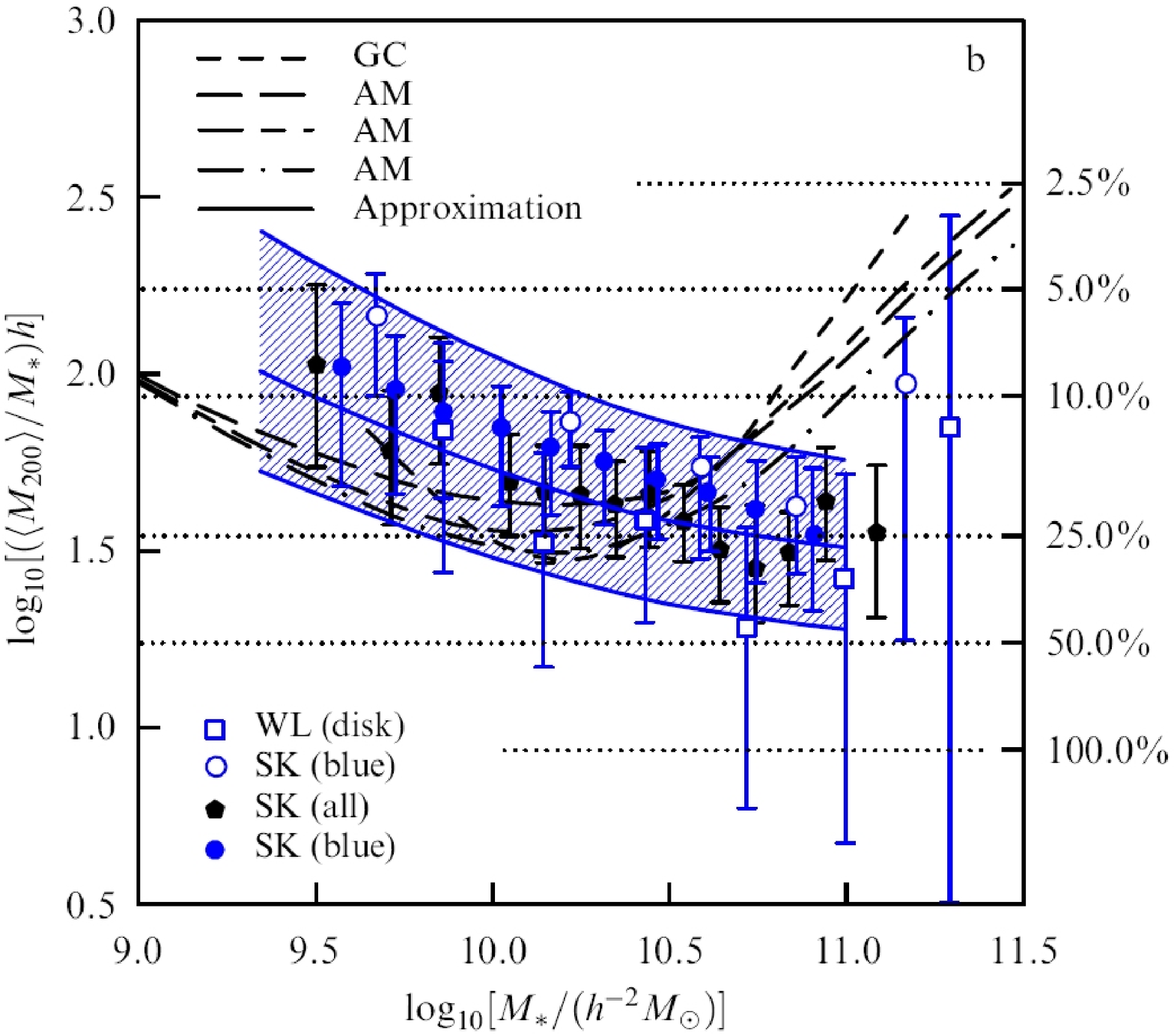}

\vskip  0.0\hsize \hskip 0.0\hsize
  \vbox{\hsize=0.999\hsize
  \caption {\hyphenpenalty=10000
Virial-to-stellar population mass ratio $M_{vir}/M_*$ for galaxies with different masses $M_*$ compiled from the literature (from \cite{2010MNRAS.407....2D...Dutton-etal-2010!kinematic-galaxies-dark-matter-haloes}). The red symbols and the strip in panel (a) and the blue symbols and the strip in panel (b) respectively relate to early-type and late-type galaxies. The methods used: SK --- galactic satellite kinematics, WL --- weak lensing, AM --- the HAM method, GC --- from parameters of galaxy groups (see \cite{2010MNRAS.407....2D...Dutton-etal-2010!kinematic-galaxies-dark-matter-haloes} for more details). The dark curves and symbols are the same in panels (a) and (b) and show the relations obtained irrespective of the galaxy type. The baryon fraction in stars relative to the expected cosmological value is plotted along the ordinates to the right. The Hubble parameter is assumed to be $H_0=100h~km~s^{-1}~Mpc^{-1}$, the mass of stars is in units $h^{-2}M_\odot$, and the dark halo mass is in units $h^{-1}M_\odot$.
 }\label{Fig-M200-Mstar} }\vskip 0.0\hsize
\end{figure}

In Fig.~\ref{Fig-M200-Mstar}, taken from paper  \cite{2010MNRAS.407....2D...Dutton-etal-2010!kinematic-galaxies-dark-matter-haloes}, ratios of the total mass of galaxies inside the virial radius to the baryonic mass (of stars) are compared to the mass $M_{vir}/M_*$,  derived in different papers for early-type (red symbols and the band ) and late-type (blue symbols and the band) galaxies. The estimates of  $M_{vir}$ are obtained by diverse methods, including the analysis of kinematics of galactic satellites, weak lensing, from parameters of galaxy groups, as well as by comparing the rate of occurrence of galaxies with different stellar mass (i.e. the luminosity function of galaxies) with that of halo with different masses (i.e. the halo mass function), as followed from cosmological collisionless N-body simulations.  The last method is referred to as the halo abundance matching (HAM). Here it is assumed that there is a unique correspondence between the two compared quantities. As seen from, there is a significant deviation between estimates obtained by various methods, but generally the estimates suggest that the ratio of the virial halo mass to the stellar mass is above 30, with this ratio being systematically higher for massive early-type galaxies, in which it can exceed 100. The relative mass of stellar matter is less than 30\% of the mean value expected in the standard cosmological model.

\begin{figure}[!h]
\centering\includegraphics[width=0.6\hsize]{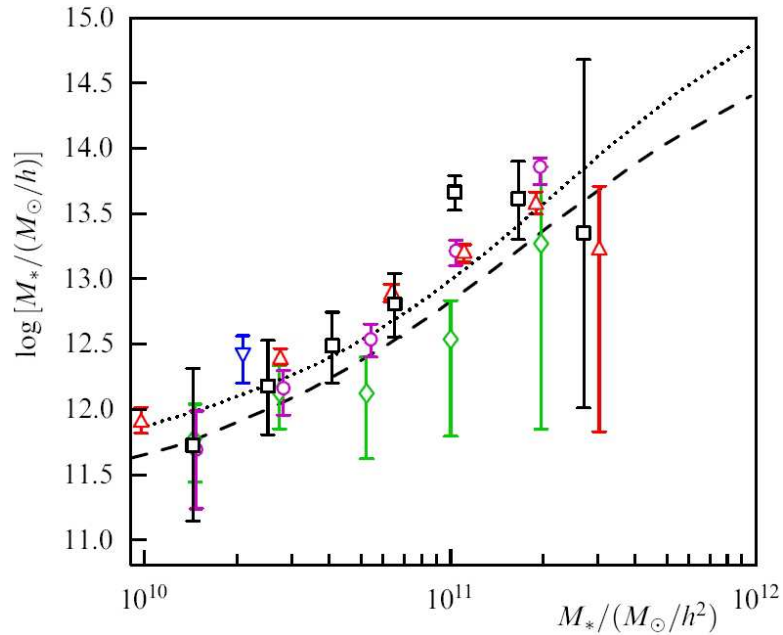}\vskip -0.\hsize
\caption{(Color online.) Comparison of the virial dark halo masses obtained from weak lensing for galaxies and galaxy groups with their stellar population mass. Various symbols show results of different studies (see  \cite{2015MNRAS.446.1356H} for more details).
}\label{fig::weak_lensing1}
\end{figure}

The increase in the virial halo mass with stellar mass in galaxies and in galaxy groups is confirmed by later paper  \cite{2015MNRAS.446.1356H} (see Fig. ~\ref{fig::weak_lensing1}), although so far there is no good agreement between estimates obtained by different methods.

To analyze the weak gravitational lensing effects around passive (red) and star-forming (blue) galaxies, paper  \cite{2014MNRAS.437.2111V...Velander-etal-2014!dark-matter-haloes-weak-lensing} used data from a large sky region  from CFHTLS comprising 154 square degrees together with the photometric survey carried out by the same telescope. Here, importantly, the gravitational effect of the baryonic component on the general potential distribution was taken into account. It was confirmed that the correlation between the stellar mass and DM mass is different for blue and red galaxies. With increasing the stellar mass  $M_*$ and luminosity of galaxies, the virial halo mass  $M_{vir}$ grows faster for objects with old stellar population ($\sim M_*^{1.5}$), than for star-forming galaxies  ($\sim M_*^{0.8}$), which apparently is related to their different formation conditions.

\subsection{Dark matter and hot gas coronae around galaxies}

 By the present time, thanks to operation of space X-ray observatories  XMM-Newton (X-ray Multi-Mirror Mission), Chandra and ROSAT (from germ. Rontgensatellit), characteristics of hot gas with a temperature of  $\sim 10^6$ K inside and around several giant elliptical galaxies and in several disk galaxies have been measured: \cite{2012ApJ...755..107D...Dai2012, 2013ApJ...772...98B...Bogdan-etal-2013!X-Ray-Coronae-Massive-Spiral-Galaxies-NGC266, 2011ApJ...737...22A, 2013ApJ...772...97B...Bogdan-etal-2013!X-Ray-Coronae-Massive-Spiral-Galaxies}.

  The X-ray coronae suggest the presence of significant dark mass, without which a galaxy could not retain the hot gas (see, for example,  \cite{2007POBeo..81....1S...Samurovic-2007!book-Dark-Matter-Elliptical-Galaxies}).  The observed hot gas parameters enables evaluating the mass necessary to retain it and measuring the total mass $M(r)$change with the distance from the galaxy center.
 For estimates, one can use the simple formula for hot gas equilibrium in gravitational field of a galaxy with account for massive halo and assuming centrally symmetric thermal gas velocities $c_g\sim \sqrt{GM/R}$. This yields for the temperature $T\sim GMm_p/(3k_Br)$ ($m_p$\, is the mass of a proton, $k_B$\,is the Boltzmann constant). For example, for $M=10^{12}M_\odot$, $r=20$\,kpc the temperature is  $T\simeq 10^6$\,К.

 Simultaneous analysis of the temperature and intensity of X-ray emission from gas coronae around bright elliptical galaxies enables constructing the total gravitational potential that suggests the presence of a massive extended dark halo. Note that in some bright elliptical galaxies a noticeable increase in $V_c(r)$ is observed  \cite{2009ApJ...693.1300D...Das-etal-2010!dark-matter-X-ray,2009A&A...502..771D...Doherty-2009!DM-E-gal}\footnote{We remind that the circular velocity is formally defined as the rotational velocity of a test particle in circular orbit. The galaxy itself can not rotate.}, whereas in disk galaxies the circular velocity $V_c(r)$ beyond the optical radius is usually constant.
 For $r\geq 2R_e$ ($R_e$ is the effective radius of a galaxy) dark mass dominates over the baryonic mass. A similar conclusion relative to the dark halo parameters was obtained from kinematics of planetary nebulae \cite{2008MNRAS.385.1729D...Lorenzi-etal-2008!Dark-matter-planetary-nebula,2011MNRAS.411.2035N...Napolitano-etal-2010!PN-dark-E-gal}.

 The presence of unresolved X-ray binaries, ultra-compact young stellar objects (YSO) and X-ray remnants of evolution of massive stars gives additional uncertainties when inferring DM properties from X-ray observations of gas around galaxies.  The hydrostatic equilibrium can be too crude approximation, especially for galaxies in which gas was influenced by active galactic nucleus.

Despite these difficulties of DM estimates from X-ray coronae of spiral galaxies, pioneer papers appeared which study a hot gas beyond the optical disks (galaxies   NGC 1961, NGC 6753 \cite{2013ApJ...772...98B...Bogdan-etal-2013!X-Ray-Coronae-Massive-Spiral-Galaxies-NGC266,2013ApJ...772...97B...Bogdan-etal-2013!X-Ray-Coronae-Massive-Spiral-Galaxies}).
The relative mass of baryons inside $(0.05-0.15) R_{vir}$ (where $R_{vir}$ is the virial radius, for these galaxies turned out to be around 0.1, which is already close to the values predicted by cosmological models. This means that in such galaxies the halo gas can contain a sizable fraction of all baryonic matter.

\subsection{Dark halo and gamma-ray emission}

An indirect evidence of the existence of weakly interactive massive particles (WIMPS) forming DM can be obtained from searches for theoretically predicted high-energy gamma-radiation or neutrino emission from annihilation of DM particles and antiparticles in rare collisions. The birth of high-energy photons from particle annihilation can significantly contribute to cosmic gamma-ray background (CGB) in a wide energy range (see review  \cite{2012PDU.....1..194B}). In principle, the annihilation radiation can be detected from regions with especially high density of DM  particles. In our Galaxy, it is the Galactic center region, but there many gamma-sources of other nature are present, which hampers detection of the annihilation radiation, if it really exists. A more suitable could be dwarf spheroidal galaxies without young stars, where DM strongly exceeds by mass the baryonic matter.

\begin{figure}[!h]
\centering\includegraphics[width=0.4\hsize]{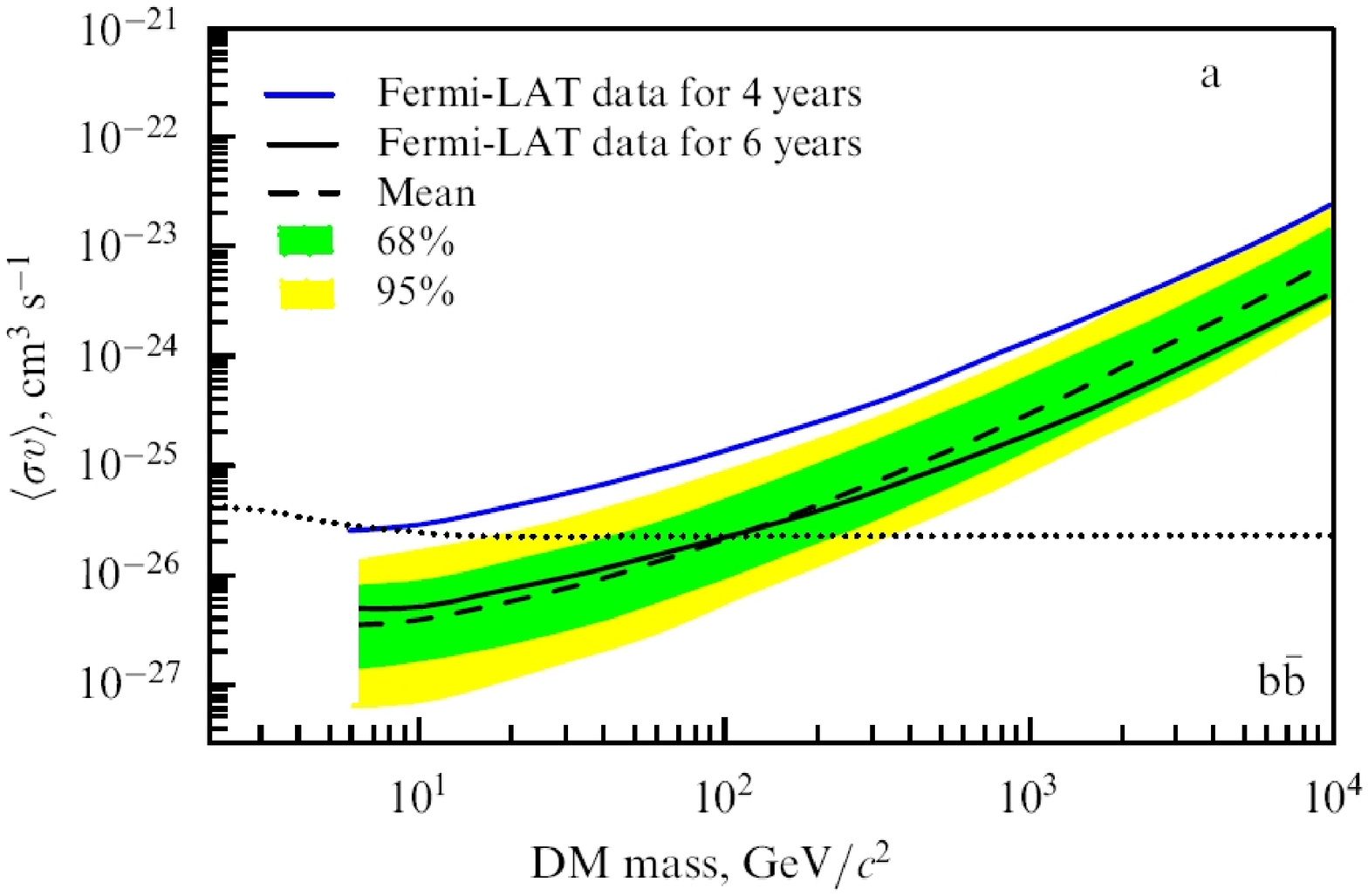} \includegraphics[width=0.4\hsize]{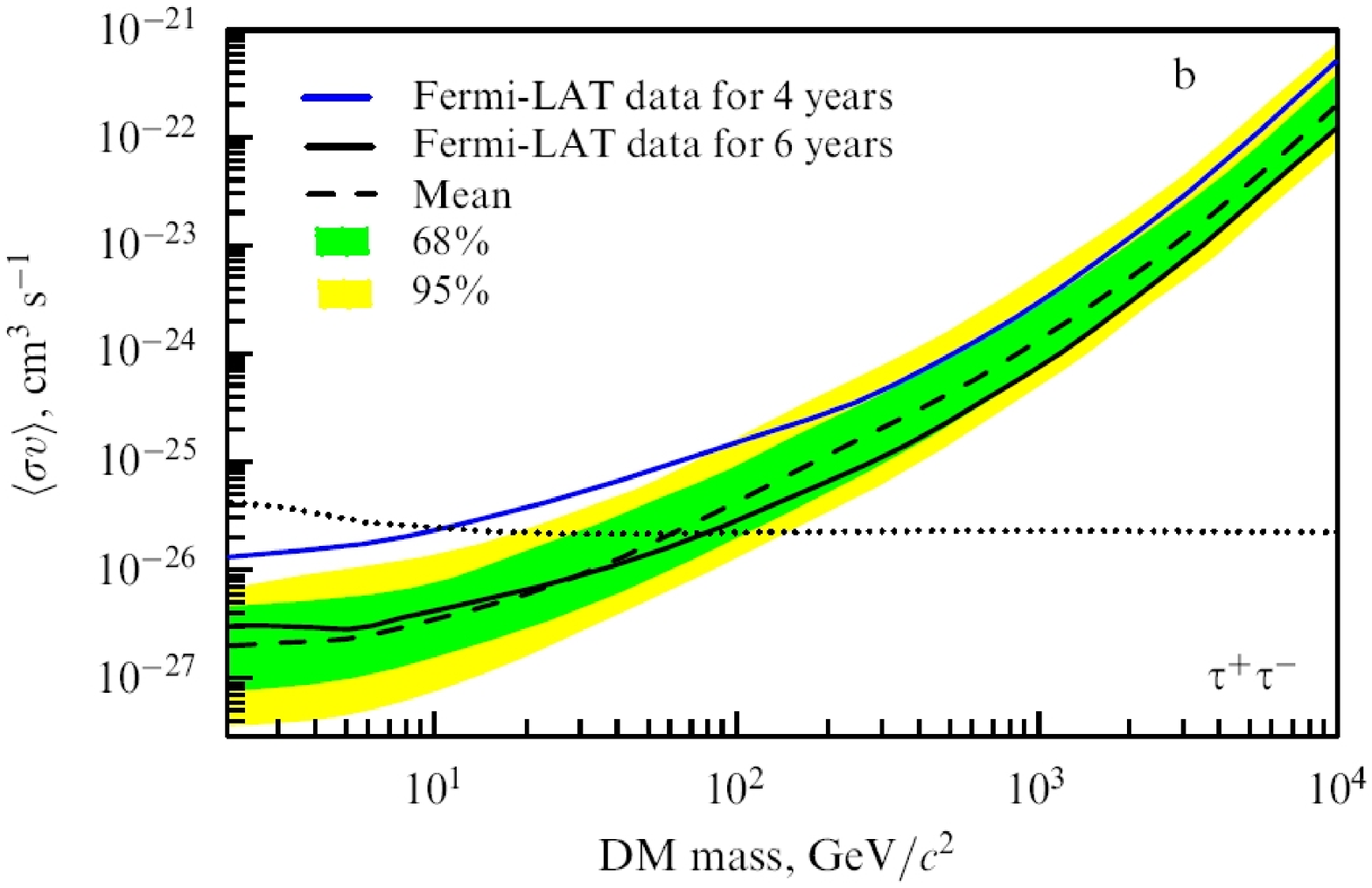}
\caption{(Color online.) Bounds on the DM particle (WIMPs) annihilation cross section as a function of its mass for two possible annihilation channels: (a) $b\bar b$ and (b) $\tau^+\tau^-$ obtained from the results of searches for excessive gamma emission in the direction of 25 dwarf spheroidal galaxies (Fermi Large Area Telescope (Fermi-LAT) data). The width of color bands corresponds to confidence levels of 68\% and 95\%. The blue solid line shows the results of previous measurements of 15 galaxies over four years. The dashed, almost horizontal, curve is the cross section calculated for an isotropic annihilation background from relic WIMPs. (From  \cite{2015PhRvL.115w1301Ackermann-Searching-DM}.)
}\label{fig::gamma_anneg1}
\end{figure}

Detection of the annihilation signal from DM particles could shed light on their nature and would provide independent constraints on dark halo parameters. This, however, is complicated by two facts.  At first, the annihilation quanta can be produced in several branched channels, which causes a big uncertainty in the expected gamma-ray flux and its spectrum. At second, the presence of the annihilation excess depends both on the model DM distribution and on background sources unrelated to DM that are difficult to be accounted for. As the rate of annihilation events per unitary volume in DM particle collisions is proportional to the product of the relative velocity of particles by their effective annihilation cross section and by their square number density, the annihilation flux (number of photons per 1 cm $^{-2}$с$^{-1}$) per second) can be expressed as
\begin{equation}
\label{equation:flux}
   F = \frac{N_{\gamma}\left<\sigma
     v\right>}{2 \, \, m_{DM}^2} \int_V \frac{\rho_{DM}^2({\bf x})}
    {4\pi \, d^2({\bf x})} \, d^3x\, ,
\end{equation}
where $N_{\gamma}$ is the number of photons in one annihilation event,
$\left<\sigma v\right>$ is the averaged product of the relative velocity by the effective annihilation cross section, $\rho_{\rm dm}$ is the DM density, $V$\, is the halo volume, $m_{dm}$  is the mass of DM particle, $d$ is the distance between the halo and observer.

An excessive $\gamma$-ray emission was detected from the Galactic center region  \cite{talk...Murgia-2014!high-energy-gamma-ray}, with the signal power being sufficiently high for the spectrum and angular resolution to be measured. However, the distinction of the annihilation signal from gamma-ray emission from other processes in the Galactic center remains difficult and uncertain problem. According to ~\cite{2006MNRAS.368.1695A}, the DM profile in the Galaxy has a cusp with logarithmic slope $>1.5$. However, later analysis of the H.E.S.S. (High Energy Stereoscopic System) data led to the opposite conclusion that the DM distribution in the Galactic center demonstrates no cusp ~\cite{2015PhRvL.114h1301A}. As a result, the data on the gamma-ray emission from the Galactic center have so far not been uniquely interpreted. Despite the possibility to explain the observations without DM, it is still preliminary to state with certainty that they contradict to model DM distributions due to uncertainties in the foreground and background sources.

Another channel of information on WIMPs annihilation is provided by searches for excessive $\gamma$-ray emission from galaxies in which DM mass strongly dominates over baryonic mass. These are dSph dwarf galaxies like satellites of our Galaxy. As the dwarf galaxies show no intensive star formation, their gamma-ray emission is almost free from the intrinsic gamma-ray background noise, which gives hope to observe more clearly DM annihilation signatures. These galaxies are studied by both space observatories and ground-based Cherenkov gamma-ray telescopes in the very high energy range. Although here the results have been negative so far, they do not exclude the existing DM models due to uncertainty of the expected flux estimates and complications related to the gamma-ray background  \cite{2015PhRvD..91f3003C}. Nevertheless, the observational data already now put certain constraints on parameters of DM particles. In Fig.~\ref{fig::gamma_anneg1}, taken from paper  \cite{2014PhRvD..90k2012A}, shown are the upper limits on the annihilation cross section for different WIMP masses calculated for two possible annihilation channels, obtained from six-years’ observations of 25 dSph galaxies by the ‘Fermi’ gamma-ray space observatory (the Fermi-LAT telescope). The nearly horizontal dotted line is the calculated annihilation cross section for the background emission from relic WIMPs assuming that they form all of DM and were in thermal equilibrium in the early Universe. In spite of the lack of the observed annihilation fluxes, it is possible to infer upper limits on the annihilation cross section and WIMP masses, although with high uncertainty.

\section{Form and structure of dark halo}

\begin{figure}[!h]
\centering\includegraphics[width=0.6\hsize]{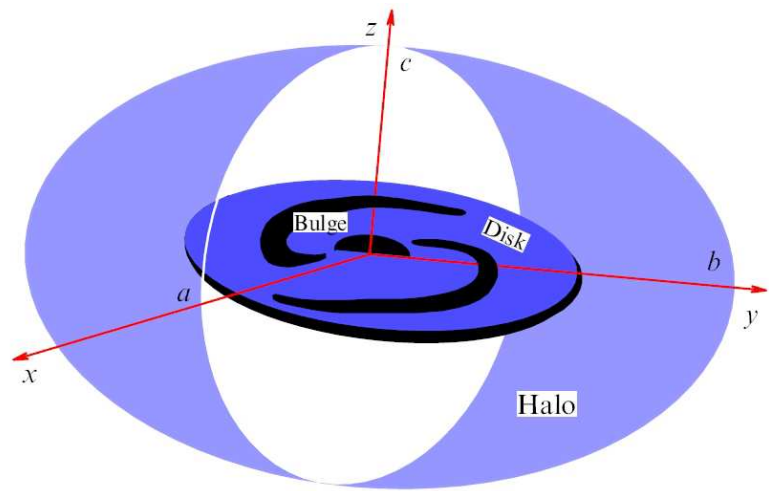}\vskip -0.\hsize
\caption{
(Color online.) Structure of a galaxy in a triaxial halo with semi-axes $a, b, c$. Shown are the halo cross section by perpendicular planes.
}
\label{fig-ShapeHalo}
\end{figure}

Cosmological models describing evolution of collisionless DM demonstrate formation of triaxial DM halos. Differences in the density profile scales along diverse directions correspond to an ellipticity  $\varepsilon$ of about $0.3$ (for a spherical halo $\varepsilon = 0$). Interaction with dissipative component changes the central ellipticity, but at the outskirts the effect of this interaction is insignificant~\cite{2004ApJ...611L..73K}.

Measuring the form and structure of dark halos can be used to test their formation models and, hence, the galaxy formation. Moreover, any statistically significant proof of non-sphericity  of DM halo argues against dark halo explanation without DM, as suggests, for example, the modified Newtonian dynamics, in which the difference from the classical Newtonian gravity field is related to the value of gravitational acceleration of particles or to the radial coordinate but does not violate isotropy.

In our Galaxy, the velocity dispersion anisotropy of halo stars can indicate a structure and possible anisotropy of the dark halo, in whose gravitational potential the stellar component had been formed ~\cite{2009ApJ...694..130M...Morrison-etal-2009!MW-Inner-Halo}. A statistical analysis of forms of disks in other spiral galaxies indirectly points to an asymmetry of the potential in the disk plane at least for some galaxies, which may suggest the motion of stars in the field of a triaxial halo. For example, according to ~\cite{2006ApJ...641..773R...Ryden-2006!Shape-Spiral-Galaxies}, in the photometric $K_s$-band the ellipticity of disks is 0.02 for late-type spirals Sc--Sd (i.e. their disks are almost circular), however it reaches 0.30 for early-type galaxies. But, despite indirect evidences for triaxial density distribution in halos, its form and ellipticity for individual galaxies have been poorly known so far, and moreover sometimes estimates of halo form are contradictory.

Some papers, for example ~\cite{2012ARep...56...16K...Khoperskov-etal-2012!Gaseous-Disks-Non-axisymmetric-Halo, 2013MNRAS.431.1230K...Khoperskov-etal-2013!stellar-disk-nonaxissymmetric-dark, 1998NewA....3..493E...El-Zant-Habler-1998!Triax-halo-disk,
2000ApJ...535L..91I,
2006ApJ...637..582Berentzen-Bar-Cuspy-Triaxial-Halos,
2006MNRAS.373.1117H...Hayashi-Navarro-2006!kinematics-disc-triaxial-halo,
2007MNRAS.377...50H...Hayashi-etal-2007!shape-gravitational-potential-halo}, studied different aspects of the effect of a non-axially symmetric gravitational potential on the dynamics of galactic disks. In particular, it was noted that the triaxia halo form complicates interpretation of the galaxy rotation curve and velocity field, especially if the dark halo dominates even in the inner galaxy (for example, in low surface brightness galaxies) ~\cite{2006MNRAS.373.1117H...Hayashi-Navarro-2006!kinematics-disc-triaxial-halo, 2008ApJ...679.1232W...Widrow-2008!Disk-Galaxies-Triaxial-Halos}. Warped galaxies can also be well described by numerical N-body models with triaxial dark halos ~\cite{2009ApJ...703.2068D...Dubinski-Chakrabarty-2009!Warps-Triaxial-Halo, 2010MNRAS.408..783R...Roskar-etal-cosmological-simulations}. Significant could be triaxial halo perturbation effects of the already existing bar ~\cite{2010MNRAS.406.2386M...Machado-Athanassoula-2010!Triax-Halo-bar, 2010ApJ...720L..62K}, \cite{2002ApJ...569L..83Athanassoula-Bar-Halo-Interaction, 2006ApJ...637..582Berentzen-Bar-Cuspy-Triaxial-Halos, 2009ApJ...697..293D}. The angular momentum exchange between the halo particles and bar can destroy the bar-structure ~\cite{2006ApJ...648..807B}.

 Nevertheless, it should be recognized that the role of halo asymmetry in the dynamical evolution of star-gas disks remains unclear. As quantitative estimates of the halo oblateness or triaxiality are sparse, in studies of their forms and effects on the disk dynamics the analysis of numerical simulations is important.

\subsection{Halo form from numerical models}

Cosmological $N$-body simulations provide a lot of information on the plausible structure of dark halos, in particular, on the form, mass distribution and kinematics of DM.  Usually, one initially assumes that in the general case dark halos have a triaxial form balanced by anisotropic velocity dispersion ~\cite{1988ApJ...327..507F, 2002ApJ...574..538J, 2006MNRAS.367.1781A...Allgood-etal-2006!shape-halo-N-body, 2007MNRAS.377...50H...Hayashi-etal-2007!shape-gravitational-potential-halo, 2007MNRAS.376..215B...Bett-etal-2007!spin-shape-dark-matter-haloes}, with the azimuthally averaged density profile of a sufficiently universal shape in a wide range of masses and redshifts (the NFW-profile in equation  (\ref{Eq-3-halo-Navarro}), see, for example, \cite{1996ApJ...462..563N, 1997ApJ...490..493N...Navarro-etal-1997!Model-halo, 2004MNRAS.349.1039N, 2010MNRAS.402...21N}).

 When studying the halo evolution it is not possible to neglect the baryonic component, which during the formation of a galaxy is concentrated to the center by forming there a star-gas system and affects the general mass distribution, including the DM profile. Models taking into account the baryonic matter feedback, including the star formation and active nucleus role in the energy powering of gas, put constraints on the dark halo parameters, in addition to those from the classical $N$-body method  \cite{1997ApJ...490..493N...Navarro-etal-1997!Model-halo, 2015MNRAS.453..469T...Tenneti-etal-2015!Galaxy-shapes-simulat}.

When stellar components of a galaxy form in the central halo, the angular momentum exchange between the dark and baryonic components must have altered the initial halo shape ~\cite{2010MNRAS.407..435Abadi-Navarro-DM-haloes}. The effective radiation cooling provides baryon concentration in the halo center to form a disk balanced by rotation. In this scenario, dark halo, via gravitational interaction with the disk, becomes more oblate with approximately constant along the radius ratio of semi-axis density distribution $s=c/a \approx 0.85$ ~\cite{2012ARep...56..664K...Khoperskov-etal-2012!cusp, 2006PhRvD..74l3522G}.

\begin{figure}[!h]
\centering\includegraphics[width=0.5\hsize]{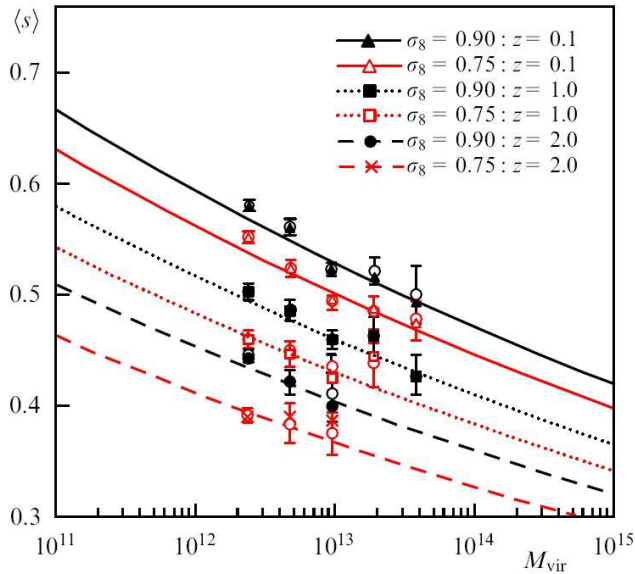}\vskip -0.\hsize

\caption{The DM halo axis ratio $s = c/a$ as a function of the halo virial mass for three redshifts $z$ and different values of the $\sigma_8$, parameter describing the fluctuation spectrum in the $\Lambda$CDM model (From \cite{2006MNRAS.367.1781A...Allgood-etal-2006!shape-halo-N-body})
 }
 \label{fig::halo_shape_vs_z_M_sigma}
\end{figure}

Different collisionless cosmological cold dark matter (CDM) models yield diverse DM halo axis ratio $s$. It depends on the initial conditions and decreases with the halo virial mass growth, and with a given mass this ratio is higher at low redshifts (see~\cite{2006MNRAS.367.1781A...Allgood-etal-2006!shape-halo-N-body} and Fig.~\ref{fig::halo_shape_vs_z_M_sigma}).  the halo form is obtained to be more close to spherical one than at early evolutionary stages. Remarkably, the halo form transformation occurs already after hierarchical DM halo formation at redshifts $z < 2$.

Cosmological models predicts a certain dependence of the  $b/a$ ratio (in the disk plane) and $c/a$ as a function of the distance to the halo center  \cite{2008ApJ...679.1232W...Widrow-2008!Disk-Galaxies-Triaxial-Halos, 2006MNRAS.367.1781A...Allgood-etal-2006!shape-halo-N-body,2007MNRAS.377...50H...Hayashi-etal-2007!shape-gravitational-potential-halo}. As a rule, the central halo is more asymmetric than at the periphery, where $b/a$, $c/a$ asymptotically tend to unity. This effect is found in most numerical cosmological models, including in those with a large number of particles $N\geq10^8$ (for example, in the Via Lactea project \cite{2011ASL.....4..297D...Diemand-Moore-2011!Review-sub-halo}).
 However, quantitative results in various models can be strongly different depending not only on the numerical model parameters, but also on the spectrum of initial perturbations  (see Fig.~\ref{fig::halo_shape_vs_z_M_sigma}).

\begin{figure}[!h]
\centering\includegraphics[width=0.5\hsize]{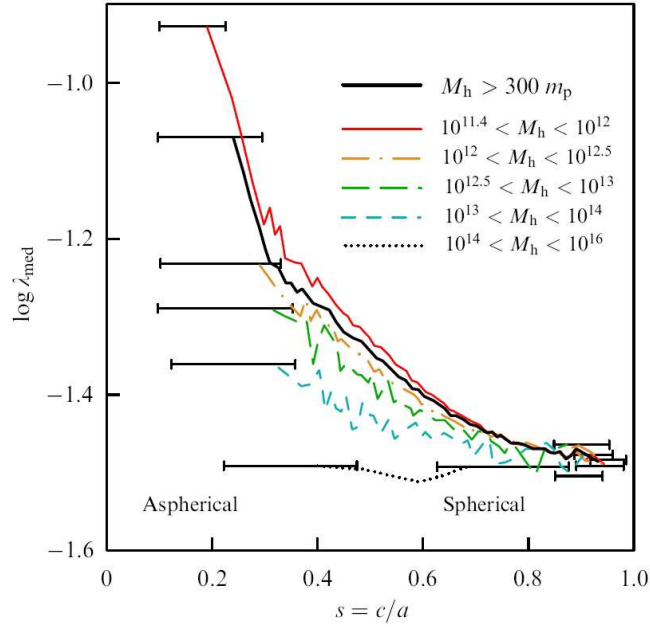}\vskip -0.\hsize
\caption{(In color online.) The halo shape at  $z=0$ vs. median spin $\lambda_{med}$  according to N-body modelling (from paper  \cite{2007MNRAS.376..215B...Bett-etal-2007!spin-shape-dark-matter-haloes}) for halos with different virial mass. The dark curve is the mean value over more than 300 particles. For each plot, the horizontal bars show the range inside which the averaging was done. There is an anticorrelation between the spin and the DM halo axis ratio, which is the strongest for aspherical halos.
 }
\label{fig::halo_shape_spin_relation}
\end{figure}

Presently, a possible halo rotation also found only from numerical cosmological simulations. DM halos, despite their being balanced as a whole not by rotation but by velocity dispersion, can have non-zero integral angular momentum. Analysis of various non-dissipative cosmological simulations suggests that more ‘spherical’  halos rotate significantly slower than non-spherical ones (see, e.g., papers \cite{2007MNRAS.376..215B...Bett-etal-2007!spin-shape-dark-matter-haloes, 2007MNRAS.378...55M} and Fig.~\ref{fig::halo_shape_spin_relation}). Note that the reliability of model relations suffers from halos being non-fully relaxed systems, and besides they can be subject to tidal interaction from nearby galaxies.

\subsection{Galactic polar rings and tidal streams as dark matter markers}

\begin{figure}[!h]
\centering\includegraphics[width=0.4\hsize]{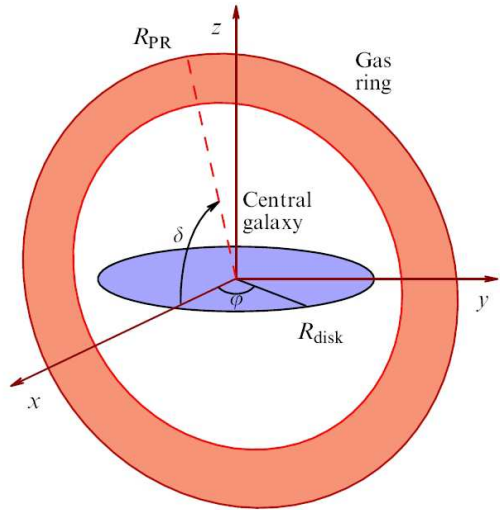}\vskip -0.\hsize
\caption{(In color online.) Structure of a polar-ring (PR) galaxy. In blue shown in the central galaxy (most frequently, S0), in red --– the polar ring mostly consisting of gas. }
\label{fig-polar-ring-gal}
\end{figure}

Polar rings are extended looped or non-closed rings of stars and gas-dust medium rotating around the galaxy center in the plane inclined (in disk galaxies) by large angle to the main galactic disk (Fig.~\ref{fig-polar-ring-gal}, \cite{1996A&A...305..763C, 1994A&A...291...57R, 2004A&A...416..889R, Brosch2010, 2015BaltA..24...76Moiseev}). There are around 300 galaxies with polar rings (PRGs) \cite{2011MNRAS.418..244M...Moiseev-etal-2011!catalogue-polar-ring-galaxies}. In rare occasions the angle $\delta$between the galaxy and polar ring can be strongly different from $90^\circ$, for example in the galaxy SPRC-7  $\delta=73^\circ$~\cite{2014MNRAS.441.2650K...Khoperskov-etal-2014!shape-DM-polar-ring}.

Polar rings around galaxies are thought to form due to close passing of a neighbor galaxy (satellite) accompanying by loss of some mass of the latter ~\cite{1990AJ....100.1489W, 2012MNRAS.425.1967S}, or by its total tidal disruption~\cite{1997A&A...325..933R}, or they result from accretion of cold gas on the galaxy from cosmological filaments ~\cite{2009ApJ...696L...6S, 2010ApJ...714.1081S}. Irrespective of the formation scenario, the existence of two connected galactic substructures rotating in mutually perpendicular planes enables estimating three-dimensional gravitational potential in the galaxy, including the mass and form of DM halo~\cite{2015BaltA..24...76Moiseev, 2013A&A...554A..11C}. Initially, the ring-disk inclination could be arbitrary. The dominance of polar rings suggests that ring planes rapidly evolve to orthogonal orientation. An important argument in favor of non-axially symmetric halo in PRGs is the stability of polar orbits exactly for such gravitational potentials~\cite{1982ApJ...263L..51S}.

The first sample including confirmed PRGs, PRG candidates and most probable PRGs  was published in~\cite{1990AJ....100.1489W}.The list of known PRGs was significantly extended based on the SDSS data  ~\cite{2011MNRAS.418..244M...Moiseev-etal-2011!catalogue-polar-ring-galaxies}.
Polar rings are mainly observed around Sa--S0 galaxies, which have no significant amount of gas at the periphery and therefore there are no complex interactions of supersonic gas streams. Due to small amount of cold gas and weak emission, the rotation curve of such galaxies is known, as a rule, only for innermost parts, but spectral observations of a ring with size many times as large as the exponential scale of the main galaxy disk offers the possibility to measure the circular velocity at the galactic periphery. To compare rotational velocities of the disk and ring, it is very important to ‘continue’ the rotation curve as long as possible from the galactic center using absorption lines in stellar spectra and to match it with the gas kinematics in the polar ring \cite{1994ApJ...436..629S, 2003ApJ...585..730I, 2014ASPC..486..221K}.

\begin{figure}[!h]
\centering\includegraphics[width=0.5\hsize]{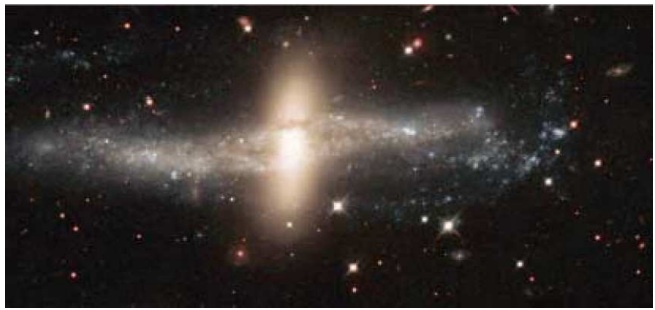}\vskip -0.\hsize
\caption{One of the most well studied polar-ring galaxy NGC\,4650A (the Hubble Space Telescope).
 }
\label{fig-NGC-4650A}
\end{figure}

Estimates of mass and form of dark halo for several polar-ring galaxies were obtained as early as in the 1980s  ~\cite{1987ApJ...314..439W...Whitmore-etal-1987!shape-DM-polar-ring-galaxies}. However, even at present the results remain quite contradictive. For example, for the best studied PRG, NGC 4650A, first measurements of the halo form based on isolines of gravitational potential suggested a nearly spherical halo with semi-axis ratio  $c/a \sim 0.86 \pm 0.21
 $~\cite{1987ApJ...314..439W...Whitmore-etal-1987!shape-DM-polar-ring-galaxies}. A similar result was obtained for another two PRGs (ESO 415-G26 and A0136-0801) from comparison of two maximum rotational velocity in the polar ring and central galaxy disk. Later, a more careful analysis of the rotation curves based on the decomposition of circular velocity in two mutually orthogonal directions for NGC 4650A (Fig.\,\ref{fig-NGC-4650A})suggested a strongly oblate halo towards the central galaxy plane~$c/a \sim 0.3 - 0.4$~\cite{1994ApJ...436..629S}.

\begin{figure}[!h]
\centering\includegraphics[width=0.6\hsize]{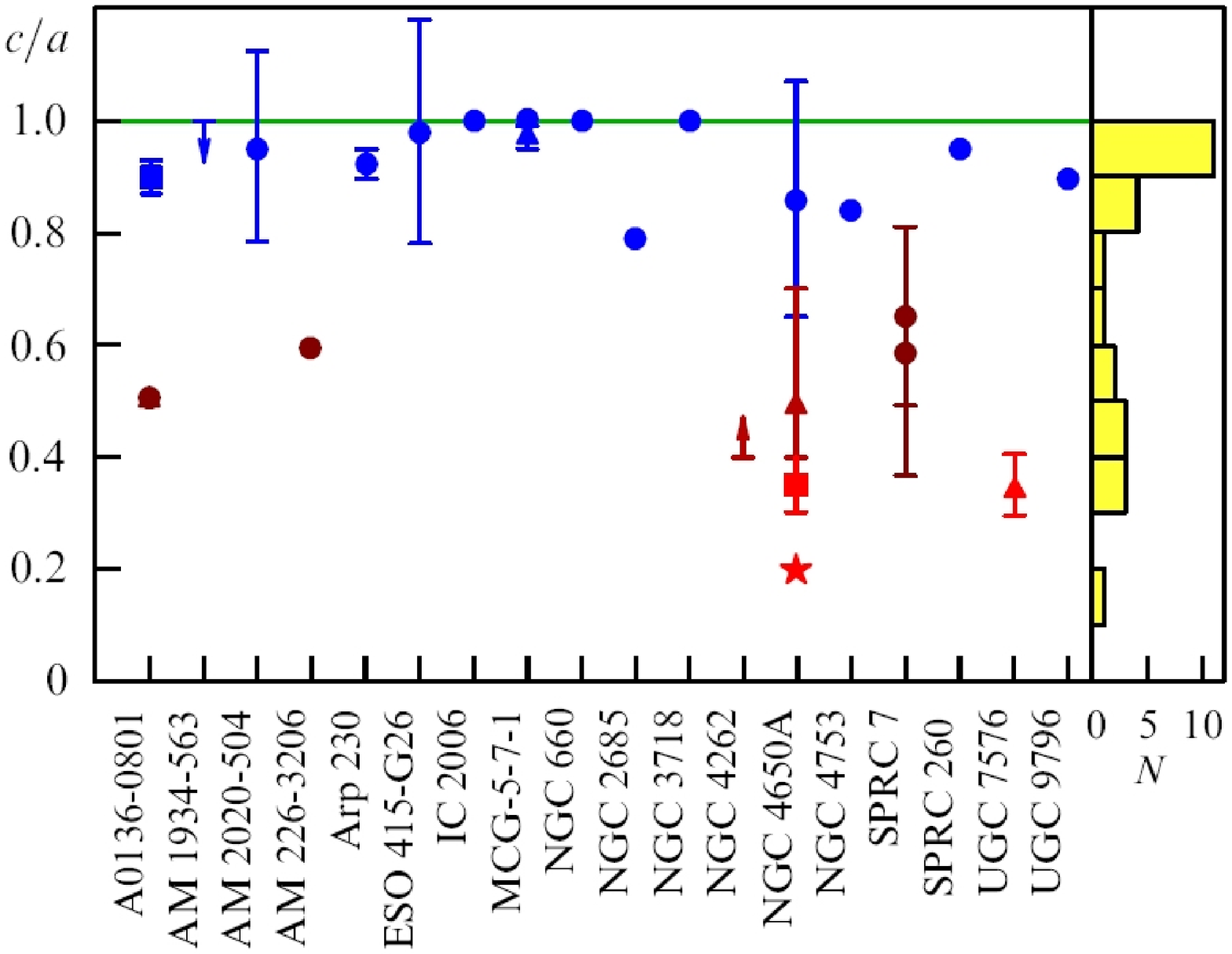}\vskip -0.\hsize
\caption{(In color online.) The halo oblateness  $c/a$ in polar-ring galaxies compiled from the literature. To the right is the number of galaxies with given axis ratio. References: A0136-0801~\cite{1994ApJ...436..629S,1995AIPC..336..141S}, AM~1934-563~\cite{2007MNRAS.382.1809B}, AM~2020-504~\cite{1992NYASA.675..207A}, AM~226-3206~\cite{1987ApJ...314..439W...Whitmore-etal-1987!shape-DM-polar-ring-galaxies},
Arp~230~\cite{2013AJ....145...34S}, ESO~415-G26~\cite{1987ApJ...314..439W}, IC~2006~\cite{1994ApJ...436..642F},
MCG-5-7-1~\cite{1996ASPC..106..168C, 2013AJ....145...34S},
NGC~660~\cite{1995AJ....109..942V}, NGC~2685~\cite{1993AJ....105.1378P}, NGC~3718~\cite{2009AJ....137.3976S...Sparke-etal-2009!Twisted-Gas-Disk},
NGC~4262~\cite{2014MNRAS.441.2650K...Khoperskov-etal-2014!shape-DM-polar-ring}, NGC~4650A~\cite{1987ApJ...314..439W...Whitmore-etal-1987!shape-DM-polar-ring-galaxies,1990ApJ...361..408S,1994ApJ...436..629S,1996A&A...305..763C}, NGC~4753~\cite{1992AJ....104.1339S...Steiman-Cameron1994},
NGC~5122~\cite{1996ASPC..106..168C}, NGC~5907~\cite{2000AstL...26..277R}, SPRC~7~\cite{2014MNRAS.441.2650K...Khoperskov-etal-2014!shape-DM-polar-ring}, SPRC~260~\cite{2013MSAIS..25...51K},
UGC~4261~\cite{1998ARep...42..439R},
UGC~7576~\cite{2002sgdh.conf..178S},
UGC~9796~\cite{2006AJ....131..828C}. (Data on NGC 5122, NGC 5907 and UGC 4261 see in the cited references.)
 }
\label{fig::polar_ring_stat}
\end{figure}

Iodice et al.~\cite{2003ApJ...585..730I}  attempted to estimate deviations of the gravitational potential from spherical form from the location of PRGs on the Tully-Fisher diagram relating the galactic rotation velocity with its luminosity. Using the observed Doppler width of the integral HI line profile as indicator of the rotation velocity, the authors ~\cite{2003ApJ...585..730I} found that the position of PRGs on the Tully-Fisher diagram are noticeably different from ordinary galaxies: most of PRGs are shifted towards larger velocities. Comparison with model disk galaxies with polar rings made it possible to interpret this feature by interaction of the ring with a halo extended along the minor disk semi-axis ~\cite{2010AIPC.1240..379I, 2008ASPC..396..483I}. For NGC 4650A, the semi-axis ratio $c/a$ is close to two ~\cite{2003ApJ...585..730I}.
  Fig.~\ref{fig::polar_ring_stat} presents the data on the dark halo form compiled from the literature.

An effective approach to determine the halo parameters is based on constructing a numerical self-consistent gas-dynamic N-body model for the disk and polar ring with taking into account of the dark halo potential. Form comparison of the results of numerical simulations with photometric and kinematic parameters of the star-gas disk of the main galaxy, all basic halo characteristic can be obtained. In some cases a sufficiently complicated dark mass distribution was found \cite{2014MNRAS.441.2650K...Khoperskov-etal-2014!shape-DM-polar-ring}. For example, the dark halo form in NGC 4262 was found to depend on radius: it is strongly flattened immediately close to the galactic disk ($c/a\simeq 0.4$), but stretched along the minor semi-axis far away from the center  ($c/a\sim 2$)~\cite{2014MNRAS.441.2650K...Khoperskov-etal-2014!shape-DM-polar-ring}.
 The similar complex character of DM distribution follows from numerical cosmological simulations of PRG formation. Paper ~\cite{2012MNRAS.425.1967S} revealed the change in the space orientation of the major halo semi-axis, which in fact means the change in the halo form with distance. For example, in the central parts of a galaxy the halo can be flattened towards the central galaxy plane, while far away from the center the halo can be flattened towards the perpendicular plane. With such a complicated halo form it is not surprising to obtain contradictory estimates of the halo oblateness in galaxies like NGC 4650A.  Confirming the changeable halo form for polar-ring galaxies can significantly change our understanding of the structure and dark matter mass both in ordinary disk and elliptical galaxies.

Another information about the halo from can be inferred from stellar tidal streams found in our Galaxy halo and in some other galaxies, which resulted from tidal disruption of dwarf galaxies (see, for example, the review by V.~Belokurov ~\cite{2013NewAR..57..100B}, as well as \cite{2007ApJ...668..221N, 2007ApJ...657L..89B...Belokurov-etal-2007!Hercules-Aquila-Cloud, 1994ApJ...436..629S...Sackett-etal-1994!dark-halo-NGC5907, 2007ApJ...671.1591I...Ibata-etal-2001!M31-halo}).  In fact, these are test halo objects, velocity change along which reflects the matter distribution far from the disk. The dynamics of destroying satellites depends on the main halo mass profile, therefore the spatial and kinematic parameters of formations observed far from the disk plane can be very important tracers of dark halo form and orientation.

Measurements of the observed form and radial velocity distribution along tidal streams can significantly constrain the possible form of equipotential halo surfaces \cite{2014ApJ...794....4P...Price-Whelan-etal-2014!MW-halo}. From observations of the Sagittarius tidal stream, Sagittarius dSph ~\cite{2007ApJ...657L..89B...Belokurov-etal-2007!Hercules-Aquila-Cloud} (Sagittarius tidal
stream, Sagittarius dSph)  the halo axis ration was derived to be $c/a\geq 0.8$, i.e. the halo can be spherical
\cite{2005ApJ...619..800J...Johnston-etal-2005!Sagittarius-Dwarf-halo-MW}, but if this stream is dynamically young,
$c/a\simeq 0.6$ is required ~\cite{2004MNRAS.351..643H...Helmi-2004!Is-dark-halo-our-Galaxy-spherical}. Later a three-dimensional N-body model of the stream motion in triaxial halo was constructed, which simultaneously explained its location and kinematics for the halo axis ratio  $c/a$=0.72, with the halo minor semi-axis lying in the stellar disk plane (!) ~\cite{2010ApJ...714..229Law-Sagittarius-Triaxial-Milky-Way-Halo}.

Studies of the dark halo structure from observations of the GD-1 tidal stream \cite{0004-637X-768-2-171} seem to be very promising. The GD-1 stream is a long and very thin formation with well distinguished density disruptions (about ten) along the length. From the numerical model it was proposed that one of the possible reasons for the density disruptions could be fly-by of a subhalo near the stream. This requires that in the Galaxy there should be about 100 subhalos with masses $\leq 10^6M_\odot$ inside the radial distance to the stream orbit apocenter (about 30 kpc)..

Studies of the stellar tidal stream in the galaxy NGC 5907 suggested  $c/a=$0.5 for the halo ~\cite{1994Natur.371..398C,2008ApJ...689..184M...Martinez-Delgado-etal-2008!NGC5907-stream}.

\subsection{Other sources of information about dark halo}

The halo form affects the distribution and velocity of motion of all its components. For example, the kinematics of globular clusters and dwarf satellites of the Andromeda Nebula suggests prolate dark halo form with major semi-axis perpendicular to the stellar disk plane ~\cite{2014ApJ...789...62Hayashi-DM-Andromeda}. Note that large stellar halo of the Andromeda Nebula extending up to several optical radii has a form excluding strong oblateness of the external halo towards the disk plane and rather suggests prolateness in the perpendicular direction \cite{2007ApJ...671.1591I...Ibata-etal-2001!M31-halo}.

However, the correspondence between the dark halo and stellar halo forms remains open. For example, the form of the latter in our Galaxy was inferred from distribution of stars near the turning point of the main sequence obtained in the survey by the Canada-France telescope on Hawaii (CFHTLS). The halo axis ratio was found to be $c/a = 0.70\pm 0.1$ (within the galactocentric distance $5-35$\,kpc), which almost coincides with the estimate $c/a=0.72$, as inferred for the dark halo from the kinematics of the Sagittarius tidal stream ~\cite{2010ApJ...714..229Law-Sagittarius-Triaxial-Milky-Way-Halo}. However, directions of halo minor semi-axis found in these papers do not coincide.

The characteristic halo axis ratio can be found from gravitational lens statistics, although this method is reliable only for axially symmetric halo models. To determine the halo form, weak gravitational lensing seems to be the most appropriate: initially, an ellipsoidal halo model is assumed and its parameters are determined from best-fit of observational data  \cite{2012A&A...545A..71V...van-Uitert-etal-2012!shapes-haloes-weak-lensing}. First such measurements of the halo ellipticity were carried out in paper~\cite{2004ApJ...606...67H}, which found the ratio of the halo ellipticity $\varepsilon$ to the brightness distribution ellipticity to be $0.77^{+0.18}_{-0.21}$, i.e. the halo turned out to be more ‘spherical’ than the visible matter distribution. The account for ellipticity of lenses made it possible to find the DM halo ellipticity  equal to 0.3. According to ~\cite{2007ApJ...669...21P...Parker-etal-2007!Masses-Shapes-DM-lens},  halos around elliptical galaxies seem to be even more oblate.

The weak lensing method was applied in paper \cite{2015arXiv150603536Clampitt-Lensing-Ellipticity-DM} to about 70 thousand giant elliptical galaxies (from the SDSS digital sky survey) with virial masses $\ge 10^{13}M_\odot$
The best-fit results correspond to a halo with axis ratio  $c/a=0.78$. The halos turn out to be more spherical than for galaxies with an order-of-magnitude smaller masses ($c/a=0.62$ according to paper  \cite{2015JCAP...01..009Adhikari-lensing-DM-shapes}). However, the account for the possible difference between symmetry axis in dark mass and galaxy brightness  distributions, which can be expected from numerical cosmological models, decreases the resulting  $c/a$, ratio and brings these estimates for galaxies with different masses in agreement.

Strong gravitational lensing was used to estimate the degree of oblateness of dark halo of individual galaxies only in several cases. For example, the strong lensing data in combination with the results of analysis of star and gas kinematics in the galaxy SDSS J2141 at redshift $z=0.14$
suggested that the galaxy has a significant (but not dominant inside the optical disk limits) DM halo mass, which is satisfactorily fitted by the NFW profile and, apparently, is slightly ‘compressed’ towards the galaxy plane $(c/a =0.75^{+0.27}_{-0.16})$ \cite{2012MNRAS.423.1073Barnabe-SWELLS-DM-lens}.

Another information on the halo form can be obtained from measuring the cold gas thickness in the disk.  The equilibrium gas layer thickness depends on both the stellar disk density and halo density and the gas itself, which increasingly contributes to the total disk gravitational potential with radial distance. Therefore, the dark halo form within the optical disk of a galaxy and nearby can be probed from the analysis of the HI gas layer thickness change with distance from the center (flaring gas layer), assuming the gas turbulent velocity dispersion isotropic. Of course, the most reliable results are obtained for edge-on galaxies, since in them the gas layer thickness can be directly measured \cite{2010A&A...515A..63O...Brien-etal-2010!dark-matter-halo-shape-UGC7321}\cite{2013arXiv1303.2463Peters-Shape-Dark-Matter-Edge-on}. As shown in paper   \cite{2011ApJ...732L...8Banerjee-DM-Traced-HI-Gas} the radial profile of the HI layer thickness in our Galaxy better agrees with a halo model prolate along the minor axis of the Galaxy than with a spherical halo model: the axis ratio reaches $c/a=2$ at $r=24$\,kpc. However, for the edge-on galaxy UGC 7321 observations of HI suggest a gravitational potential distribution close to that of spherical isothermal halo model  \cite{2010A&A...515A..63O...Brien-etal-2010!dark-matter-halo-shape-UGC7321}. The radial dependence of the HI vertical scale in M31 is bet-fit by an oblate halo model with axis ratio $c/a=0.4$ \cite{2008ApJ...685..254B...Banerjee-Jog-2008!M31-HI-Shape-halo}.

Thus, for diverse galaxies contradictory results have been obtained. Note that the methods of determination of the halo shape from flat components should be applied with care, since the results are model-dependent and little sensitive to the halo density distribution at large distances from the disk plane. In addition, for periphery regions of the gas disk the hydrostatic equilibrium assumption can be too crude.

Thus, the existing methods of the dark halo shape determination suggest plausible deviation of the halo density from spherical symmetry, and not only near the disk plane but also far beyond its limits, although quantitative estimates of the halo form for individual galaxies are not always in agreement with each other, and the whole picture remains quite controversial. Nevertheless, gradually we came to the understanding that can have a triaxial spatial form and, moreover, the orientation of the disk and halo axes can differ significantly (see models from paper \cite{2013MNRAS.434.2971D}).

Let us enumerate possible processes capable of affecting the halo shape.  Dissipationless cosmological models predict triaxial halos whose form reflects anisotropic character of matter compression during their formation. The account for hydrodynamic processes in baryonic matter (gas) which forms dissipative medium inside a collisionless halo must decrease the triaxiality effect by making the halo shape more spherical (see, for example,  \cite{2010MNRAS.407..435Abadi-Navarro-DM-haloes}). At the same time, the formation of massive disks at halo centers inevitably leads to the halo flattening towards the galaxy plane. The efficiency of these processes is different at diverse space scales, which complicates the picture. In addition, accretion of matter onto the halo from intergalactic space, as well as escape of gas from galactic disks, continues after the galaxy formation, which affects the general mass distribution. The most general statement is that the mass distribution in the halo can be anisotropic and different at diverse distances from the center. It will be also correct to accept that galaxies can have various halo shapes.

\subsection{Subhalo}

Let us also draw attention to the possible structure of dark halos: extended galaxy halos can include, in addition to spatially distributed DM, smaller structures that could form during initial perturbations growth (see  \cite{2007ApJ...671.1135K...Kuhlen-etal-2007!shapes-orientation-subhalo, 2006ApJ...649....1D..Diemand-etal-2006!sub-halo} and references therein to earlier papers). High-resolution cosmological models demonstrate systems of substructures also known in the literature as subhalo, minihalo, microhalo (which can be called ‘halolets’). The typical number of such ‘halolets’ with   $\gtrsim 10^{-6}M_{vir}$ (about $10^6M\odot$)in a large halo with size  $R_{vir}=400$\,kpc in the Via Lactea simulations exceeds  $10^4$  \cite{2006ApJ...649....1D..Diemand-etal-2006!sub-halo}, and in the Via Lactea II, GHALO (Galactic dark matter halo) models their number can be as high as  $10^5$ 200 kpc region. These gravitationally bound formations represent in turn triaxial structures with $a\ne b\ne c$, slowly changing over the cosmological time. Their relative scales are characterized by the axis ratios  $\langle q \rangle=b/a = 0.8-0.9$, $\langle s \rangle=0.6-0.9$,  and they have a complicated spectrum of masses, sizes, spin and internal kinematics  \cite{2011ASL.....4..297D...Diemand-Moore-2011!Review-sub-halo}. There are traces of internal structure of the ‘halolets’ themselves, therefore it is possible to believe that in a dark halo, a complex hierarchical system of scales –- from several hundred kiloparsec to fractions of a kiloparsec –- is formed.

 The halo mass functions inferred from numerical structure formation models are in agreement with the spectrum of masses (internal velocities) of galaxies and their systems, however the observed number of dwarf galaxies-satellites in the Milky Way and other large galaxies turns out to be to small compared to the number of more massive galaxies \cite{2011MNRAS.415L..40Boylan-Kolchin-massive-MW-subhaloes, 2014MNRAS.444..222Garrison-Kimmel-fail-Local-Group}. Therefore, the problem of low-mass subhalo detection remain actual, providing additional verification of cosmological models. Being a mixture of dark and baryonic matter, subhalos can both contain stars and be fully starless. In the first case they must be observed as dwarf satellites of DM-dominated galaxies, while in the second case they can be difficult to discover. A large number of fully dark dwarf galaxies almost free from baryonic matter may exist: gas can be lost in the course of evolution due to ionization  or sweeping-out when moving in a medium, and stars had no possibility to form. If this is confirmed, the precision of the very term ‘galaxy’ will be needed.

 The dynamic to baryonic mass ratio in very low-mass dwarf galaxies amounts to several hundred solar units, which is likely to related to the lack of conditions for active star formation  in these objects at all stages of their evolution due to low density or small amount of gas. When cold gas is present, subhalos can be observed as compact HI regions with very low and even undetectable optical luminosity, but they can be difficult to separate from tidally formed clouds observed near interacting galaxies without analyzing their internal motions \cite{2015AJ....149...72C...Cannon-2015}.  Amazingly, recently in one of the HI clouds near the dwarf galaxy Leo T traces of regular gas rotation were discovered enabling the determination of its dynamical mass  ($\sim 10^8 M_\odot$),  which is several dozen times as high as the observed gas mass. There are no signatures of stars in the cloud \cite{2015A&A...573L...3A...Adams-2015}, i.e. the mass-luminosity ratio is extremely high: stars in this structure, apparently, had not formed at all.

 Inconsistency of the observed number of dwarf satellites to the CDM-model predictions for subhalos is an open and actively discussed issue (see, for example, \cite{2013MNRAS.428.1696Vera-Ciro-dark-haloes-dwarf-MW},\cite{2015MNRAS.453.2133Brook-DM-galaxy-populations},\cite {2015arXiv151108741Papastergis-dwarf-galaxies} and references therein). The problem is partially relaxed allowing for the effect of baryonic matter and star formation on the DM density profile and distribution inside a subhalo \cite{2015arXiv151108741Papastergis-dwarf-galaxies}.One of the proposed solutions is to use a cosmological model with ‘warm’ dark matter, in which low-mass subhalos will have a low central density, which changes the relation between the observed and predicted number of dwarf satellites  with  given velocity of internal motions \cite{2012MNRAS.420.2318Lovell-haloes-satellite-galaxies-warm-DM-universe}. It is also not excluded that the number of subhalos is indeed very high, but in the vast majority of cases they do not contain sufficient amount of baryons to be detectable by radiation.

\section{Dark halo and structure details of the disk}

\subsection{Dark halo and galactic bars}

A significant fraction of S-galaxies belongs to SB type, in centers of which an oval-like stellar structure called bar is observed. The size of a typical stellar bar varies in a wide range $r_{bar}\simeq (0.1\div 0.5)R_{opt}$, i.e. sufficiently long bars go well outside the region where the bulge gravitation dominates, and hence the bars are controlled by the disk and halo gravitational field. Stellar bars form due to gravitational instability of the so-called bar-mode. The main progress in its studies is obtained from N-body numerical modelling pioneered in paper  \cite{1973ApJ...186..467O...Ostriker-Peebles-1973!Nbody}, in which the disk stability condition relative to the global bar-mode was formulated: $T_{rot}<0.28 T$ (where $T$ is the total kinetic energy of stars, $T_{rot}$ is the kinetic energy of rotation). Gravitational field of the halo decreases the effect of the disk self-gravitation by hampering the bar formation owing to the gravitational instability suppression.

 As for a collisionless gravitating system $T_{\textrm{rot}}/T \sim V^2/(V^2 + 2\,c^2_r)$, to within an order of magnitude we have $c_r \sim V$, i.e. without spheroidal component of a galaxy the disk should be very hot to stabilize the bar-mode.  The bar-mode stabilization conditions depend on many factors (the spatial distribution of matter density, kinematical parameters, the bulge and its properties)  \cite{Fridman-Khoperskov-2011!book}, however under other equal conditions a more massive halo stabilizes the bar-mode at smaller stellar velocity dispersion in the disk.

Well-distinguished bars are observed in almost one third part of disk galaxies, and with account for weaker bars best-seen in IR images, the fraction of barred galaxies exceeds 50\%. This evidences for long life of galactic bars. Nevertheless, a bar, by rigidly rotating inside a galaxy, can evolve and can be destroyed by gravitational interaction with slowly rotating components (halo and/or bulge). This interaction is similar to the dynamical friction of a massive body moving relative to the surrounding stars. By the loss of angular momentum, the bar length  $r_{bar}$ should increase with time, and its angular velocity  $\Omega_{bar}$ should decrease and with time the bar should ‘dissolve’, by dynamically heating up the ambient inner disk (see, for example, \cite{2000ApJ...543..704Debattista-Dynamical-Friction-DM-Bar, 2013MSAIS..25...45Combes-DM-evolution-galaxy-disks}). The bar, however, will not be long-lived at a high matter density in the galactic center (the result of scattering of stellar orbits), which could initially exist or appear, for example, due to motion of part of gas towards the disk center.

By allowing for not only the effect of a halo on the disk and its structure, but also the back reaction of bar on the halo, i.e. by passing from a ‘rigid’ to ‘live’ halo model, the picture becomes not so simple: although at the initial stage of the bar growth its formation occurs more rapidly in the self-gravitating disk, at the non-linear stage, due to resonance interaction effects with the halo, the bar reaches a large amplitude, if the halo mass is comparable to or even somewhat exceeds the disk mass \cite{2002ApJ...569L..83Athanassoula-Bar-Halo-Interaction}. In other words, the increase of the bar amplitude inside a ‘live’ axially symmetric halo occurs even in the case where the disk is not self-gravitating: it is related to the angular momentum transfer from the bar to a slowly rotating or non-rotating halo. The triaxiality of the  ‘live’ dark halo further complicates the picture of its interaction with the bar and can destroy the bar that have angular velocity different from that of the halo \cite{2006ApJ...637..582Berentzen-Bar-Cuspy-Triaxial-Halos}.

To conclude, the bar effect on the formation and time of existence of the bar in a galactic disk seems to be obvious and is supported by numerical simulations, but the bar-interaction process turns out to be enough complicated, so that no unique correspondence between halo parameters and the presence of a bar in galaxies is observed.

\subsection{Dark halo triaxiality and spiral structure of the disk}

The lack of central symmetry in dark matter distribution, prolateness or triaxiality of halo discussed in Section 5, pose question on the influence of such features on the galactic disk dynamics  \cite{1998NewA....3..493E...El-Zant-Habler-1998!Triax-halo-disk, 2002ApJ...574L..21B...Bekki-Freeman-2002!Spiral-Structure-Triaxial-Halo, 2003ApJ...586..152M...Masset-Bureau-2003!Spiral-Structure-NGC2915-Triax-Halo}. Different aspects of the non-axial symmetry of gravitational potential on the disk dynamics were considered in papers \cite{2006ARep...50..785T...Tutukov-Fedorova-2006!dark-haloes-spiral-patterns, 2012ARep...56...16K...Khoperskov-etal-2012!Gaseous-Disks-Non-axisymmetric-Halo, 2013MNRAS.431.1230K...Khoperskov-etal-2013!stellar-disk-nonaxissymmetric-dark, 2006MNRAS.373.1117H...Hayashi-Navarro-2006!kinematics-disc-triaxial-halo, 2007MNRAS.377...50H...Hayashi-etal-2007!shape-gravitational-potential-halo}.
The halo triaxiality was invoked to explain gas velocity field in galaxies, in particular, in LSB-galaxies  \cite{2006MNRAS.373.1117H...Hayashi-Navarro-2006!kinematics-disc-triaxial-halo, 2008ApJ...679.1232W...Widrow-2008!Disk-Galaxies-Triaxial-Halos}.
 The central bar generation and features of its dynamics in the presence of a non-axially symmetric halo have been studied in detail (see, for example, \cite{2010MNRAS.406.2386M...Machado-Athanassoula-2010!Triax-Halo-bar, 2007ApJ...671..226H...Heller-etal-2007!Triaxial-Halos-Disks, 2008ApJ...687L..13R...Romano-Diaz-etal-2008!Triax-halo-disk}.
 The disk modeling in a triaxial halo also provides explanation to the nature of observed warps at the galactic disks periphery \cite{2009ApJ...703.2068D...Dubinski-Chakrabarty-2009!Warps-Triaxial-Halo, 2010MNRAS.408..783R...Roskar-etal-cosmological-simulations}.

\begin{figure}
  \centering\includegraphics[width=0.8\hsize]{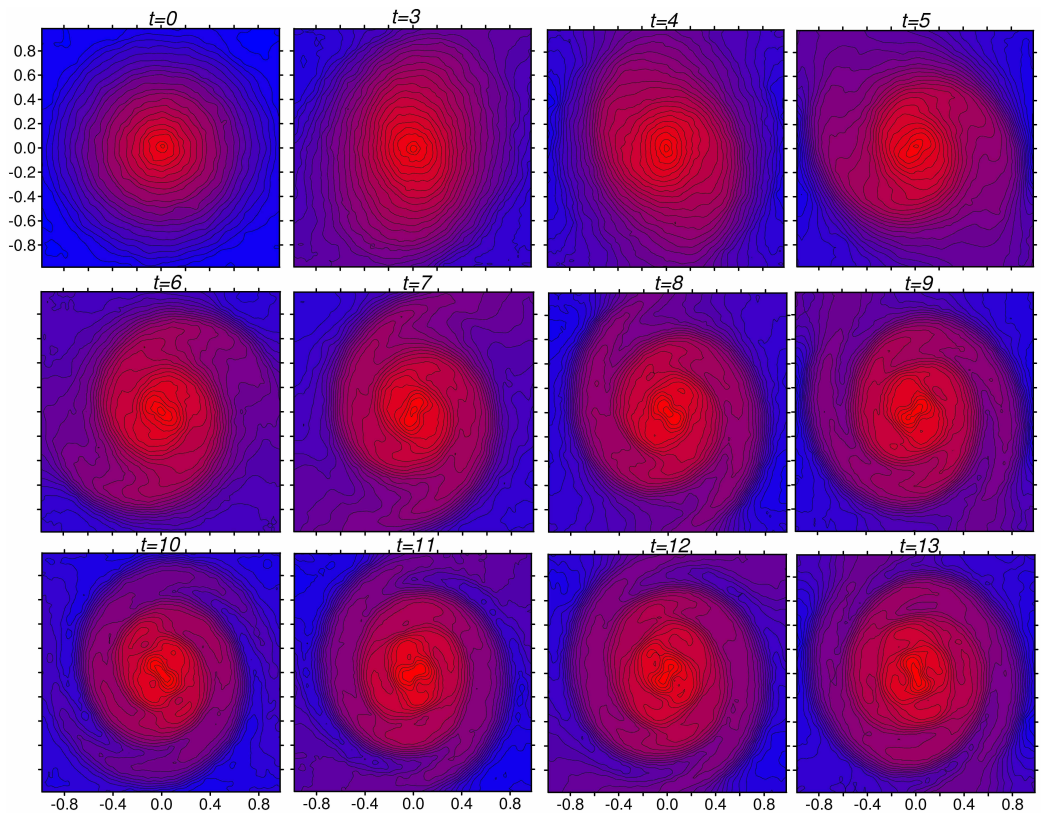}
  \caption {(In color online.) Isolines of the logarithm of stellar disk surface density  in the N-body model immersed in the gravitational field of a non-axially symmetric dark halo at different times  $\Sigma_*(r,\varphi)$ for $\varepsilon =1-b/a= 0.2$
}\label{fig-TriaxHaloStar-density}
\end{figure}

 Of special interest is the possibility to explain the appearance and support of a regular spiral structure of galaxies due to the disk interplay with a non-axially symmetric halo. Numerical simulations clearly showed how non-axially symmetric gravitational potential of a massive halo in the disk plain leads to the formation of two-arms spiral pattern in the gaseous \cite{2012ARep...56...16K...Khoperskov-etal-2012!Gaseous-Disks-Non-axisymmetric-Halo} and stellar \cite{2013MNRAS.431.1230K...Khoperskov-etal-2013!stellar-disk-nonaxissymmetric-dark}. disks. Fig.~\ref{fig-TriaxHaloStar-density} illustrates this process for a stellar disk model. Images shown in this Figure correspond to different time from the beginning of the experiment (in the adopted units, the period of rotation of the disk outer parts is equal to three). The stellar disk of the model galaxy is assumed to be sufficiently hot to exclude the development of gravitational instability both in the disk plane and in the perpendicular direction. Under the same initial conditions but without halo asymmetry the disk remains axially symmetric over several tens rotation periods (about 10 billion years).

In spite of noticeable halo asymmetry (the ellipticity is $\varepsilon=0.2$) and related perturbation in the form of the two-arm spiral, at late evolutionary stages the disk as a whole conserves symmetric shape.  In the process if the spiral structure formation and later during many rotational periods of the disk, azimuthally averaged disk parameters (the surface density, three components of the local stellar velocity dispersion  $c_r$, $c_\varphi$, $c_z$, the rotational velocity $V$, and the disk thickness  $h_z$) do not exhibit  any significant changes: there is neither disk heating nor radial density redistribution. Thus the original source of the spiral structure generation is preserved by sustaining a long-lived spiral pattern. The lack of disk heating is an important difference from the spiral wave dynamics caused by gravitational instability, which is accompanied by increasing velocity dispersions $c_r$,
  $c_\varphi$, $c_z$ \cite{2013MNRAS.431.1230K...Khoperskov-etal-2013!stellar-disk-nonaxissymmetric-dark}.

\begin{figure}
  \centering\includegraphics[width=0.4\hsize]{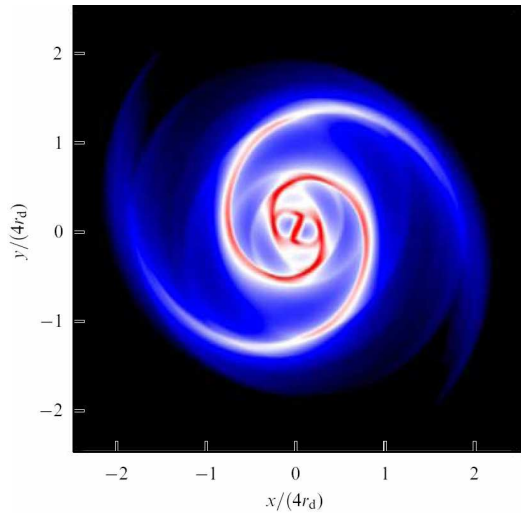}
  \caption {Surface gas density from modelling of the gas disk dynamics in a non-axially symmetric halo potential.
}\label{fig-TriaxHaloGas-density}
\end{figure}

A non-axially symmetric halo significantly affect both the stellar and gas subsystem dynamics.  In numerical models, a long-lived quasi-steady spiral structure forms  \cite{2012ARep...56...16K...Khoperskov-etal-2012!Gaseous-Disks-Non-axisymmetric-Halo}. The two-arm wave encompasses almost the entire gas disk but the very center, in which more complex structures are observed (Fig.~\ref{fig-TriaxHaloGas-density}). For the typical disk and halo parameters, the wave amplitude increases to the non-linear level by forming a system of global shocks. A massive asymmetric halo is able to generate non-linear waves in the gas component even for small ellipticity $\varepsilon \simeq 0.01$,  but the time of growth of the wave amplitude in this case increases by amounting to eight periods of rotation of the disk periphery parts. The spiral pattern geometry depends on the rotation curve $V(r)$, the DM halo density distribution, the radial surface density profiles $\Sigma_g(r)$ and the sound velocity  $c_s(r)$.

\begin{figure}[htbp!]
\centering\includegraphics[width=0.7\hsize]{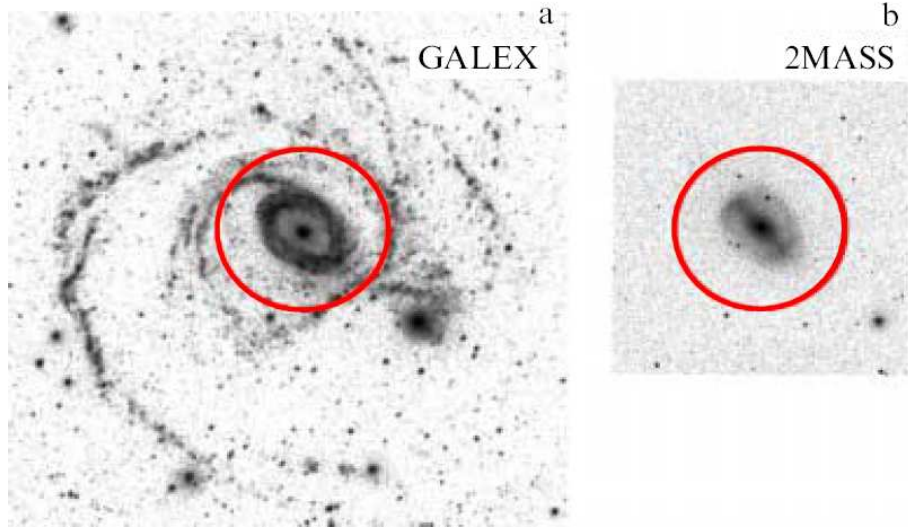}\vskip -0.\hsize
\caption{
The galaxy  NGC\,1512 \cite{2015BaltA..24..119Butenko}: Images (a) in the ultraviolet diapason taken by the GALEX (Galaxy Evolution Explorer) telescope and (b) in the near IR from 2MASS (Two Micron All Sky Survey) survey, which characterize the stellar component distribution. The red circle corresponds to  $R_{opt}$.
 }
\label{fig-ngc1512}
\end{figure}

  A triaxial halo can be also responsible for another phenomenon typical in some galaxies  (NGC 1512, NGC 2841, NGC 2915, NGC 3198, NGC 3359, NGC 5055, NGC 5236, NGC 6744, NGC 6946, NGC 7793), in which an expended sufficiently regular global spiral structure is observed in the gas component far beyond the stellar disk limits (Fig.\,\ref{fig-ngc1512}). Such an external spiral pattern is revealed by HI, UV and even H$_\alpha$ \cite{2009MNRAS.400.1749Koribalski-Gas-dynamics-NGC1512, 2012ApJ...753..138Radburn-Smith-Outer-disk-NGC7793}.
The gas density beyond the radius  $R_{opt}$, is likely to be insufficient to form a spiral wave due to gravitational instability. In this case a massive non-axially symmetric halo can be the efficient generator of the density wave in the external gas subsystem of the galaxy, as suggested by numerical gas-dynamic simulations \cite{2015BaltA..24..119Butenko}.

\subsection{The central cusp problem}

The widely known central cusp problem stems from comparison of the results of numerous cosmological galaxy formation models with observations. The problem is that in the numerical models he spatial density $\rho$ in halo centers diverges by forming the so-called central cusp, whereas observations, as a rule, do not demonstrate dramatic central density increase. The density profile in a dark halo at small  $r$ is usually approximated as $\rho \propto r^\alpha$ with $-1.5 < \alpha < -1$. We remind that for the most frequently used NFW profile~\cite{1996ApJ...462..563N}:
\begin{equation}\label{eq::NFW}
\Oo \rho \propto \frac{\rho_0}{ (r/r_s)(1 + (r/r_s))^2}\,,
\end{equation}
where $r$ is the distance from the center, $\rho_0$ and $r_s$ are model parameters describing the radial halo profile. Clearly, in this case  $\alpha = -1$. If the radial density profile reliably gives $\alpha > -1$, it is referred to as the ‘core’-type distribution. To infer the DM profile from observations, usually rotation curves (or velocity dispersion) in the central parts of galaxies are invoked.  Here dwarf galaxies or low-surface brightness galaxies are preferable, in which DM can significantly contribute to the rotational velocity both at the periphery and in the center and therefore the DM contribution can be  more easier to separate from stellar-gas components.

In the vast majority of cases the rotation curves and stellar population  models for inner disks disagree with $\alpha = -1$; as a rule, the curves  $V(R)$ are better fit with a pseudo-isothermal model with a ‘core’-type profile (see, for example,  ~~\cite{2001AJ....122.2396D...deBlok-etal-2001!Rot-LSB, 2001AJ....122.2396D...deBlok-etal-2001!Rot-LSB, 2012MNRAS.420.2034Salucci-Dwarf-kinematics-spiral, 2011ApJ...742...20Walker-Mass-Profiles-Dwarf, 2008AJ....136.2648D...deBlok-etal-2008!Rotation-Curves-THINGS, 2015MNRAS.454.2092Onorbe-cusps-cores-dwarf, 2015ApJ...808..158Burkert-Structure-Dark-Halo-Core-Dwarf} and references therein). A sufficiently complete review of the problem can be found in paper \cite{2008IAUS..244...44Wyse-Observed-Properties-DM}.

Several scenarios of the formation and evolution of galaxies, which are in agreement with the cosmological galaxy formation scheme, have been proposed. Two possibilities are considered in the literature: either the cusp emerges but rapidly smoothens due to baryonic matter redistribution, or for some reasons it does not appear at all.

The first approach assumes that the cusp forms but rapidly disappears due to gravitational interaction between DM and gas, since during intensive star formation in the galaxy center the energy supplied by massive stars to gas (supernova explosions, stellar winds, radiation pressure) leads to the gas outflow from the center. This scenario is supported by numerical hydrodynamic calculations allowing for the feedback between baryonic and dark matter \cite{2011MNRAS.415.1051Brook-Hierarchical-formation-galaxies, 2014MNRAS.437..415Di-Cintio-DM-stellar-halo-cusps-cores, 2012MNRAS.422.1231Governato-Cuspy, 2012ARep...56..664K...Khoperskov-etal-2012!cusp}. For the flattening of the central density gradient crucial are the star formation intensity at the time after the cusp formation, as well as the total mass and degree of central concentration of matter. For galaxies with stellar mass of order of $10^9 M_\odot$ the deep central potential well will hampers the gas outflow and the cusp smearing out, and in low-mass dwarf galaxies with stellar masses of $10^6 M_\odot$ and below the cusp can be preserved because of insufficiently high star formation rate. The most effective cusp destruction occurs in galaxies with intermediate stellar masses of several hundred million solar masses \cite{2015MNRAS.453.2133Brook-DM-galaxy-populations}. The bulge formation makes the central density profile somewhat steeper, however even for massive galaxies with bulges the modulus of the logarithmic slope  $\alpha$ in the center still remains less than one \cite{2014JCAP...01..047Del-Popolo-Cusps}.

In the second approach, the problem solution is sought for in certain non-trivial DM properties and initial conditions for density perturbations growth. For example, the transition from cold DM to warm DM is assumed \cite{2001ApJ...556...93Bode-Ostriker-Halo-Formation-Warm-Dark}, or a model of DM based on the existence of hypothetic ultralight axions  in which quantum-mechanical pressure prevents the cusp formation is used \cite{2015MNRAS.451.2479Marsh-DM-cusp-core-problem}. All these hypotheses meet certain theoretical difficulties and, mainly, are difficult to test. Doroshkevich et al. \cite{2012PhyU...55....3D} proposed another solution. In the frame of the standard cosmological model, the authors showed analytically that random motions of matter in collapsing protohalos ‘heat up’ DM at the collapse stage by suppressing the cusp formation. In the later paper \cite{2013arXiv1306.3210Semenov-multicomponent-dark-matter-models} to solve the cusp problem a modified version of DM was proposed, in which DM can  be inhomogeneous and include particles with strongly different masses (for example, sterile neutrinos in addition to massive particles). Such a mixture of cold and warm DM changes the spectrum of growing perturbations by suppressing the appearance of low-scale fluctuations, and in principle can solve both the cusp problem and the insufficient amount of low-mas halos (dwarf galaxies) compared to expected in the $\Lambda$CDM-model (see Section  5.5).

\section{Statistical relations between dark halo mass and other parameters of galaxies}

Search for and analysis of statistical relations between dark halo mass and other observable characteristics of galaxies play the key role in disclosing the nature of DM. These characteristics include, for example, the luminosity, the mass of stellar population, the surface brightness or color index of a galaxy (the difference of stellar magnitudes in different photomertrical filters), etc. Let us consider the most important of these characteristics.

\subsection{Baryonic Tully-Fisher relation}

The close relation between the dark halo mass within optical limits and stellar mass of a galaxy is clearly manifested in the relationship between the galaxy rotation velocity and luminosity, which was first discovered in paper \cite{1969AJ.....74..859Roberts-Properties-Spiral-Galaxies}, and later dubbed after Tully and Fisher \cite{1977A&A....54..661T...Tully-Fisher-1977}, who guessed to use it for galaxy distance estimates. This relationship connects the circular velocity at large distances from the galaxy centers, which is primarily determined by the mass and degree of concentration of dark halo, with the integral luminosity of galaxy, which directly depends on the mass of stellar population.

Since the discovery, the Tully-Fisher (TF) relation was many times reproduced in papers by various authors for luminosities in different spectral bands and for rotation velocities estimated  both from the HI line width and more precisely from the rotation curves (see, for example, \cite{2001ApJ...563..694V...Verheijen-2001}). Formally, this dependence can be applied for galaxies without disks (ellipticals) by taking for them the rotation velocity estimates from dynamical models. The TF relations constructed separately for galaxies of diverse morphological types show that early-type galaxies (for example, S0) have a lower luminosity  with the same circular velocities than spirals (see, for example,  \cite{2007ApJ...659.1172D...DeRijcke-etal-2007,2013MNRAS.432.1010C...Cortesi-etal-2013}), which is quite expected, at least qualitatively, since the aging of a stellar system leads to its luminosity decrease. Indeed, for a baryonic TF (BTF) relation, in which the baryonic mass (the total mass of stellar mass and gas) is used instead of luminosity, the difference in the location of early- and late-type galaxies in the TF diagram disappears (see \cite{2007ApJ...659.1172D...DeRijcke-etal-2007}).This correlation is tighter than for the classical TF relation (see, for example, \cite{2014AJ....147..134Z...Zaritsky-2014}), and bears deeper physical sense, because it is independent of the stellar population properties. Here one should bear in mind the difficulty in the baryonic mass estimating in massive galaxies due to its being mostly concentrated in stars, and the integral stellar mass depends on the mass-luminosity ratio used. This estimate is not only affected by the adopted model of evolution of stars with different masses, but also by the form of the initial stellar mass function, which is especially poorly known for low-mass stars that mostly contribute to the stellar mass in galaxies.

The BTF-relation in principle can be used to test the MoND predictions (see \cite{2012AJ....143...40M...McGaugh-2012} and references therein).  The slope of the BTF dependence (in logarithmic scales) is, according to different estimates, between 3.5 and 4; MoND predicts the coefficient 4. In paper  \cite{2015ApJ...802...18M...McGaugh-Schombert-2015} based on reliable optical and near-IR multicolor photometry and rotation velocities on flat parts of galaxy rotation curves, the slope of the BTF relation does not contradict to the MoND prediction, and the conclusion is confirmed that low-mass and dwarf galaxies lie along one linear BTF-dependence. However, a smaller slope close to 3.5 was obtained in paper \cite{2014AJ....147..134Z...Zaritsky-2014}, for galaxies from a statistically large sample (903 galaxies); in that paper, the BTF-relation was obtained with a quite small dispersion of points (0.18 dex in the logarithmic scale). The location of dwarf galaxies on the diagram, however, is crucial to estimate the slope of this dependence. In many dwarf galaxies  the gas mass dominates over stars, therefore the baryonic mass weakly depends on the stellar mass estimates, and the location of gas-rich dwarfs on the BTF-diagram is used for its calibration (assuming a linear logarithmic dependence). This supports the possibility to quite accurately determine the baryonic mass in galaxies, and hence the dark halo mass, from the rotation velocity.

\begin{figure}
\includegraphics[width=12cm,keepaspectratio]{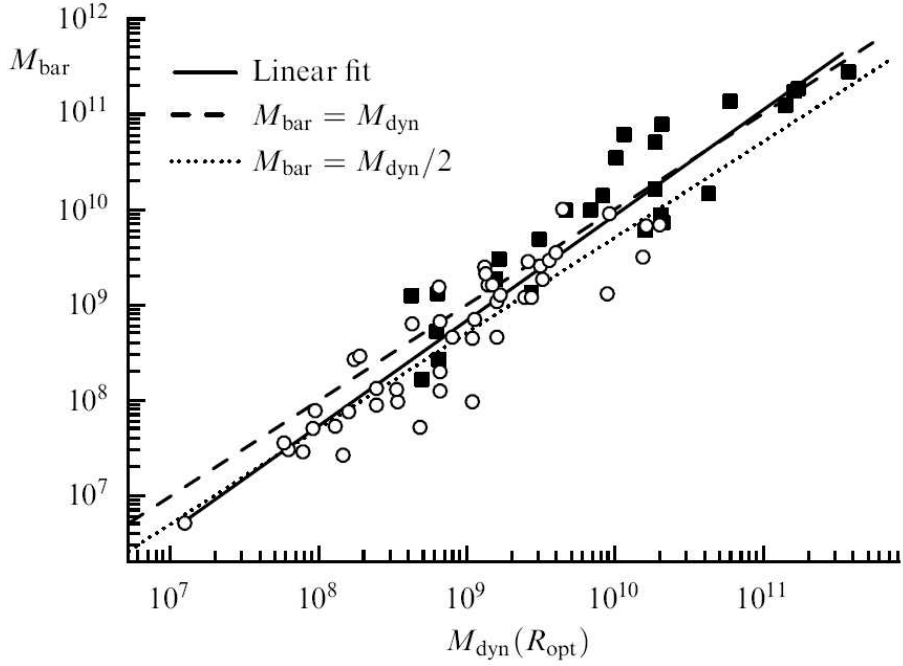}
 \caption{Matching of the baryonic and total mass (in units of  $M_\odot$) inside the optical radius (from data in  \cite{2015ApJ...802...18M...McGaugh-Schombert-2015}). The solid curve is the leas-mean square fit. The dotted and dashed lines show the expected dependences for the case of equally massive dark and baryonic matter and in the lack of dark matter, respectively. The filled and open symbols relate to galaxies in which the baryonic mass is dominated by stars and gas, respectively.
}
\label{masses}
\end{figure}

The BTF-diagram can be easily constructed from the relation between the dynamic mass within the optical radius, $M_{dyn} = V^2R_{opt}/G$, which includes both baryonic and non-baryonic components of a galaxy (see Fig. \ref{masses}, constructed using the BTF-relation  \cite{2015ApJ...802...18M...McGaugh-Schombert-2015}). The galaxy sample used in  \cite{2015ApJ...802...18M...McGaugh-Schombert-2015}, ] included both massive galaxies (filled symbols) and gas-dominated dwarfs (open symbols). The dashed and dotted lines in Fig.\,\ref{masses} show the dependences expected in the absence of DM and for the case of equal dark and baryonic masses, respectively. The mass estimates are accurate to within the factor 2-3, therefore in fact the correlation between the compared quantities must be very high. Galaxies located above the dashed line are due to random errors, since the dynamic mass cannot be smaller than baryonic one. The figure suggests that dwarf and massive galaxies follow close relations, and the mass ration weakly changes from dwarfs to massive systems: $\log(M_{bar})=(-1.02\pm 0.43) + (1.09\pm 0.04)\log(M_{dyn}(R_{opt}))$, with the correlation coefficient $R=0.94$. Thus, $M_{dyn}$ and $M_{bar}$ can be considered to be almost linearly related. As $M_{dyn} = M_{bar}+M_{dm}$, the dark to baryonic mass ratio within the optical radius, $\eta = M_{dyn}/M_{bar}$ on average is equal to 1.5-2 for low-mass galaxies and  $\sim 1$ for massive systems, which suggests a decrease in DM contribution to the total mass with galaxy mass growth. However, large dispersion of points on the diagram does not allow precise estimates of $\eta$.

Physical reasons for tight (with small dispersion) TF and BTF relations have been actively debated up to now. In principle, the dependence can follow from the virial theorem relating the integral mass and rotation velocity, however here, putting aside the MoND theory, it requires a certain ‘tuning’ between the masses and radial scales of the disk and dark halo (see, for example,  \cite{2013IAUS..289..296G...Giovanelli-2013,2012ISRAA2012E..12Z...Zaritsky-2012} and references therein).  It is very important that a BTF-relation close to the observable one can be inferred from cosmological numerical models of galaxy formation and evolution using the Newtonian gravity \cite{2013MNRAS.434.3142A...Aumer-2013}, although the results remain model-depending.

\begin{figure}
\includegraphics[width=11cm,keepaspectratio]{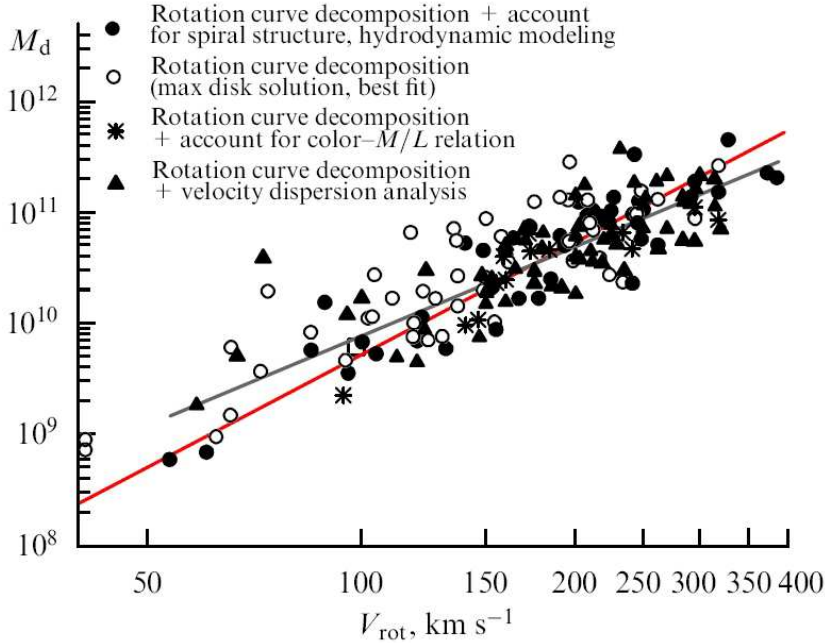} \caption{(In color online.) The baryonic TF relation obtained by dynamical methods of the disk mass estimates. The grey line is the least-mean square fit. The red line is the dependence obtained in \cite{2005ApJ...632..859M...McGaugh-2005!Baryonic-Tully-Fisher}, where disk masses were derived from photometry. Different symbols show the results obtained by diverse methods.
}
\label{btf}
\end{figure}

As the stellar baryonic component mass estimated via the $M/L$ ratio depends on the adopted stellar population model and has some uncertainties (the initial stellar mass function in the first place), it is important to check the existence of the BTF-relation constructed without using a stellar population model. Fig.\,\ref{btf} presents such a dependence as inferred from disk mass estimates obtained by different dynamical methods. The grey line shows a least-mean square (LMS) fit: $\log(M_{d})=(4.5\pm 0.3) + (2.67\pm 0.15)\log(V)$, with the correlation coefficient  $R=0.85$ (the estimates obtained by the maximum disk method were ignored). Most of the dispersion is due to uncertainties in the disk mass estimates. Although the dependence demonstrate a larger dispersion than in the case of the disk mass estimates from photometrical data, the results obtained by both methods are in agreement.  However, on may notice a tendency of higher dynamical disk estimates in galaxies  obtained by the maximum disk method (see the discussion above in Section 3).

\subsection{Dark matter and morphological type of galaxy}

 Of separate interest is the relation between dark halo masses within optical radii and the morphological type of galaxies. Their correlation can suggest that the morphological types depend, in particular, on the joint evolution of the dark halo and baryonic matter in galaxies.

 As shown in paper \cite{2014MNRAS.445.3512S...Saburova-del-Popolo-2014}, the column density of dark halos (the product of the central spatial density by the halo core radius) turns out to be higher for brighter galaxies and for earlier morphological types (Sa-Sbc), for which, as a rule, the bulge contribution into the luminosity is more significant. The bulge formation, therefore, can be directly or indirectly related to the halo mass and central density. This issue is unsolved and requires more studies. Interestingly, a similar increase in the column density was found with growth of the total (virial) halo mass \cite{2009arXiv0911.1774B...Boyarsky-2009}. The column density of a dark halo is weakly sensitive to the effect of baryonic matter on the halo spatial density distribution, which is difficult to account for, and therefore it is convenient to compare with numerical simulations \cite{2009arXiv0911.1774B...Boyarsky-2009}.

\begin{figure}
\includegraphics[width=11cm,keepaspectratio]{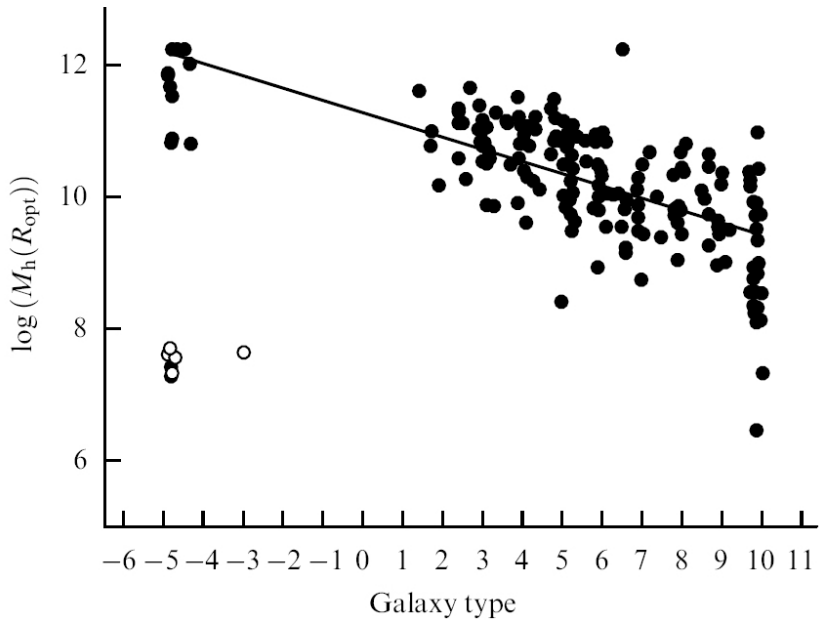}
 \caption{Matching of the dark halo mass inside the optical radius and the galaxy morphological type $t_{dV}$. White dots show dwarf spheroidal galaxies.
}
\label{mht}
\end{figure}

 In Fig. \ref{mht} the dark halo mass within the optical limits is matched with the galaxy morphological type according to paper \cite{2014MNRAS.445.3512S...Saburova-del-Popolo-2014}. The morphological type is given in the numerical coding by de Vaucouleurs (types S0, Sa, Sb, Sc, Sd and Irr correspond to $t = -1, 1, 3, 5, 7$ and $>7$). The open symbols show dwarf spheroidal galaxies (dSph).  In the Figure, they form a separate group, which reflects their lower luminosity compared to that of spiral galaxies. The Figure suggests that except for dwarf spheroidals, for earlier morphological types the dark halo mass tends to increase: $\log(M_{h}(R_{opt}))=(11.3\pm 0.1) - (0.18\pm 0.01)t$ with the correlation coefficient 0.68. This relation is largely due to the statistical dependence between the morphological type and luminosity of galaxies, because the baryonic to dark mass ratio remains virtually constant (to within a factor of $\sim 2$) along the morphological sequence. Paper \cite{2014MNRAS.445.3512S...Saburova-del-Popolo-2014} also noted the dependence between the halo mass and color index of a galaxy – more red systems with mostly evolved stellar population and low gas content (which predominantly belong to earlier morphological types) have on average higher DM mass within $R_{opt}$.

As noted above, the highest dark to baryonic mass ratio occurs in dwarf galaxies, especially in dSph (see Section 2.3).  But in gas-rich dwarf dIrr, which unlike dSph have rotating gas disks, the dark halo can comprise most mass: for many dIrr dwarfs the rotation curve continues increasing even at significant distances from the center, beyond the optical disk limits (see, for example,  \cite{2010MNRAS.404.2061E...Elson-etal-2010, 2008MNRAS.383..809B...Begum-etal-2008, 2015AJ....149..180O...Oh-et-al-2015}).

\begin{figure}
\includegraphics[width=0.5\hsize]{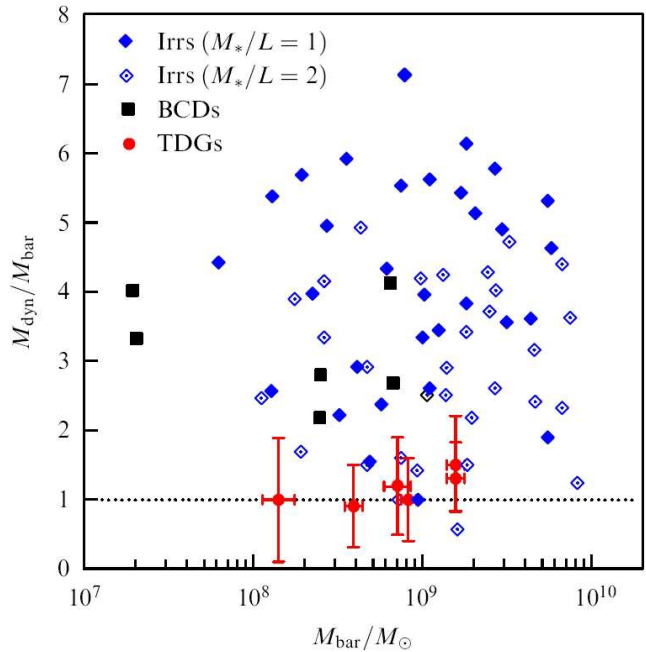}
 \caption{(In color online.) Matching the baryonic mass (the horizontal axis) with the dynamic to baryonic mass ratio inside three radial scales for dwarf galaxies of different types.  \cite{2015arXiv150905404L...Lelli-etal-2015}.
}
\label{tidal}
\end{figure}

Generally, dwarf galaxies form inhomogeneous class of objects, and the DM contribution to their total mass varies in a wide range (see Section 2).  Fig \ref{tidal}, taken from paper  \cite{2015arXiv150905404L...Lelli-etal-2015}) compares the  $M_{bar}/M_{dyn}$ ratio with baryonic mass inside three radial disk scales for red irregular galaxies (Irrs), blue compact galaxies (BCGs) and tidal dwarf galaxies (TDGs). The latter are of special interest, since they were formed from matter that earlier belonged to disks of interacting galaxies. Therefore, it can be expected that TDGs are almost DM free.

The last conclusion, apparently, are supported by observations, although they need further testing. Paper \cite{2015arXiv150905404L...Lelli-etal-2015} used the results of optical and HI observations of six galaxies with gas disks to reliably conclude that the dynamic and baryonic mass of these dwarf coincide: from the location on the BTF-diagram, they rotate tow times as slow as ordinary dwarf galaxies suggesting a low DM content. However, it is unclear how well the measured gas velocities correspond to those of equilibrium circular rotation in these non-stationary systems.

For massive elliptical galaxies, the dark halo estimates require a more sophisticated analysis than for disk galaxies, because these galaxies, for very rare exceptions, have almost no cold gas, and stars there move in different planes, which complicates their kinematic mass estimates. The initial data for the mass or stellar density determination as a function of the galactocentric distance $r$ can include: a) the dispersion of radial velocities of stars  and its change along  $r$, b) radial distribution of brightness and temperature of X-ray emission from hot gas filling a galaxy; c) effects of weak, and in some cases of strong gravitational lensing (see Sections 3.3. and 4).

The analysis of weak and strong lensing suggests that the total density profiles of elliptical galaxies are well fit by a combination of the stellar spatial density profile corresponding to de Vaucouleurs surface brightness distribution ($\ln\,I(r) = -(r/R_e)^{1/4}+\textrm{const}$), and the NFW dark halo profile suggested by cosmological modelling (see  \cite{2007ApJ...667..176G...Gavazzi-et-al-2007-SLACS}). The DM mass fraction found in \cite{2007ApJ...667..176G...Gavazzi-et-al-2007-SLACS}, is $27\pm 4$\% within one effective radius  $R_e$, that contains (by definition) half integral luminosity of a galaxy. Although there are different opinions about this ratio, all present estimates suggest a significant stellar mass domination over DM inside $R_e$. Observations of X-ray emission from hot gas in elliptical galaxies, as well as lensing data, suggest the NFW dark halo density profile beyond $R_e$. Here the ratio of the total mass to luminosity decreases by an order of magnitude  when increasing from $r=1R_e$ to $r=10R_e$ (according to the Chandra observatory data), which evidences of a large DM mass fraction in galaxies \cite{2006ApJ...646..899H...Humphrey-etal-2006}.

The richest sample (up to 98000 objects) from the SDSS survey was utilized to estimate the dynamic and stellar (i.e. baryonic) mass within the effective radius $R_e$ in paper  \cite{2015MNRAS.446...85N...Nigoche-Netro-etal-2015}. The stellar population mass was found from the luminosity and color index for different stellar mass functions, and the dynamic mass was estimated from the velocity dispersion at the effective radius using the virial relation.  It was found that inside $R_e$ the DM mass fraction does not exceed several per cents. This is lower than estimates found by other methods, however the resulting mass estimate depends on the adopted mass function and assumptions about its universality (see the discussion in that paper).

We remind that below we mainly considered inner galaxy parts. Inside large radial distances the DM content is preferably derived from lensing observations and X-ray data, which suggest the $M/L$  ratio increase from the center up to several dozen solar units at a distance of several $R_e$, so that the ratio of the total (virial) halo mass of galaxies to their integral luminosity in the V-band amounts to 100  \cite{2007ApJ...667..176G...Gavazzi-et-al-2007-SLACS}. The radial density profile of E-galaxies (including DM) in a wide range of distances $r$ is almost isothermal ($\rho \sim r^{-2}$) although far away from the center it can be satisfactorily fit with the NFW profile \cite{2007ApJ...667..176G...Gavazzi-et-al-2007-SLACS}.

Comparison of DM content in disk and elliptical galaxies is difficult because in both cases the result is strongly dependent on radial distance from the galaxy center and on the galaxy luminosity: the relative DM mass in both types of galaxies increases with distance from the center and is higher in dwarf systems. However, the comparison of the relative DM content in spiral and elliptical galaxies inside the effective radius reveals that it is likely to be lower for elliptical galaxies. Indeed, estimates show that the role of DM for elliptical galaxies at $r\le R_e$  is comparatively small (see above). For spiral galaxies this not so: their radius $R_e$ is only slightly (by $\sim 22$\%) smaller than the radius at which the exponential disk contribution to the rotation curve is maximal, with its contribution to the rotational velocity is  $0.6-0.7$ of the observed rotation velocity (see Section 3), which suggests a noticeable contribution of a non-disk dark component to the total mass. However, the situation can strongly differ from galaxy to galaxy which demonstrate diverse rotation curves within $2 - 3$ radial scales.

\subsection{Correlation between the dark matter and  neutral hydrogen density}

An interesting feature noted as early as in the beginning of the 1980s  \cite{1981AJ.....86.1791B...Bosma-1981}, and confirmed later (see \cite{2012MNRAS.425.2299S...Swaters2012} and references therein) is that the DM density projection on the galaxy disk plane proves to be on many cases proportional to the gas density distribution, mainly comprised in neutral hydrogen. Based on this finding, a hypothesis was proposed that DM represents a gas which is difficult to observe and distributed in the form of small, dense and very cold molecular clouds not related to star formation regions whose total mass many times exceeds that of the observed gas (see, for example,  \cite{1994A&A...285...79P...Pfenniger-Combes-1994!dark-matter-cold-gas}). The ‘dark’ gas that radiates neither in radio nor optical actually exists in our Galaxy thus forming the baryonic part of DM in galaxies (see, for example, \cite{2015ARA&A..53..199G...Grenier-2015,2011A&A...536A..19P...PlanckCollaboration-2011}). Indirectly, such a gas is manifested in weak gamma-ray emission generated by collisions of cosmic-ray protons with gas atoms or in excessive thermal dust radiation in this medium. The ‘invisibility’ of the gas can be due to both destruction of CO molecules whose radio emission is usually used to estimate the molecular gas content, and a very low temperature of the medium. Part of the gas can be unobservable due to large optical depth of the clouds, which is confirmed by observations \cite{2014ApJ...796...59Fukui-CO-Clouds}.

Generally, the dark gas can constitute a sizable fraction of the total mass of the cold gas in our and nearby galaxies (see the discussion in в \cite{2014ApJ...796...59Fukui-CO-Clouds, 2014MNRAS.437.3072K...Kasparova-2014}),]), but it is definitely insufficient to explain DM effects.  In addition, if DM indeed concentrates in a plane layer, its thickness must be very high for such a self-gravitating disk to be gravitationally stable. However, as shown in \cite{2013AstL...39..291Z...Zasov-Terekhova-2013!neutral-hydrogen-dark-mass}, the correlation between the DM column density and neutral hydrogen can be explained without invoking exotic DM forms. This correlation can arise because gas disks in most galaxies are in about marginally stable state, which dictates a certain radial gas density, which turns out to have similar shape with the expected column density distribution (projected onto the disk plane) for a pseudo-isothermal  spherical halo.

\subsection{Dark matter and gas metallicity relation}

The dark halo effect on the chemical evolution of galaxies appear in the first place as a dependence between the star and gas metallicity and galaxy (or its stellar population) mass (the mass-metallicity or luminosity-metallicity dependence). Low-mass galaxies have a poorer metal content compared to massive galaxies. Clearly, this correlation results primarily from different star formation rates which control metal ejection into the interstellar medium, but the result is also related to the galaxy gas exchange with the surrounding medium which depends  on the dark halo properties. Accretion of gas on the disk, the possibility of gas ejections from the galaxy due to stellar activity, as well as the efficiency of gas mixing in the disk also depend on the halo mass.  A massive halo, on the one hand, serves as hot gas reservoir, which, by  cooling, supplies low-metallicity gas to the galaxy, and on the other hand, produces a potential well preventing gas ejection from the galaxy, so that the gas enriched with heavy elements returns to the disk, although may be at different distance from the center. Therefore, the galaxy mass-metallicity relation can be considered as the manifestation of the relation between the halo mass and chemical evolution of a galaxy.

\begin{figure}
\includegraphics[width=0.5\hsize]{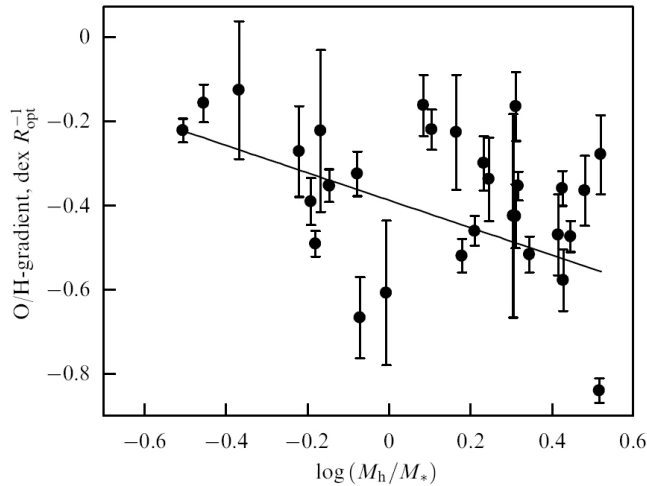}
 \caption{Matching of radial gradient of the oxygen abundance with the dark halo to stellar mass ratio inside the optical radius. (From paper  \cite{2015arXiv151008924Z...Zasov-Saburova-Abramova-2015}).
}
\label{mhgradoh}
\end{figure}

 Since the halo to stellar or baryonic mass ratio varies in a sufficiently wide range, it is interestingly to clarify the role of the relative halo mass (within the optical limits) $M_h/M_*$. Тесной связи между параметрами, описывающими химический состав газа и No tight correlation between parameters describing the chemical abundance of the gas as $M_h/M_*$ has been discovered, however it was found that galaxies with low relative halo mass, as a rule, demonstrate also small (by modulus) radial gradients of the oxygen abundance (O/H) (Fig.\,\ref{mhgradoh} \cite{2015arXiv151008924Z...Zasov-Saburova-Abramova-2015}). The possible explanation of such a relation, if confirmed by larger statistical data, is in a more efficient radial gas mixing in galaxies with the lowest dark halo mass fraction, i.e. in disks residing in shallower DM potential wells.

The most intensive gas mixing occurs in tightly interacting systems. The interaction can affect the halo masses: numerical simulations \cite{2011MNRAS.418..336L...Libeskind-etal-2011} show that halos can be partially destroyed. The efficient gas mixing, which results in flat radial profiles of the oxygen abundance, can also be related to turbulent viscosity or accretion of gas  with low angular momentum (see  \cite{2014ApJ...796..110E}, although the efficiency of these processes in galaxies with low-mass halos remains unclear).

\subsection{Virial halo mass --- baryonic (stellar) mass relation and the baryon deficit problem}

 In Sections 7.1-7.4, if not specially assumed, the halo mass was considered within the optical boundaries of a galaxy  $R_{opt}$. This boundary is taken sufficiently arbitrarily, because galaxies extend beyond this limit, and the relative halo mass continues gradually increase with radial distance  $r$ above $R_{opt}$. The total, or virial halo mass  $M_{vir}$ within the virial radius $R_{vir}$ is about one order of magnitude as high as  $M_h$ (see Section 4). We remind that the virial radius is defined as the distance inside which the mean density of the total (baryonic + dark) matter is several times as high as the present critical density of the Universe. The ratio of these densities is usually taken to be 200. If the halo is not strongly oblate, the virial mass $M_{vir} = V_{vir}^2R_{vir}/G$, where $R_{vir}$ is the virial radius, $V_{vir}$ is the circular velocity at $R_{vir}$. The virial mass so defined is uniquely related to  $R_{vir}\gg R_{opt}$ and $V_{vir}$, which enables the estimation of  $M_{vir} $ from kinematic measurements.

As noted in Section 3, the flat part (plateau) on the rotation curve is explained by assuming a pseudo-isothermal halo for which $V_c= \textrm{const}$ at large  $r$. Unlike the circular velocity in this model, the circular velocity  $V^{(h)}_c(r)$ in a NFW halo, which is usually invoked in calculations, at large distances $r$ from the center passes through a smooth maximum and then starts decreasing. Therefore, although the velocity  $V_{vir}$ is frequently assumed to be equal to the observed rotational velocity of disks in the plateau region, in the general case  $V_{vir}$ can noticeably, by 20-30\%, deviate from the observed velocity. The relation between the observed galaxy rotation velocity at the plateau and the velocity at the virial radius put certain constraints on the halo structure and its density change during the galaxy formation process  \cite{2007ApJ...654...27D}.

The problem of the  $M_{vir}/M_*$ ratio or of the virial to baryon mass ratio $M_{vir}/(M_*+M_{gas})$ considered in many papers (see Section 4).  The estimate of the relative baryonic mass is interesting primarily because the dark to baryonic mass ratio is assumed to be known on cosmological scales, since it can be derived from cosmological data with high accuracy: the mass fractions of dark and baryonic matter should be around 30\% and 5\% of the critical density, respectively, and their ration should about 6:1.  Matching of this value with really measured inside the virial radius of galaxies brings information on the different character of evolution of collisional (baryonic) and collisionless (dark) matter, on the star formation efficiency during galaxy formation, and on what fraction of baryonic matter, initially mixed up with DM, remains undiscovered.

The method of comparison of mass functions (or circular velocities)  of  galaxies and model halos (HAM) suggests that the deficit of baryons is especially significant for low-mass galaxies and is minimal for systems like our Galaxy with the halo mass  $M_{vir}\approx 10^{12}M_\odot$, however the  $M_{vir}/M_*$ ratio for any masses does not exceed 25-30\% of the cosmological value  \cite{2012ApJ...759..138P}. We remind that inside the optically observed galaxy the situation is opposite: there the baryonic and dark mass are comparable, whereas on average on cosmological scales the DM density must be about six time as high as the baryonic mass. The nature of this baryonic ‘deficit’ is apparently different in massive and low-mass galaxies. In dwarf systems the baryonic deficit is most likely due to the loss of gas from the galaxy at the early epoch when it dominated the galaxy baryonic mass. In massive galaxies it is due to a significant amount of gas having high temperature and residing outside the disk filling the dark halo volume up to the virial radius (see below).

The virial mass and hence the  $M_{vir}/M_*$ ratio in galaxies can be estimated not only by the weak lensing method or HAM, both of which has its own difficulties, but, which is important, using alternative approaches. For example, in galaxies with rotation curves measured far away from the center, it is possible to extrapolate the rotation curves using the NFW profile of DM halo  \cite{2015PASJ..tmp..266S...Sofue-2015}. Masses $M_{vir}= M_{vir}$ for our Galaxy and the Andromeda Nebula in that paper found by this method turned out to be  $(81\pm 7)\cdot10^{10}M_\odot$ and $(193\pm 15)\cdot10^{10}M_\odot$ respectively. The mean mass ratio of baryonic components (disk and bulge) to the halo viral mass here is $6.2 \pm 1.8$\,\%.

\begin{figure}
\includegraphics[width=11cm,keepaspectratio]{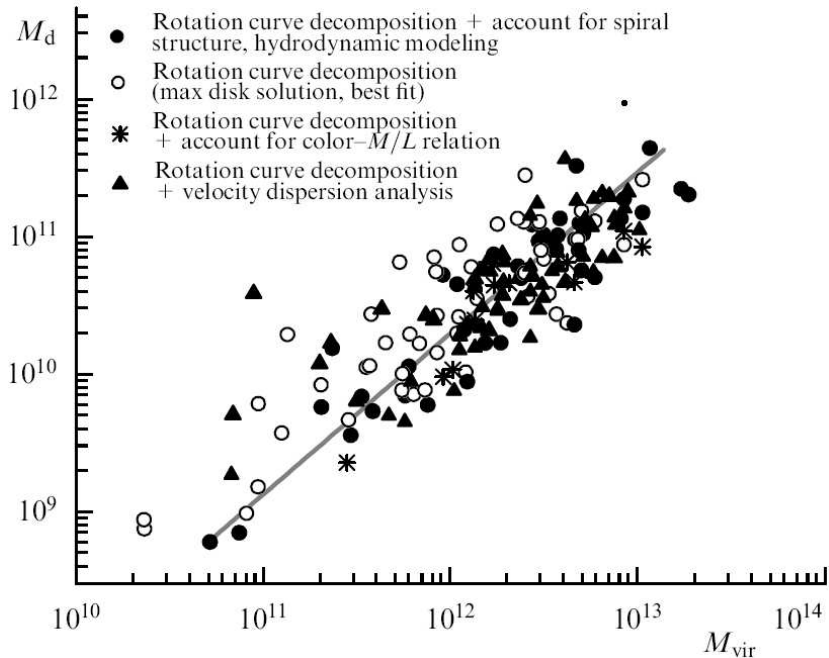}
 \caption{Comparison of the virial halo mass with the total disk mass. The
solid line is the least-mean-square best fit. Different symbols show the disk
mass estimates by various methods based on kinematical and photometrical data.
}
\label{mvirmd-fig18}
\end{figure}

In the disk galaxies  where most of the baryonic matter (stars and interstellar medium) reside, as a rule, in the disk, whose mass can be estimated by several methods mentioned above.  To compare different approaches, in Fig.\,\ref{mvirmd-fig18} we match virial masses  $M_{vir}$ and disk masses  $M_d$, calculated by photometric and kinematic methods listed below:

\begin{itemize}
\item  by the photometric method in which to decompose the rotation curve, constrains on the disk and bulge surface density are taken from the observed color index and the $\!M/L$\,---color relation derived by the population synthesis \cite{Kassin2006}).
\item   by the method based on the disk marginal stability condition using stellar velocity dispersion estimates \cite{1993A&A...275...16B...Bottema-1993!Stellar-kinem-many, 2001ARep...45..180K};
\item by the modelling of the rotation curve using the best-fit model or maximum disk model \cite{Yoshino2008},  \cite{2009ARep...53..801S...Saburova-etal-2009!Spiral-galaxies-mass-to-light-ratios}, \cite{Moriondo1998},  \cite{Dicaire2008},  \cite{Ryder1998}, \cite{Barnes2004});
\item  methods based on hydrodynamic modelling of gas motion in spiral arms  \cite{2003ApJ...586..143K...Kranz2003}) and on the spiral density wave condition  \cite{Athanassoula1987}).
\end{itemize}

To pass from the mass inside the optical galaxy to the virial mass the rotational velocity at the galaxy disk periphery was assumed to be proportional the virial rotation velocity  ($v_{vir}/v \approx 0.79$ according to \cite{Reyes2012}), and the mass  $M_{vir}$ was assumed to directly follow from  $V_{vir}$ (see, for example, \cite{Alam2002}).

Most of the galaxies shown in Fig. relate galaxies with comparatively high mass and active star formation. The solid line in the Figure  is the least-mean square approximation: $\log(M_{vir})=(3.2\pm 0.4) + (0.85\pm 0.04)\log(M_{d})$ (коэффициент корреляцииthe correlation coefficient is  $0.85$). Despite significant dispersion, the results obtained by different methods are consistent with the general dependence that is somewhat different from the linear law  ($(M_{vir}\sim M_d^{0.85})$) in a wide range of baryonic mass (disk mass) of galaxies – from $10^9M_\odot$ to $\sim 3\cdot10^{11} M_\odot$. The main uncertainty is due to galaxies with low-mass halos.

 Thus, the virial (predominantly dark) mass differs from the baryonic mass in the disk by 1.5-2 orders of magnitude, which is much larger than the cosmological dark to baryonic mass ratio. The baryon deficit problem on the virial radius scale can be strongly relaxed or fixed at all by allowing for difficult-to-observe gas in the galaxy halo far beyond its optical boundaries. The mass of hot gas with a temperature of about K observed in the inner halo of some massive galaxies is manifestly small to explain the baryon deficit. However, large gas masses far beyond the optical galaxy are ‘groped’ indirectly from ultraviolet absorption lines in spectra of remote quasars observed at small angular separations from comparatively nearby galaxies. The redshift of an absorption line is close to that of the galaxy may suggest that the absorption is directly related to the gas in an extended galactic halo. In this way, using the spectrograph of the Hubble Space Telescope, absorption lines of OVI \cite{2011Sci...334..948T...Tumlinson-2011}, and $Ly\alpha$ \cite{2015ApJ...813...46B...Borthakur-2015} were measured for several dozen low-redshift galaxies.The absorption lines were observed at distances up to 150-250 kpc from the galaxies (of order of their virial radius!). The absorbing gas has a temperature of  $10^5$K, and is gravitationally connected to the galaxies. The analysis showed that the mean gas density decreases monotonically with $r$, and the total mass is likely to many times exceed the mass of interstellar  gas in the galaxies themselves. For our Galaxy, for example, calculations by Fraerman et al. \cite{2016arXiv160200689Faerman-Hot-Galaxy-Coronae}, show that the missing baryonic mass of order of  $10^{11}M_\odot$ could be comprised in the hot gas of an extended corona, which shows up in observations of oxygen ions absorption lines. It should be also borne in mind that a significant part of galaxies belong to groups and clusters which also have dark halos and gas with the virial temperature; the mass of this gas in giant clusters can exceed the total stellar mass of the cluster galaxies.

The mass of a sufficiently massive group or cluster of galaxies can also be estimated using an original method that gives the total system mass beyond the virial radius, if present. The method is based on the examination of galactic radial velocities (the Hubble flows) near groups and clusters and on the analysis of the difference of this velocity field from that expected for the classical Hubble expansion law, according to which the relative velocities of galaxies are proportional to distances between them. In other words, it is inspected how the gravitational interaction of the galaxy system on the surrounding galaxies slows down  the Hubble expansion (the Hubble flow) near the given system.  To do it, however, a lot of galaxies with distances known independently of the Hubble law (for example, from stars with known luminosity) should be present in the direction to the chosen galaxy group or cluster.

The galactic velocity field near the group or cluster makes it possible to find the conventional radius of the ‘zero’ velocity sphere $R_0$, inside which the gravitational field of the given galaxy system had time since the beginning of the cosmological expansion to fully stop the cosmological expansion. Outside this sphere galaxies continue receding from the system, and with distance their recession velocities asymptotically tend to the Hubble law. The estimate of $R_0$ enables the integral mass of the galactic system. This method was first applied to estimate the mass of the Local group of galaxies in papers by Lynden-Bell  \cite{1981Obs...101..111Lynden-Bell-local-group} and Sandage  \cite{1986ApJ...307....1Sandage-redshift-distance-relation}. Increase in the number of galaxies with distance determined not from the redshift made it possible to apply this method for many systems (see  \cite{2009MNRAS.393.1265K-Karachentsev-Hubble-flow, 2014ApJ...782....4Karachentsev-Virgo-Cluster} and references therein).

It is important that the estimates of masses, as well as of the density distributions inside big space volumes, require allowance for the cosmological repulsion force (dark energy), decreasing the gravitational interaction of sufficiently remote bodies (see
\cite{2014ApJ...782....4Karachentsev-Virgo-Cluster, 2015MNRAS.449.2069C...Chernin-2015, 2013A&A...553A.101Chernin-Dark-energy} for more detail). Without this effect, the mass inside  $R_0$ will be strongly underestimated. The mass of several galactic systems, for which the zero-velocity radius $R_0$ was measured with account for the repulsion force, turned out to be close to their virial mass $M_{vir}$, in spite of the virial radius $R_{vir}$  being much smaller than  $R_0$ (see \cite{2014ApJ...782....4Karachentsev-Virgo-Cluster} and references therein).  For example, for the closest to us massive Virgo cluster the zero-velocity radius was found to be around 7 Mpc, which is several times as large as the virial radius of the cluster, and the mass within  $R_0$ is $(8.0\pm2.3)\cdot 10^{14} M_\odot$ --- в хорошем согласии с оценкой вириальной массыin good agreement with the  virial mass estimate  \cite{2014ApJ...782....4Karachentsev-Virgo-Cluster}. This suggests that beyond $R_{vir}$, and hence in the space between galactic systems, there is little dark matter.

The last conclusion agrees with the galactic mass measurements by the weak lensing method. This the mostly certainly follows from the paper by Bahcall and Kulier  \cite{2014MNRAS.439.2505Bahcall-dark-matter}, which analyzed the DM distribution from weak lensing data by averaging (stacking) of more than 100 thousand galaxy systems – from small groups up to clusters – with redshifts  $0.1<z<0.3$. It was found that at all scales, from several hundred kpc to several ten Mpc, the stellar mass fraction is approximately constant (about 1\% of the total mass), and this value does not depend on the radius of the volume being considered. The mass-luminosity $M/L$ ratio beyond several hundred kpc from the system’s center stops changing significantly. The authors conclude that the integral masses of groups and clusters of galaxies, mainly dominated by DM, are equal to the sum of mass of individual galaxies, including their dark halo masses. Little DM is present where there are no galaxies.

The effect of gravitational fields of galaxy systems on the Hubble flow in the local Universe around our Galaxy is clearly demonstrated in papers by Makarov and Karachentsev ~\cite{2011MNRAS.412.2498M, 2012AstBu..67..123K}, where the location of about 11 thousand comparatively nearby galaxies on the Hubble diagram velocity-distance inside a 50 MPc sphere around the Local group of galaxies was investigated. This volume contains a lot of galactic systems with diverse scales (mainly groups).  The authors estimated the mean density of matter related to the galaxy systems of different multiplicity (the matter must be mostly dark) within the 50 Mpc distance and found it to be about three times as low as the expected 28\% of the mean density of matter in the Universe.  What fraction of DM resides beyond the galactic systems is an open issue, since it is not excluded that the Local group lies inside a giant low-density volume (inside a void) several Mpc in diameter.

\subsection{Dark halo and supermassive black holes}

At the centers of most high-luminous galaxies there are massive compact objects – nuclear stellar clusters and supermassive black holes (SMBH). The latter can be responsible for activity of galactic nuclei, if powerful accretion onto the black hole takes place. The SMBH formation, their role in the parent galaxy evolution and the relation to the integral galactic parameters and kinematic features are actively discussed \cite{2002ApJ...578...90F...Ferrarese-2002, 2010ARep...54..578C...Cherepashchuk-Afanasev-etal-2010!supermassive-black-holes}. The existence of powerful quasars at high redshifs suggests that SMBH in these quasars should have been formed and have increased up to several hundred million solar masses very rapidly, over already first billion years of the cosmological expansion, when the process of galaxy formation had not been completed.

The dynamics and mass concentration in galaxies or non-disk galactic components (bulge, halo) play fundamental role in the formation and growth of SMBHs. This clearly follows from the well-known tight correlation between the SMBH mass and stellar velocity dispersion in the bulge, as well as the disk angular velocity in the central  1-2 kpc region (see the discussion in \cite{2002ApJ...578...90F...Ferrarese-2002, 2013ARep...57..797Z...Zasov-2013}). As the central velocity dispersion correlates, although not very strongly, with the circular  velocity at large  $r$, it is natural to expect that the correlation between the SMBH masses and the rotation velocity of the galaxy, which in turn characterizes the virial halo mass  $M_{vir}$.

Dark halo of galaxies indeed could play large role  in the SMBH mass growth, since the halo gravitational field determines the depth of the potential well in which galactic gas concentrates and where stellar bulge and the black hole are formed. The possible relation between  $M_{vir}$ and the central SMBH mass  $M_{bh}$ has been discussed in many papers (see, for example, \cite{2004JETP...98....1I...Ilyin-2003, 2003ApJ...582..559V...Volonteri-2011, 2012MNRAS.423.2533B...Barrause-2012, 2013ARA&A..51..511Kormendy-Ho-Supermassive-BH}). However, the test of the dependence of $M_{bh}$ on the maximal galactic rotational velocity $V_c$ gave no compelling results, although on the statistical level the dependence certainly takes place, because the most massive black holes do not find in low-luminous galaxies with comparatively small halos. The tight correlation between  $M_{bh}$and the rotational velocity obtained in some papers is likely to be the result of using indirect estimates of mass and velocities based on their empirical dependence on the velocity dispersion. In the first place, this relates to the empirical correlation between  $M_{bh}$ and the central velocity dispersion \cite{2002ApJ...578...90F...Ferrarese-2002}, which proves to be especially strong for elliptical galaxies. Direct measurements of  $V_c$ and $M_{bh}$, however, showed that the relation between these quantities is rather loose (see, for example,  \cite{2015ApJ...803....5S...Sabra-2015,2013ARep...57..797Z...Zasov-2013}, and references therein). According to Kormendy and Bender, \cite{2011Natur.469..377K...Kormendy-2011}, the mass $M_{bh}$ in disk galaxies tightly correlates only with the bulge parameters, and the correlation with the velocity  $V_{max}$ arises only as a result of the relation between the circular rotation velocities of bulge and disk (the bulge-disk conspiracy). In addition, as noted in the review by Kormendy and Ho \cite{2013ARA&A..51..511Kormendy-Ho-Supermassive-BH}, if the central black hole mass were tightly related to the dark halo mass, the very massive black holes should be found in the central galaxies of the clusters, since the clusters have halos with masses several orders of magnitudes as heavy as around separate galaxies.

When matching the central black hole masses with DM mass, it should be borne in mind that the virial mass estimate from the galaxy rotation velocity is not fully correct: the directly measured disk rotation velocities could differ from the circular velocity at the virial radius, and therefore they do not uniquely assess the halo mass $M_{vir}$. The authors  \cite{2009ApJ...704.1135B...Bandara-2009}, by using the galactic masses as derived from strong gravitational lensing for 43 galaxies (mainly of early type), found the relation between  $M_{bh}$ and the virial halo mass  $M_{vir}$ , $\log(M_{bh}) = (8.18 \pm 0.11) + (1.55 \pm 0.31)(\log(M_{vir})-13.0)$, but in paper  indirect estimates as inferred from the empirical $M_{bh}$, а оценки, полученные из эмпирической зависимости <<$M_{bh}$ --- central velocity dispersion were utilized. Later Bohdan and Goulding  \cite{2015ApJ...800..124B...Bogdan-2015} obtained the similar relation: $M_{bh}\sim M_{vir}^{1.8^{+0.7}_{-0.8}}$ for more than three thousand elliptical galaxies, but they also used empirical relations and not direct estimates: $M_{bh}$ was derived from stellar velocity dispersion and $M_{vir}$ --- from the dependence of this mass on the galactic X-ray luminosity  $L_X\sim M_{vir}^{2.4}$, found in paper \cite{2006ApJ...652L..17M...Mathews-2006}.

In any case, the SMBH formation is a complex process, and whatsoever role DM played there, the resulting SMBH mass should significantly depend on the evolution of baryonic components of galaxies, which is different for high-mass and low-mass galaxies (see review  \cite{2013ARA&A..51..511Kormendy-Ho-Supermassive-BH}).

The physical connection between dark halo and SMBH masses also emerges in numerical or theoretical models of the SMBH mass growth (see, for example, \cite{2010MNRAS.405L...1B...Booth-2010}, \cite{2003ApJ...593...56D...DiMatteo-2003}). Here, as follows from theoretical model calculations, the SMBH mass is determined not only by the dark halo mass, but to a significantly degree by its central density \cite{2010MNRAS.405L...1B...Booth-2010}.

In general, the problem of the dark halo effect on the black hole mass growth in galactic nuclei is far from being solved. The relation between the halo and SMBH masses, which is apparently weak at present, could be much stronger at the epoch of galactic youth, when the central black holes had been forming \cite{2011ApJ...737...50V...Volonteri-2011}.

\section{Conclusion. Dark matter and processes in galaxies}

The main purpose of the present review has been to show that the conclusion on the presence of dark, directly unobservable, matter is based on many facts and therefore is sufficiently reliable, in spite of DM nature being unclear. The role of dark halos in the evolution of galactic disks is one of the hot topics in the physics of galaxies.

To conclude, we stress that the most important (but not unique) parameter of a dark halo is its mass. All DM mass estimates are reduced to determining the total mass (density) – local or inside certain radius – and to subtracting the observable matter (stars and gas) from the obtained mass (density). Remind the main methods of DM mass estimates in galaxies (without estimates from empirical relations) discussed above:

\begin{itemize}
\item

 The construction of dynamical galactic models based on measurements of rotation curves and stellar velocity dispersion with account for photometrical data to assess the stellar mass fraction in the integral mass and radio data in HI and CO lines to assess the gas mass.

\item

 The use of data on radial velocities and velocity dispersion of halo objects (halo stars and globular clusters, dwarf satellites, tidal stellar streams), as well as hot halo gas temperature measurements. Here the virial relationship between the kinetic and gravitational energy of the component is assumed.

 \item

 The use of the condition of the disk gravitational stability relative to perturbations in the disk plane and of the disk stability to warp perturbations.

\item

 The construction of galactic polar ring models, as well as of models of tidal formations of interacting galaxies based on measurements of the velocity and structure of the tidal formations.

\item

The application of methods of strong and weak gravitational lensing to estimate the mass of the lensing galaxy.

\item

The construction of equilibrium models of stellar disks and cold gas layers in the disk plane with account for data on the rotational velocity and velocity dispersion of stars and gas.

\item

The use of the density waver propagation condition, as well as of gas-dynamic calculations of gas crossing the spiral arm, to estimate the disk density  \cite{2003ApJ...586..143K...Kranz2003, Athanassoula1987}.

 \end{itemize}

The DM dominance over ordinary matter (consisting of atoms) on large spatial scales follows from analysis of dynamics of gas and stars in both galaxies and galaxy systems. It is supported by gravitational lensing methods and is consistent with the cosmological concept of large-scale structure formation, which started more than 13 bln years ago. The total density of stars and gas in galaxies being much lower than the baryonic matter density suggested by the standard cosmological Big Bang model implies that certain fraction of dark matter must be comprised in the missing baryons (apparently, in hot gas), however most dark matter must be of a non-baryonic origin (see Section 1).

The natural question arises as to how the existence of DM affects processes inside galaxies? Some aspects of this issue were discussed above. Here, to conclude, we briefly consider the possible channels of DM influence (more precisely, of its gravity) on processes inside massive halos.

In the inner parts of most galaxies with normal brightness, both disks and ellipticals, the baryonic matter dominates, and only a small fraction of matter can be related to DM. Therefore, DM in the inner parts has no effect on both stellar and gas motions, star formation and stellar evolution processes, and activity of galactic nuclei  \footnote{Note, however, that one possible DM candidate particles are the so-called asymmetric weakly interacting particles (the symmetry is in unequal amount of particles and anti-particles), which, like WIMPs, have a very small interaction cross-section with baryons. A tiny fraction of these particles can, by being scattered when flying through stars, lose the kinetic energy and to store during a long time in the very center of stars thus affecting the process of energy generation and transport (see  \cite{2014MPLA...2940001C} and references therein).}. Nevertheless, the DM mass and distribution can affect a number of processes determining the observable properties of galaxies and their evolution. The most obvious features are as follows.

\begin{itemize}

\item[1.] A massive dark halo creates a deep potential well in which a galaxy resides. In the case of elliptical galaxies this leads to a larger velocity dispersion of stars and to the possibility of retaining the observed hot gas inside the galaxy. For disk galaxies this, first of all, means higher linear and angular rotational velocities of stars and gas, which make the disk more stable to gravitational perturbations. If the stellar disk is close to the marginal stability, the rapid rotation corresponds to the lowest velocity dispersion of stars, which results in  the smallest vertical scale of the stellar disk (see Section 2). In turn, this renders the gas layer in the disk plane more compressed and dense. With the same surface density, the mean volume density of gas turns out to be higher than in the absence of halo, which affects the local star formation rate (see the discussion of the relation of the star formation rate with gas volume density in  \cite{2012AstL...38..755A}).

\item[2.] At the galaxy formation stage, the DM concentration to the center of the system could help rapid SMBH growth. Accretion onto SMBH is responsible for the phenomenon of active galactic nucleus or quasar. A powerful energy release from the active nucleus can affect the total gas content and subsequent evolution of the galaxy. The degree of the dark halo concentration also influences on the disk formation from baryonic matter which initially, before the energy dissipation, was mixed with DM. An anomalously low central density and very large radial scale of dark halo could result in the formation of very rarefied disks. Such a scenario was proposed, for example, for the giant galaxy Malin-2 in paper  \cite{2014MNRAS.437.3072K...Kasparova-2014}.

\item[3.] The halo can play an important role by determining the formation and maintaining large-scale structures in the disk (bar, spiral arms). A non-axially symmetric outer halo, apparently, is responsible for generation of long-lived density waves in the disk thus explaining the regular spiral structure even in galaxies where the disk has a large margin of gravitational stability (see Section 3).  This is not only mechanism of maintaining long-lived spiral arms, but it could be quite a widespread. If the mass of a disk is much smaller than that of a halo, in whose gravitational field it resides (for example, in LSB-galaxies), the halo not only stabilizes the growth of local gravitational perturbations due to rapid rotation, but also prevents the formation of small-scale wave structure by suppressing the swing amplification mechanism of wave growth \cite{2014MNRAS.439..929G}. The massive halo played a big role during the formation of massive (globular) clusters  in young forming galaxies. This is suggested by the correlation between the number or the total mass of globular clusters in a galaxy and the mass of its dark halo (see \cite{2015ApJ...806...36Harris-DM-Halos} and references therein).

\item[4.] The presence of massive halos increases the relative velocity of galaxies  in systems (pairs, groups, clusters), which reduces effects of tidal interactions between them that distort the form of the galaxies and velocity field of gas and stars. The allowance for the halo mass can cardinally change estimates of trajectories of galaxies which are close to each other, of their orbital periods and collision frequency. A clear example is the nearby system of apparently gravitationally bound galaxies: Large and Small Magellanic Clouds. In these galaxies, proper motions of stars are measured, so known are not only radial (along the line of sight), as usually, but also three-dimensional velocity vectors. This offered the unique possibility to estimate possible masses of their extended halos, which were found to be significantly exceeding the total mass of stars and gas, and also to precise their orbits and evolution ~\cite{2015arXiv151103346B...Besla-15}. Dark matter filling a galaxy cluster, over long time intervals can ‘pull over’ galaxies in the inner parts of the cluster by dynamical friction  helping their higher central concentration and formation of central cD-galaxies \cite{2010ARep...54..704Dremova-dynamical-evolution}, although the efficiency of this process remains an open issue.

\item[5.] A massive halo facilitates the merging or destruction of satellites flying into the halo due to dynamical friction and tidal interactions. The dynamical friction shortens large semi-axis of orbits of the most massive dwarfs, and tidal forces partially or fully destroy them thus favoring additional DM and baryonic (including gas)  mass inflow into the inner regions of the ‘host’ galaxy. Extended stellar streams related to destroying dwarf systems are definitely observed in our and other galaxies, and velocity of stars in these streams reflect the mass and density of halo in which they are found. Dwarf systems merged with the galaxy supply low-metallicity gas thus effecting the chemical evolution of the galaxy and its gas content.

\item[6.]	Like dwarf galaxies, similarly important can be numerous subhalos (minihalos), i.e. low-mass, including starless, formations partially or fully consisting of non-baryonic DM and therefore unobservable or difficult discoverable in the optical. Their number remains an open issue. The observed dwarf satellites of galaxies are frequently considered to be low-mas subhalos, in which some amount of stars could form. However, the number of subhalos predicted by model calculations many times exceed the directly observed number of dwarf satellites with similar values of directly measured circular velocities (see Section 5).

Dark subhalos, if present in a sufficient amount, can differently manifest themselves in the ‘host’ galaxy. Dark subhalos, for example, cam significantly increase the velocity dispersion of stars   $c_z$, especially at the disk periphery, can excite long-lived vertical oscillations in the disk or even stimulate formation of large-scale structures in the disk (see the discussion in paper   \cite{2015MNRAS.450..266W...Widrow-2015}). By crossing the disk gas layer, the subhalos, if they contain at least a few percent gas mass, can be responsible for the appearance of low-density ‘holes’ in the gas distribution and enhanced star formation at their edges, which is confirmed by hydrodynamic simulations \cite{2012ApJ...746...10K...Kannan-2012}. The authors note that similar formation are indeed observed in galaxies, but their number is too high for the subhalos to be the only reason for their formation. Unfortunately, the role of subhalo is difficult to assess with certainty, since their number and space density remain unknown.

\item[7.] The dark halo mass and its central density determines the character of gas exchange between the galaxy and its surroundings. Apparently, this is the most important effect of halos in the galaxy evolution. A deep potential well caused by halo hampers gas ejection (or the galactic wind) from the disk into the ambient medium, which results from the activity of young stars or nuclear activity of the galaxy. By leaving the galaxy, the gas can remain within the halo virial radius and by cooling can partially return to  the galaxy to continue participating in the star formation process. On the other hand, the massive halo favors accretion of the intergalactic gas, although the gaseous medium filling the halo hinders the accreting gas falling in the central halo part, i.e. into the optically observed galaxy.

 The directly observed gas accretion rate onto our and other galaxies proves to be lower than required to sustain star formation, however the accretion rate is difficult to estimate: most of the infalling gas is rarefied and ionized, which makes its discovery at large distances from the disk a difficult task. Nevertheless, there are a lot of arguments suggesting that gas accretion onto our and other massive galaxies indeed occurs, and the accretion rate is comparable to the star formation rate, especially several billion years ago, and largely determines evolution of the galaxies and their gas components (see, for example, \cite{2014ASPC..480..211C...Combes-2014, 2013MNRAS.435..999D...Dekel-2013, 2014A&ARv..22...71S...Sanchez-2014}).

The galactic halos accrete gas (or accreted gas in the past) from the intergalactic space. The accretion flows must contain both gas and DM. However, while DM is collisionless and mixes with dark halo DM, the fate of gas is not so clear. The infalling gas streams can directly impinge the galactic disk as gas clouds or individual flows, but the situation is also possible when the infalling gas enters the galaxy with a time lag as long as one billion year and more, depending on the depth of the halo potential well.

Theoretical calculations and numerical models of galaxy formation imply that accretion onto the disk must occur differently in galaxies with low-mass and high-mass halos (see, for example, model calculations by van de Voort et al.  \cite{2011MNRAS.414.2458V...Voort-2011}, or review  \cite{2014A&ARv..22...71S...Sanchez-2014} and reference therein). In massive halos, accreting gas heats up to temperatures close to virial values (of order of $10^6~K$). The accreting gas flows in the halo pass through the shock front and dissolve in the hot gas. In a star-forming galaxy, gas falls from the inner halo regions after cooling. This is the so-called hot accretion mode. Most of the hot gas has the characteristic cooling time exceeding the Hubble age of the Universe, i.e. cannot fall into the galaxy disk at all.

 In low-mass galaxies the gas temperature in the halo is lower and the cooling time is shorter than in massive galaxies.  Therefore, flows of relatively cold and low-metal intergalactic gas by entering the halo do not heat up to the virial temperatures. They fragment and fall in the rarefied halo gas by reaching its central parts. Gas accretion rates onto the galaxy in this case are determined not by its cooling rate but the accretion from the intergalactic space. This regime is called the cold accretion  mode. Both accretion modes can be present in galaxies, and their roles can interchange with time. The cold mode is more efficient, and presently it dominates in galaxies, only if their halo mass is not too high (not significantly above our Galaxy halo), although at large redshifts this accretion apparently dominated in the mass growth of galaxies. In principle, both accretion modes can co-exist.

 The cold accretion regime can explain the presence of gas and star-forming regions sometimes observed in lenticular galaxies. Usually, such galaxies have insignificant amount of cold gas, in distinction from other disk galaxies, therefore gas in some of them is likely to be of external origin. However, there is concurrent with accretion mechanisms, the galaxy merging or supply of chemically processed gas from nearby galaxies. Measurements of kinematic features and metallicity of the acquired by the galaxy enables the distinction of the most probable gas supply mechanism (see the discussion in \cite{2014MNRAS.439..334Ilyina-Silchenko-Afanasiev}). From the analysis of isolated lenticular (S0) galaxies Katkov et al. \cite{2015AJ....150...24Katkov-Kniazev-Silchenko} concluded that the morphological type and observed gas content of disk galaxies that have no comparable-mass neighbors can be fully determined by a significant cold gas accretion or by merging of small gas-rich satellites.  Clearly, in the case of accretion, as well as during merging, the gas supply into the galaxy and its passing through halo to the disk is fully controlled by the dark halo gravitational field.

 Tight interaction between nearby galaxies, as well as blowing of galaxies by intergalactic gas flows, strongly complicates the gas exchange between the galaxy with the surroundings. These processes affect the mass and physical state of the gas filling its halo, and thus influence on the gas supply and evolution in galactic disks. If accretion onto a galaxy stops for some reason, the lack of ‘fresh’ gas inflow quenches star formation in several billion years.

\end{itemize}
Thus, the galaxy evolution is determined not only by the mass and structure of galaxies but also by the degree of concentration and shape of their dark halos. It is very important to quantitatively explain (which is the task for not a near future) how dark matter affected processes in galaxies at different redshifts and in different surroundings – from isolated galaxies to galaxies is close groups and rich clusters – and what was the role of DM in the formation of galaxies of diverse morphological types.

The review made use of the results of numerical simulations carried out on the ‘Lomonosov’ and ‘Chebyshev’ supercomputers of the Research Computing Center of the M.V. Lomonosov Moscow State University. The work was supported by the Russian Foundation for Basic Research (grants 14-22-03006, 15-02-06204, $NNIO_a$ 15-52-12387, 15-52-15050, 16-02-00649, 16-32-60043), as well as by the grants of the President of the Russian Federation (MK-4536.2015.2), Russian Science Foundation (15-12-1007) (Section 6.2 of the review) and the State Task of the Ministry of Education and Science of RF  (2.852.2017/4.6).

\bibliography{references}

\end{document}